%% file: main.tex
\documentclass[prd, 11pt, %twocolumn,
showpacs, floatfix, nofootinbib
]{revtex4-2}
\usepackage[utf8]{inputenc}
\usepackage{amsmath,amsfonts,mathtools}
\usepackage{amssymb}
\usepackage{epsfig}
\usepackage{graphicx}
\usepackage{color}
\usepackage{hyperref}
\hypersetup{colorlinks=true,linkcolor=blue,citecolor=blue,urlcolor=black}
\usepackage{comment}
\usepackage{cancel}
\usepackage{bbold}
\usepackage{booktabs}
\usepackage{vwcol} % for multiple columnsverbatim
\usepackage[normalem]{ulem}
\usepackage[dvipsnames]{xcolor}
\usepackage{pbox}
\usepackage{multirow,bigdelim} % for braces in tables
\usepackage[dvipsnames]{xcolor}

\definecolor{plotGreen}{rgb}{0,0.666667,0}
\definecolor{plotBlue}{rgb}{0,0,1}
\definecolor{plotMagenta}{rgb}{1,0,1}
\definecolor{plotRed}{rgb}{1,0,0}

\DeclareMathOperator{\sign}{sign}

\newcommand{\SU}{\mathrm{SU}}

\newcommand{\LR}{$\mathrm{SU}(3)_c \times \protect\linebreak[0]\mathrm{SU}(2)_L \times \protect\linebreak[0]\mathrm{SU}(2)_R \times \protect\linebreak[0]\mathrm{U}(1)_{B-L} \ $}

\newcommand*\xbar[1]{\hspace{0.1em}%
	\hbox{%
		\vbox{%
			\hrule height 0.5pt % The actual bar
			\kern0.4ex%         % Distance between bar and symbol
			\hbox{%
				\kern-0.1em%      % Shortening on the left side
				\ensuremath{#1}%
				\kern-0.1em%      % Shortening on the right side
			}%
		}%
	}\hspace{0.1em}%
}

\def\slog{\mathrm{lg}\,}

\def\textdef{\textit}

\begin{document}

%\numberwithin{equation}{section}
%\tableofcontents
%\newpage
%----------------------------------------------------------------------------------
\title{The trouble with the minimal renormalizable $\mathrm{SO}(10)$ GUT}
\preprint{}
\pacs{}
%----------------------------------------------------------------------------------
\author{Kate\v{r}ina Jarkovsk\'a}\email{jarkovska@ipnp.mff.cuni.cz}
\affiliation{Institute of Particle and Nuclear Physics,
Faculty of Mathematics and Physics,
Charles University in Prague, V Hole\v{s}ovi\v{c}k\'ach 2,
180 00 Praha 8, Czech Republic}
\author{Michal Malinsk\'{y}}\email{malinsky@ipnp.mff.cuni.cz}
\affiliation{Institute of Particle and Nuclear Physics,
Faculty of Mathematics and Physics,
Charles University in Prague, V Hole\v{s}ovi\v{c}k\'ach 2,
180 00 Praha 8, Czech Republic}
\author{Vasja Susi\v c}\email{susic@ipnp.mff.cuni.cz}
\affiliation{Institute of Particle and Nuclear Physics,
Faculty of Mathematics and Physics,
Charles University in Prague, V Hole\v{s}ovi\v{c}k\'ach 2,
180 00 Praha 8, Czech Republic}
%----------------------------------------------------------------------------------

%%%%%%%%%%%%%%%%%%%%%%%%%%%%%%%%%%%%%%%%%%%%%%%%%%%%%%%
% ABSTRACT
%%%%%%%%%%%%%%%%%%%%%%%%%%%%%%%%%%%%%%%%%%%%%%%%%%%%%%%

\begin{abstract}

We scrutinize the physical viability of the minimal non-supersymmetric $\mathrm{SO}(10)$ GUT 
with the scalar sector $\mathbf{45}\oplus\mathbf{126}\oplus\mathbf{10}_{\mathbb{C}}$,
in which the unified symmetry is broken by the former two representations, and a realistic 
Yukawa sector is supported by the last two. Alongside the known issue of a relatively low GUT scale (and thus overly fast proton decay) encountered in minimally fine-tuned scenarios, we identify a very general problem of the model: the inability to properly accommodate a Standard-Model-like low-energy Higgs doublet in the perturbative regime.  
\end{abstract}
\maketitle
%\tableofcontents
%\newpage

%%%%%%%%%%%%%%%%%%%%%%%%%%
\section{Introduction}
%%%%%%%%%%%%%%%%%%%%%%%%%%
The renormalizable $\mathrm{SO}(10)$ Grand Unified Theory (GUT) in which the unified gauge symmetry breaking is triggered by the 45-dimensional adjoint scalar, followed by the rank reduction imposed by a 126-dimensional 5-index antisymmetric scalar representation, has recently received quite some attention~\cite{Bertolini:2009es,Bertolini:2012im,Bertolini:2013vta,Kolesova:2014mfa,Brennan:2015psa,Babu:2016bmy,Babu:2016cri,Graf:2016znk,Schwichtenberg:2018cka,DiLuzio:2020qio,Ohlsson:2020rjc,King:2021gmj,Jarkovska:2021jvw} as a candidate for a minimal potentially realistic and {\em calculable} grand-unified model of possible baryon-number violating (BNV) phenomena like proton decay. 

What is meant by calculability here is namely the capacity of the model to admit relatively good estimates of the relevant proton decay widths which could be (at least in principle) comparable with the data from the new experimental facilities with detection masses in the hundreds of kiloton range (like Hyper-K~\cite{Abe:2011ts,Hyper-Kamiokande:2018ofw} and DUNE~\cite{DUNE:2015lol,DUNE:2016hlj}) that will get on-line by the end of this decade.   
The defining feature here is the robustness of the framework (and the theoretical calculations within) to different types of theoretical uncertainties emanating from several basic sources. These include the limited information about the relevant hadronic matrix elements, little grip on the flavour structure of the charged-current interactions involved, the proximity of the GUT and Planck scale, etc., see e.g.~\cite{Aoki:2017puj,Kolesova:2016ibq}. 

While the first two of the listed issues can be at least partly avoided by improving the corresponding QCD calculations (first) or focusing on flavour-blind observables (second), there is no simple way to avoid the emergence of uncertainties associated to Planck-scale-induced $d\geq 5$ operators. The most dangerous are those operators which smear the mass estimates of the GUT-scale vectors and scalars that mediate BNV processes. The prominent culprit among these is the $d=5$ structure $F_{\mu\nu}F^{\mu\nu}\Phi/\Lambda$ (with $\Phi$ denoting a GUT-breaking scalar), which inflicts out-of-control shifts in the GUT-scale matching conditions and hence in estimates of mediator masses~\cite{Hill:1983xh,Shafi:1983gz,Calmet:2008df}.   
Remarkably, the minimal renormalizable $\mathrm{SO}(10)$ GUT advocated above is free from these concerns at leading order, owing to the general absence of the corresponding $d=5$ cubic adjoint invariant. However, this beautiful feature is ``compensated'' by the need to consider the model at the fully quantum level, since it has no consistent classical counterpart~\cite{Buccella:1980qb,Yasue:1980fy,Anastaze:1983zk,Babu:1984mz}. 

In general, the relative rigidity of the ${\bf 45}\oplus {\bf 126}$ Higgs sector admits only two minimally fine-tuned~\cite{delAguila:1980qag,Mohapatra:1982aq} $\mathrm{SO}(10)$ symmetry breaking patterns potentially compatible with unification and proton-lifetime constraints~\cite{Bertolini:2009qj}. They feature either a $\mathrm{SU}(4)_C\times \mathrm{SU}(2)_L\times \mathrm{U}(1)_R$ or $\mathrm{SU}(3)_c\times \mathrm{SU}(2)_L\times \mathrm{SU}(2)_R\times \mathrm{U}(1)_{B-L}$ intermediate stage.  However, a recent detailed analysis~\cite{Jarkovska:2021jvw} revealed that only the former variant can be reasonably implemented as a consistent perturbative scenario, and even then only in a very narrow domain of the parameter space.
Moreover, the corresponding prediction for the GUT scale falls well below $10^{15}\,\mathrm{GeV}$, and thus the concerns about proton lifetime are reiterated.

Nevertheless, even with such a strong observation at hand, there are still several routes to explore before proclaiming this scenario completely unphysical. (i) First, the unification constraints in~\cite{Jarkovska:2021jvw} have been implemented at one-loop level and a dedicated higher-loop analysis may in principle change them. It is known that this is unlikely to happen in the class of minimally fine-tuned scenarios~\cite{Gipson:1984aj,Deshpande:1992em,Bertolini:2009qj}, but this option cannot be entirely ignored if the extended survival hypothesis~\cite{Dimopoulos:1984ha} was not adhered to and some of the naturally heavy scalar multiplets were brought well into the desert~\cite{Bertolini:2013vta,Kolesova:2014mfa}.  (ii) Second, the results of the one-loop analysis of the ${\bf 45}\oplus {\bf 126}$ Higgs model in~\cite{Jarkovska:2021jvw} cannot be passed on when it comes to physics, as there is a need for at least one more scalar representation (typically, a $\mathbf{10}$ of $\mathrm{SO}(10)$) to accommodate a realistic Yukawa sector. Along with that, a new source of quantum effects that were not taken into account in~\cite{Jarkovska:2021jvw} emerges from the $\mathbf{10}$.

Needless to say, due to the complexity of computing the one-loop spectrum even in the $\mathbf{45}\oplus\mathbf{126}$ simplified model~\cite{Graf:2016znk,Jarkovska:2021jvw}, a general analysis of all the options under point (i) above (i.e., all possible combinations of accidentally light thresholds) would be extremely tedious and unlikely to bring any decisive result. However, as we argue in this paper, there seems to be a deeper and universal issue plaguing even the fully physical $\mathbf{45}\oplus \mathbf{126}\oplus \mathbf{10}_\mathbb{C}$ variant of the model when it comes to its low-energy effective structure, namely, there is a {\em strong tension between the need to attain a light SM Higgs-like doublet and the perturbativity of the setting}  irrespective of what one does with the rest of the scalar sector. 

The study is organized as follows:
In Section~\ref{sect:themodel}, we recapitulate the salient features of the minimal potentially realistic Higgs sector containing $\mathbf{45}\oplus \mathbf{126}\oplus \mathbf{10}_\mathbb{C}$, where the first two representations are involved in the symmetry breaking of GUT- to SM-symmetry, while the complex scalar $\mathbf{10}_\mathbb{C}$ together with the $\mathbf{126}$ admits a realistic Yukawa sector. The quantitative conditions for this to work are formulated in Sect.~\ref{sec:Yukawa-sector}, and then scrutinized at tree-level analytically in Section~\ref{sec:tree-level-fine-tuning} and at one-loop level numerically in Section~\ref{sec:one-loop-fine-tuning}. We conclude in Section~\ref{sec:conclusions-and-outlook}. The technical details of the analysis are deferred to a set of Appendices.

%%%%%%%%%%%%%%%%%%%%%%%%%%
\section{The $\mathbf{45} \oplus \mathbf{126} \oplus \mathbf{10}$ model\label{sect:themodel}}
%%%%%%%%%%%%%%%%%%%%%%%%%%

In what follows we consider the $\mathrm{SO}(10)$ GUT model with the $\mathbf{45} \oplus \mathbf{126} \oplus \mathbf{10}_\mathbb{C}$ scalar sector, which is a minimal extension of the Higgs model from \cite{Jarkovska:2021jvw} by a complex $\mathbf{10}_{\mathbb{C}}$, thus potentially allowing for realistic Yukawa matrices. We stress that the extra $\mathbf{10}$ of $\mathrm{SO}(10)$ must be complex (equivalent to two copies of a real $\mathbf{10}$) --- a point we discuss in more detail later in Sect.~\ref{sec:Yukawa-sector}. In addition, the three generations of the Standard Model fermions (including a right-handed neutrino in each family) are embedded into three copies of the spinorial $\mathbf{16}_F$ in the usual way \cite{Langacker:1980js}, while the gauge fields are present in the adjoint $\mathbf{45}_G$. The model under consideration is essentially the minimal renormalizable potentially realistic $\mathrm{SO}(10)$ GUT model.

For future reference, we list the scalar field content of this model in terms of SM irreducible representations in Table~\ref{tab:particle-content}. Each SM representation may be present with a multiplicity $\#$ greater than one, and we specify the origin of each copy in terms of $\mathrm{SO}(10)$ irreducible representations of scalars labeled as
\begin{align}
    \mathbf{45}&\sim \phi_{ij},&
    \mathbf{126}&\sim \Sigma_{ijklm},&
    \mathbf{10}_\mathbb{C}&\sim H_{i}. \label{SO10scalars}
\end{align}
In Eq.~(\ref{SO10scalars}) we wrote the $\mathbf{45}$ as a real anti-symmetric matrix, the $\mathbf{126}$ as a complex self-dual anti-symmetric $5$-index tensor, and the $\mathbf{10}_\mathbb{C}$ as a vector with complex components. All Latin indices run from $1$ to $10$, and are assumed to refer to the real basis of the defining representation $\mathbf{10}$, cf.~Appendix~A of \cite{Antusch:2019avd} for more details.
The complex conjugates of $\Sigma$ and $H$ are denoted by $\Sigma^\ast$ and $H^\ast$, respectively. 

Let us note that the main difference between  Table~\ref{tab:particle-content} here and an analogous table in~\cite{Jarkovska:2021jvw} is the presence of the complex $\mathbf{10}_\mathbb{C}$, which adds 2 scalars transforming as $(1,2,+\tfrac{1}{2})$ of the SM and 2 copies of $(\bar{3},1,+\tfrac{1}{3})$. From now on, we shall refer to the representations of this type as SM \textit{doublets} and \textit{triplets}.

\begin{table}[htb]
	\caption{The field content of the scalar sector $\mathbf{45}\oplus\mathbf{126}\oplus \mathbf{10}_\mathbb{C}$ of the $\mathrm{SO}(10)$ model under consideration in terms of the corresponding SM components. For each SM representation $R$ its reality/complexity ($\mathbb{R}/\mathbb{C}$), multiplicity $\#$ and $\mathrm{SO}(10)$ origin are indicated. The ``WGB'' column specifies the number of massless copies corresponding to would-be Goldstone modes when $\mathrm{SO}(10)$ is completely broken down to $G_{321}\equiv\SU(3)_c\times SU(2)_L\times U(1)_Y$.	\label{tab:particle-content}}
	\renewcommand{\arraystretch}{1.2}
	\vspace{0.1cm}
    \centering
	\begin{tabular}{l@{\qquad}c@{\quad}c@{\quad}c@{\qquad}l}
		\toprule
		\makebox[1.5cm][l]{$R\sim G_{321}$}
		&$\mathbb{R}/\mathbb{C}$
		&\makebox[0.5cm][r]{$\#$}
		& WGB
		&$\subseteq \mathrm{SO}(10)$\\
		\midrule
		$(1,1,0)$			&  $\mathbb{R}$  &  $4$ & $1$  &  $\phi$, $\phi$, $\Sigma$, $\Sigma^{\ast}$ \\ 
		$(1,1,+1)$			&  $\mathbb{C}$  &  $2$ & $1$ &$\phi$, $\Sigma$ \\
		$(1,1,+2)$			&  $\mathbb{C}$  &  $1$  	&  $0$  & $\Sigma$ \\
		$(1,2,+\tfrac{1}{2})$		&  $\mathbb{C}$  &  $4$ & $0$ 	&  $\Sigma$, $\Sigma^\ast$, $H$, $H^\ast$ \\
		$(1,3,-1)$			&  $\mathbb{C}$  &  $1$  	& $0$  & $\Sigma$ \\
		$(1,3,0)$			&  $\mathbb{R}$  &  $1$  	&  $0$  & $\phi$ \\
		$(3,2,-\tfrac{5}{6})$		&  $\mathbb{C}$  &  $1$ & $1$  & $\phi$ \\
		$(3,2,+\tfrac{1}{6})$		&  $\mathbb{C}$  &  $3$ & $1$& $\phi$, $\Sigma$, $\Sigma^\ast$ \\
		$(3,2,+\frac{7}{6})$		&  $\mathbb{C}$  &  $2$  	& $0$ &  $\Sigma$, $\Sigma^\ast$ \\
		$(3,3,-\tfrac{1}{3})$		&  $\mathbb{C}$  &  $1$  	& $0$  & $\Sigma$ \\
		$(\bar{3},1,-\frac{2}{3})$	&  $\mathbb{C}$  &  $2$ & $1$ &  $\phi$, $\Sigma$ \\
		$(\bar{3},1,+\tfrac{1}{3})$	&  $\mathbb{C}$  &  $5$  	& $0$  & $\Sigma$, $\Sigma$, $\Sigma^\ast$, $H$, $H^\ast$ \\
		$(\bar{3},1,+\tfrac{4}{3})$	&  $\mathbb{C}$  &  $1$  	&  $0$ & $\Sigma$ \\
		$(6,3,+\tfrac{1}{3})$		&  $\mathbb{C}$  &  $1$  	& $0$  & $\Sigma$ \\
		$(\bar{6},1,-\tfrac{4}{3})$	&  $\mathbb{C}$  &  $1$  	&  $0$  & $\Sigma$ \\
		$(\bar{6},1,-\tfrac{1}{3})$	&  $\mathbb{C}$  &  $1$  	&  $0$  &  $\Sigma$ \\
		$(\bar{6},1,+\tfrac{2}{3})$	&  $\mathbb{C}$  &  $1$  	& $0$    & $\Sigma$ \\
		$(8,1,0)$			&  $\mathbb{R}$  &  $1$  	& $0$  &  $\phi$ \\
		$(8,2,+\frac{1}{2})$		&  $\mathbb{C}$  &  $2$ & $0$	&  $\Sigma$, $\Sigma^\ast$ \\
		\bottomrule
	\end{tabular}
\end{table}
%	
	
%%%%%%%%%%%%%%%%%%%%%%%%%%
\subsection{The scalar potential and $\mathrm{SO}(10)$ breaking}
%%%%%%%%%%%%%%%%%%%%%%%%%%

%%%%%%%%%%%%%%%%%%%%%%%%%%
\subsubsection{The potential}
%%%%%%%%%%%%%%%%%%%%%%%%%%
The tree-level scalar potential of the model in the unbroken phase can be written sector by sector as
\begin{align}
V_0(\phi,\Sigma,\Sigma^\ast,H,H^\ast)=V_{45}(\phi)+V_{126}(\Sigma,\Sigma^\ast)+V_{\textrm{mix}}(\phi,\Sigma,\Sigma^\ast)
+\tilde{V}_{10}(H,H^{\ast})+\tilde{V}_{\mathrm{mix}}(\phi,\Sigma,\Sigma^\ast,H,H^\ast),\label{eq:scalar-potential-0}
\end{align}
where 
\begingroup
\allowdisplaybreaks
\begin{align}
    %%%%%%%%%%%%%%%%%%
    \begin{split}
    V_{45}&=-\frac{\mu^2}{4}(\phi\phi)_0+\frac{a_0}{4}\,(\phi\phi)_0(\phi\phi)_0+\frac{a_2}{4}(\phi\phi)_2(\phi\phi)_2,\label{eq:scalar-potential-V45}\\
    \end{split}\\
    %%%%%%%%%%%%%%%%%%
    \begin{split}
    V_{126}&=-\frac{\nu^2}{5!}(\Sigma\Sigma^\ast)_0+\frac{\lambda_0}{(5!)^2}\,(\Sigma\Sigma^\ast)_0(\Sigma\Sigma^\ast)_0 +\frac{\lambda_2}{(4!)^2}(\Sigma\Sigma^\ast)_2 (\Sigma\Sigma^\ast)_2 +  \\
    &\quad + \frac{\lambda_4}{(3!)^2(2!)^2}(\Sigma\Sigma^\ast)_4(\Sigma\Sigma^\ast)_4 +\frac{\lambda_4'}{(3!)^2}(\Sigma\Sigma^\ast)_{4'}(\Sigma\Sigma^\ast)_{4'} + \\
    &\quad + \frac{\eta_2}{(4!)^2}(\Sigma\Sigma)_2(\Sigma\Sigma)_2 + \frac{\eta_2^\ast}{(4!)^2}(\Sigma^\ast\Sigma^\ast)_2(\Sigma^\ast\Sigma^\ast)_2,\\
    \end{split}
    \label{eq:scalar-potential-V126}\\
    %%%%%%%%%%%%%%%%%%
    \begin{split}
    V_{\textrm{mix}}&=
	\frac{i\tau}{4!}(\phi)_2(\Sigma\Sigma^\ast)_2
	+\frac{\alpha}{2\cdot 5!}(\phi\phi)_0(\Sigma\Sigma^\ast)_0 
	+\frac{\beta_4}{4\cdot 3!}(\phi\phi)_4 (\Sigma\Sigma^\ast)_4 
	+\frac{\beta_4'}{3!}(\phi\phi)_{4'}(\Sigma\Sigma^\ast)_{4'} +\\
	&\quad + \frac{\gamma_2}{4!}(\phi\phi)_2 (\Sigma\Sigma)_2
	+\frac{\gamma_2^\ast}{4!}(\phi\phi)_2 (\Sigma^\ast\Sigma^\ast)_2,\\
    \end{split} \label{eq:scalar-potential-Vmix}
\\
%\end{align}
%\endgroup
%\begingroup
%\allowdisplaybreaks 
%\begin{align}
    %%%%%%%%%%%%%%%%%%
    \begin{split}
    \tilde{V}_{10}&=-\xi^2(H^\ast H)_0 - \xi'{}^{2}(HH)_0 - \xi'{}^{\ast 2}(H^\ast H^\ast)_0 + \\
	&\quad + h_{4}\,(HH)_0 (HH)_0 +h_{4}^\ast\,(H^\ast H^\ast)_0 (H^\ast H^\ast)_0  + h_{3}\,(HH)_{0} (HH^{\ast})_{0} +h_{3}^\ast\,(H^{\ast}H^{\ast})_{0} (H^{\ast}H)_{0} +\\
	&\quad +{h_{2}\,(H^\ast H)_0 (H^\ast H)_0 
	+h'_{2}\,(H^\ast H^\ast)_0 (H  H)_0},\\
    \end{split} \label{eq:scalar-potential-Vtilde10}\\
    %%%%%%%%%%%%%%%%%%
    \begin{split}
    \tilde{V}_{\textrm{mix}}&=
	\frac{\kappa_0}{2}\,(H^\ast H)_0(\phi\phi)_0
	+ \kappa_2\;(H^\ast H)_2(\phi\phi)_2 + 
	\frac{\kappa'_{0}}{2}\,(HH)_0(\phi\phi)_0
	+ \frac{\kappa'_{0}{}^{\ast}}{2}\,(H^{\ast}H^{\ast})_0(\phi\phi)_0 + \\
	&\quad + \kappa'_{2}\,(HH)_2(\phi\phi)_2
	+ \kappa'_{2}{}^{\ast}\,(H^{\ast}H^{\ast})_2(\phi\phi)_2 + \\
	&\quad + \frac{\zeta}{4}(\phi\phi)_4(H\Sigma)_4 
	+ \frac{\zeta^\ast}{4}(\phi\phi)_4(H^\ast \Sigma^\ast)_4 + \frac{\zeta'}{4}(\phi\phi)_4(H^\ast\Sigma)_4 
	+ \frac{\zeta'{}^\ast}{4}(\phi\phi)_4(H \Sigma^\ast)_4+ \\
	&\quad +\frac{\rho_{0}}{5!}(H^\ast H)_0(\Sigma\Sigma^\ast)_0
	+ \frac{\rho_{2}}{4!}(H^\ast H)_2(\Sigma\Sigma^\ast)_2  
	+ \frac{\rho'_{0}}{5!}(HH)_0(\Sigma\Sigma^\ast)_0
	+ \frac{\rho'_{0}{}^{\ast}}{5!}(H^{\ast}H^{\ast})_0(\Sigma\Sigma^\ast)_0 + \\
	&\quad +\frac{\psi_2}{4!}(HH)_2(\Sigma\Sigma)_2
	+\frac{\psi_2^\ast}{4!}(H^\ast H^\ast)_2(\Sigma^\ast\Sigma^\ast)_2 
	+ \frac{\psi_1}{4!}(HH^\ast)_2(\Sigma\Sigma)_2
	+\frac{\psi_1^\ast}{4!}(H^\ast H)_2(\Sigma^\ast\Sigma^\ast)_2 + \\
	&\quad +\frac{\psi_0}{4!}(H^\ast H^\ast)_2(\Sigma\Sigma)_2
	+\frac{\psi_0^\ast}{4!}(HH)_2(\Sigma^\ast\Sigma^\ast)_2  + i\tau' (\phi)_2(HH^\ast)_2 +\\
	&\quad + \frac{\varphi}{3!} (H \Sigma)_4(\Sigma\Sigma^\ast)_4 
	+ \frac{\varphi^\ast}{3!} (H^\ast \Sigma^\ast)_4(\Sigma\Sigma^\ast)_4
	+ \frac{\varphi'}{3!} (H^\ast \Sigma)_4(\Sigma\Sigma^\ast)_4 
	+ \frac{\varphi'{}^\ast}{3!} (H \Sigma^\ast)_4(\Sigma\Sigma^\ast)_4. \\
    \end{split} 
    \label{eq:scalar-potential-Vtildemix}
\end{align}
\endgroup
Note that the definitions of the ``old'' parts of the potential corresponding solely to the $\mathbf{126}\oplus \mathbf{45}$ sector are identical to those entertained in~\cite{Jarkovska:2021jvw}, while
the new parts involving the $\mathbf{10}_\mathbb{C}$ are denoted by a tilde.

To avoid any confusion, we explicitly wrote complex conjugate terms where applicable. The notation for contractions is a straightforward extension of the one already specified in \cite{Jarkovska:2021jvw}: the subscript of an object in parentheses indicates the number of uncontracted indices (placed symmetrically at the end of each factor), e.g.\footnote{Note that $4$-index objects have two possible contractions, namely,
$(X)_{4}(Y)_{4}\equiv X_{ijkl}Y_{ijkl}$ and 
    $(X)_{4'}(Y)_{4'}\equiv X_{ijkl} Y_{ikjl}$.
}
\begin{align}
    (H^{\ast}H)_{0}&\equiv H^{\ast}_{i} H_{i},&
    (H^{\ast}H)_{2}&\equiv H^{\ast}_{i} H_{j},&
    (H\Sigma)_{4}&\equiv H_{m}\Sigma_{mijkl},
\end{align}
\begin{align}
    (\Sigma\Sigma^\ast)_{0}&\equiv\Sigma_{ijklm}\Sigma^{\ast}_{ijklm},&
    (\Sigma\Sigma^\ast)_{2}&\equiv(\Sigma\Sigma^\ast)_{mn}=\Sigma_{ijklm}\Sigma^{\ast}_{ijkln},&
    (\Sigma\Sigma^\ast)_{4}&\equiv(\Sigma\Sigma^\ast)_{lmno}=\Sigma_{ijklm}\Sigma^{\ast}_{ijkno}.
\end{align}
As a cross-check for the completeness of the scalar potential, we confirmed that all invariants in Eqs.~(\ref{eq:scalar-potential-V45})-(\ref{eq:scalar-potential-Vtildemix}) are linearly independent, while their number matches the counting by the standard Hilbert series methods~\cite{Lehman:2015via,Henning:2017fpj} (assisted by the Mathematica package LieART 2.0 \cite{Feger:2012bs,Feger:2019tvk}, which provided the weight vectors for the $\mathrm{SO}(10)$ representations). 

In total, the list of dimensionless parameters in the scalar potential now consists of 15 real and 14 complex couplings, from which 6 and 12 are new compared to the Higgs model~\cite{Jarkovska:2021jvw}, respectively. For further convenience we list all couplings, both dimensionless and dimensionful, in Table~\ref{tab:potential-parameters}.

\begin{table}[htb]
\caption{The parameters of the scalar potential of Eq.~(\ref{eq:scalar-potential-0}) arranged according to their novelty (i.e., whether they have already been present in~\cite{Jarkovska:2021jvw} or not), reality/complexity ($\mathbb{R}/\mathbb{C}$) and their mass dimension $D$. \label{tab:potential-parameters}}
\vspace{0.1cm}
\centering
\begin{tabular}{r@{\qquad}l@{\qquad}l@{\qquad}l}
    \toprule
    & $D=0$ & $D=1$ & $D=2$\\
    \midrule
    old $\mathbb{R}$& $a_0,a_2,\lambda_0,\lambda_2,\lambda_4,\lambda'_4,\alpha,\beta_4,\beta'_4$
        &$\tau$&$\mu^2,\nu^2$\\
    new $\mathbb{R}$& $h_2,h'_2,\kappa_0,\kappa_2,\rho_0,\rho_2$
        &$\tau'$&$\xi^2$\\
    old $\mathbb{C}$& $\gamma_2,\eta_2$
        &&\\
    new $\mathbb{C}$& $h_4,h_3,\kappa'_0,\kappa'_2,\zeta,\zeta',\rho'_0,\psi_2,\psi_1,\psi_0,\varphi,\varphi'$
        &&$\xi'^2$\\
    \bottomrule
\end{tabular}
\end{table}

%%%%%%%%%%%%%%%%%%%%%%%%%%
\subsubsection{Symmetry breaking}
%%%%%%%%%%%%%%%%%%%%%%%%%%

Let us now turn to the \textdef{spontaneous symmetry breaking} (SSB) of the GUT symmetry. The scalar sector contains two real SM singlets in the $\mathbf{45}$ and one complex SM singlet in the $\mathbf{126}$, with no extra SM singlets in the added $\mathbf{10}_\mathbb{C}$. The SSB thus proceeds identically as in the pure Higgs model~\cite{Jarkovska:2021jvw}. 

In short, the SM singlet fields obtain non-zero \textdef{vacuum expectation values} (VEVs) defined as 
\begin{equation}
\langle (1,1,1,0)_{45} \rangle \equiv \sqrt{3} \; \omega_{BL},\qquad
\langle (1,1,3,0)_{45} \rangle \equiv \sqrt{2} \, \omega_{R},\qquad
\langle (1,1,3,+2)_{126} \rangle \equiv \sqrt{2} \, \sigma,
\label{eq:VEV-definitions}
\end{equation}
where the numbers in parentheses refer to their \LR transformation properties and the subscripts $45$ and $126$ denote their $\mathrm{SO}(10)$ origin. The redefinition of the overall phase of $\Sigma$ can make $\sigma$ real and positive, while a sign redefinition of $\phi$ allows one to choose $\omega_{R}$ positive.

It is convenient to define the \textdef{dimensionless universal VEV ratio}
\begin{align}
\chi:= \frac{\omega_{BL} \omega_{R}}{|\sigma|^2}, \label{eq:chi-definition}
\end{align}
which is a quantity that appears in the solution of the stationarity conditions for the SM vacuum, as well as in loop corrections to the effective one-loop scalar masses. Perturbativity requires $|\chi| \lesssim 1$, thus greatly constraining the set of viable breaking patterns, see \cite{Jarkovska:2021jvw}. As mentioned in the Introduction, we focus only on the regime \hbox{$|\omega_{BL}| \ll |\sigma| \ll |\omega_R|$} (to be referred to as $\omega_{BL}\to 0$) featuring the $SU(4)_C\times SU(2)_L\times U(1)_R$ intermediate symmetry, abbreviated as $G_{421}$, which turns out to be much more promising than any of the other options, see~\cite{Jarkovska:2021jvw}. 

The large hierarchy between the VEVs in Eq.~(\ref{eq:VEV-definitions}) allows us to imagine the SSB as a two stage process:
\begin{align}
\mathrm{SO}(10)
\quad\xrightarrow{\mathmakebox[2em]{\omega_{R}}}\quad 
\SU(4)_{C}\times\SU(2)_L\times\mathrm{U}(1)_R 
\quad\xrightarrow{\mathmakebox[2em]{\sigma}}\quad 
\SU(3)_c\times \SU(2)_L\times\mathrm{U}(1)_{Y},
\end{align}
where, phenomenologically, $\omega_R$ is the GUT-breaking scale and $\sigma$ plays the role of the seesaw scale for neutrinos. The $\omega_{BL}$ VEV is irrelevant for the symmetry breaking here and obtains only an induced value $\omega_{BL}=\chi|\sigma|^{2}/\omega_{R}$.

The dimensionful parameters $\lbrace \mu, \nu, \tau \rbrace$ are related to the $\lbrace \omega_{BL}, \omega_R, \sigma \rbrace$ VEVs via the vacuum stationarity conditions, which take the tree-level form
\begin{align}
\mu^2 & = (8 a_0 + 2 a_2) \omega_{R}^2 
+ 2 (a_2 \chi + 2\alpha + 2\beta_4') |\sigma|^2 , \label{eq:mu} \\
\nu^2 & = 2 (\alpha + 3 \beta_4' + a_2 \chi) \omega_{R}^2 
+ ( 4 \lambda_0 + 12 \beta_4' \chi + 5a_2 \chi^2)  |\sigma|^2 ,
\label{eq:nu} \\
\tau & = (4 \beta_4' + a_2 \chi) \omega_R,\label{eq:tau}
\end{align}
where $O\left(\tfrac{\omega_{BL}}{\omega_R}\right)$ terms were neglected in the $\omega_{BL}\to 0$ regime. 

One can thus consider the VEVs $\omega_{R}$ and $\sigma$, along with the ratio $\chi$, to be the input parameters, and take the $\lbrace \mu, \nu, \tau \rbrace$  parameter values computed from the vacuum conditions, as in~\cite{Jarkovska:2021jvw}. Note that with the addition of the $\mathbf{10}_\mathbb{C}$, 
new mass scales are introduced through the $\lbrace \xi,\xi',\tau'\rbrace$ parameters. Since, at the tree level, these influence only the {\em doublets} and {\em triplets} (see subsection~\ref{sec:mass-spectrum}), the aforementioned interpretation of $\omega_{R}$ and $\sigma$ can be retained, assuming that none of the new dimensionful parameters is parametrically larger than $\omega_{R}$.

%%%%%%%%%%%%%%%%%%%%%%%%%%
\subsubsection{The tree-level mass spectrum \label{sec:mass-spectrum}}
%%%%%%%%%%%%%%%%%%%%%%%%%%

Since there are no GUT or intermediate-scale VEVs in the newly introduced $\mathbf{10}_\mathbb{C}$, all new couplings in Table~\ref{tab:potential-parameters} impact only those tree-level scalar-mass-matrices which include a state from  $\mathbf{10}_{\mathbb{C}}$. Consequently, almost all scalar mass matrices of the $\mathbf{45}\oplus\mathbf{126}\oplus\mathbf{10}_{\mathbb{C}}$ model at stake are identical to those of the $\mathbf{45}\oplus\mathbf{126}$ Higgs model~\cite{Jarkovska:2021jvw}, with the only exceptions of the SM doublets $(1,2,+\tfrac{1}{2})$ and triplets $(\bar{3},1,+\tfrac{1}{3})$. Similarly, the gauge boson masses 
are identical in these two cases as well.

For those tree-level mass matrices that are common to both scenarios, the interested reader is referred to \cite{Bertolini:2012im,Graf:2016znk} (with minor corrections included in the latter); their approximate forms in the $\omega_{BL}\to 0$ limit are given in~\cite{Jarkovska:2021jvw}. The enlarged matrices for doublets and triplets, in which the vacuum of Eqs.~\eqref{eq:mu}--\eqref{eq:tau} has been inserted (and only the dominant contributions in the $\omega_{BL}\to 0$ regime were retained), read
\begin{align}
M^2(1,2,+\tfrac{1}{2})_{\text{tree}} = \left(
\begin{smallmatrix}
\left(\frac{\beta_4 }{2}-9 \beta_4'-3 a_2 \chi \right) \omega_R^2 & -2 \gamma_2 \omega_R^2 & -\sqrt{6} \zeta'  \chi |\sigma| ^2 & -\sqrt{6} \zeta   \chi |\sigma|^2 \\
%%%%%%%%%%%%%%%
-2  \gamma_2^* \omega_R^2& \left(\frac{\beta_4}{2}-\beta_4'-a_2 \chi \right) \omega_R^2 &  \sqrt{6}\left( 16 \phi ^* +  \zeta ^* \chi \right)|\sigma|^2 & \sqrt{6}\left(16 \phi'^{*} +    \zeta'^* \chi\right)|\sigma|^2  \\
%%%%%%%%%%%%%%%55
-\sqrt{6} \zeta'^{*} \chi   |\sigma| ^2 & \sqrt{6}\left(16  \phi +\zeta  \chi \right)|\sigma| ^2  & -\xi^2 -\frac{\tau' \omega_R}{24} +\left(2 \kappa_0+\kappa_2\right)
\omega_R^2 &-2
\xi'^2 + 2\left( 2\kappa_0'+ \kappa_2'\right) \omega_R^2 \\
%%%%%%%%%%%%%%%%%%%%%
-\sqrt{6}  \zeta ^* \chi  |\sigma| ^2  &  \sqrt{6}\left(16  \phi'+ \zeta' \chi \right) |\sigma| ^2 & -2
\xi'^{2 *} + 2\left( 2\kappa_0'^*+ \kappa_2'^*\right) \omega_R^2  & -\xi^2 +\frac{\tau' \omega_R}{24} +\left(2 \kappa_0+\kappa_2\right)
\omega_R^2 
\end{smallmatrix}
\right),\label{eq:doubletmassmatrix}
\end{align}
\begin{align}
M^2(\overline{3}, 1 +\tfrac{1}{3})_{\text{tree}}=\left(
\begin{smallmatrix}
    \left(\beta_4 -4 \beta_4'-2 a_2 \chi \right)  \omega_R^2 & 
    4 \gamma_2 \omega_R^2 & 
    2\sqrt{2}(\beta_{4}\chi-8\lambda'_4)\,|\sigma|^{2}& 
    -\sqrt{2} \zeta \omega_R^2 & 
    -\sqrt{2} \zeta'  \omega_R^2 \\
%%%%%%%%%%
    4\gamma_2^\ast \omega_R^2  & 
    \left(\beta_4 -4 \beta_4' -2 a_2 \chi \right) \omega_R^2 & 
    0 & 
    -\sqrt{2} \zeta'^\ast \omega_R^2& 
    -\sqrt{2} \zeta^\ast \omega_R^2\\
%%%%%%%%%
    2\sqrt{2}(\beta_{4}\chi-8\lambda'_{4})|\sigma|^{2}&  
   0 & 
    \left(2\beta_4 -4 \beta_4'- 2a_2 \chi \right) \omega_R^2 & 
    -4(8\varphi+\zeta\chi)|\sigma|^{2}  & 
    -4(8\varphi'+\zeta'\chi)|\sigma|^{2}  \\
%%%%%%%%%%%
    -\sqrt{2}  \zeta^\ast \omega_R^2 & 
    -\sqrt{2}  \zeta'  \omega_R^2 & 
    -4(8\varphi^\ast+\zeta^\ast\chi)|\sigma|^{2} & 
    -\xi^2 +2 \kappa_0 \omega_{R }^2 &
    -2 \xi'^{*2}+ 4 \kappa_0'^\ast \omega_R^2 \\
%%%%%%%%%%
    -\sqrt{2} \zeta'^\ast \omega_R^2 & 
    -\sqrt{2} \zeta \omega_R^2 & 
    -4(8\varphi'^\ast+\zeta'^\ast\chi)|\sigma|^{2} & 
    -2 \xi'^{2}+ 4 \kappa_0' \omega_R^2 & 
    -\xi^2 +2 \kappa_0 \omega_{R }^2\\
\end{smallmatrix}
\right), \label{eq:tripletmassmatrix}
\end{align}
\noindent
where the basis of the row states is defined as follows:
\begin{align}
    \text{for }(1,2,+\tfrac{1}{2}):&\quad\{
        (15,2,+\tfrac{1}{2})_{\Sigma},(15,2,-\tfrac{1}{2})^\ast{}_{\Sigma^\ast},(1,2,+\tfrac{1}{2})_{H},(1,2,-\tfrac{1}{2})^\ast{}_{H^\ast}\},\label{eq:basis-doublets}\\
    \text{for }(\bar{3},1,+\tfrac{1}{3}):&\quad\{
        (6,1,0)_{\Sigma},(6,1,0)^\ast{}_{\Sigma^\ast},(\overline{10},1,0)_{\Sigma},(6,1,0)_{H},(6,1,0)^\ast{}_{H^\ast}\}.\label{eq:basis-triplets}
\end{align}
The basis states are unambiguously specified by their intermediate $\SU(4)_C\times\SU(2)_L\times\mathrm{U}(1)_R$ and GUT symmetry origin in the parentheses and in the subscripts, respectively; their ordering is consistent with the one suggested by Table~\ref{tab:particle-content}. The basis for column states is conjugate to that of the basis for rows.

Note that in this basis the upper-left $2\times 2$ block of Eq.~\eqref{eq:doubletmassmatrix} and the upper-left $3\times 3$ block of Eq.~\eqref{eq:tripletmassmatrix} depend only on the old couplings. These sub-blocks are identical to the Higgs model expressions~\cite{Jarkovska:2021jvw}, while all other entries (the enlarged part) involve only the new couplings. Furthermore, in the limit\footnote{Note that the $\sigma\to 0$ limit is properly taken by keeping the universal ratio $\chi$ in Eq.~(\ref{eq:chi-definition}) fixed.} $\sigma\to 0$ one recovers the intermediate $G_{421}$ symmetry with  the mixing occurring solely between the same type of representations within the bases of Eq.~\eqref{eq:basis-doublets} and \eqref{eq:basis-triplets}.  In particular, the doublet mass matrix becomes block diagonal in this limit.

The tree-level spectrum, however, is qualitatively not  the end of the story in the model under consideration. A key observation was made in \cite{Yasue:1980fy,Anastaze:1983zk} that certain scalar states cannot be made simultaneously non-tachyonic at tree-level unless the breaking proceeds through an intermediate (flipped) $\mathrm{SU}(5)$ symmetry. The problematic states include the SM triplet $(1,3,0)$ and octet $(8,1,0)$, along with the singlet $(1,1,0)$ as pointed out in \cite{Graf:2016znk}. Given their specific behaviour in various limits, these states were dubbed the \textdef{pseudo-Goldstone bosons} (PGBs), and we shall refer to them as such from now on. 

It is very instructive to consider the PGBs in the strict $\omega_{BL}\to 0$ regime. In such a case the field $(\overline{3},1,-\tfrac{2}{3})$, sharing the same parent $(15,1,0)$ multiplet of $G_{421}$ with $(8,1,0)$ and $(1,1,0)$, also joins the list of states whose masses acquire a particularly simple form. Taking the limit $\sigma\to 0$, we find the dominant parts of the tree-level masses of these fields to be
\begin{align}
M^2_S(1,3,0) & = \phantom{+} 4 a_2 \omega_R^2, \label{eq:tripletmass} \\
M^2_S(8,1,0) & = -2a_2 \omega_R^2, \label{eq:octetmass}\\
M^2_S(1,1,0)_3 &= -2a_2 \omega_{R }^2, \label{eq:signletmass}\\
M^2_S(\overline{3},1,-\tfrac{2}{3})_2 &= -2a_2 \omega_{R }^2.\label{eq:tripletpgbmass}
\end{align}
Note that subscripts are used when multiple states with the same $G_{321}$ quantum numbers are present, and they label the relevant mass eigenstates in an ascending mass order.
Observe also that no choice of $a_2$ (except for the singular $a_2=0$ case) can make all these states simultaneously non-tachyonic. 

Remarkably, this situation can be remedied by loop corrections (as first discussed in \cite{Bertolini:2009es}) if one takes $|a_2| \ll 1$, so that the tree-level PGB masses become suppressed. However, this also means that any in-depth analysis of the model requires access to the one-loop scalar spectrum; for that sake the numerical methods developed for the $\mathbf{45}\oplus\mathbf{126}$ Higgs model in~\cite{Jarkovska:2021jvw} have been extended to the $\mathbf{45}\oplus\mathbf{126}\oplus\mathbf{10}_\mathbb{C}$ case.

%%%%%%%%%%%%%%%%%%%%%%%%%%
\subsection{The Yukawa sector \label{sec:Yukawa-sector}}
%%%%%%%%%%%%%%%%%%%%%%%%%%

\subsubsection{General considerations}

With both the $\mathbf{126}$ and $\mathbf{10}_\mathbb{C}$ at play, the renormalizable Yukawa Lagrangian consists schematically of the terms~\cite{Bajc:2005zf} (with the $\mathrm{SO}(10)$ indices and Lorentz structures suppressed)
    \begin{align}
    \label{eq:YukawaLagrangian}
    \mathcal{L}_Y = \mathbf{16}_F^a \left( Y_{10}^{ab}\;\mathbf{10}_\mathbb{C} + \tilde{Y}_{10}^{ab} \;\mathbf{10}^\ast_\mathbb{C} + Y_{126}^{ab}\;\mathbf{126}^\ast \right) \mathbf{16}_F^b + h.c.,
    \end{align} 
where $\mathbf{16}^{a}_F$ contains fermions, $a,b= 1,\ldots,3$ denote family indices, and the matrices $Y_{10}$, $\tilde{Y}_{10}$ and $Y_{126}$ are complex symmetric $3\times 3$ matrices in family space.

The electrically neutral components of the SM weak doublets with hypercharges $\pm 1/2$ acquire VEVs in the electroweak broken phase. They are labeled by 
    \begin{align}
        v^u_{10} &= 
        \left \langle (1,2,+\tfrac{1}{2})_{10} \right \rangle, \quad
        v^d_{10} = \left \langle \right (1,2,-\tfrac{1}{2})_{10}\rangle,\quad
        v^u_{126} = \left \langle (1,2,+\tfrac{1}{2})_{\overline{126}} \right \rangle, \quad 
        v^d_{126} = \left \langle \right (1,2,-\tfrac{1}{2})_{\overline{126}}\rangle.
    \end{align}
The notation above is straightforward: the subscripts denote their $\mathrm{SO}(10)$ origin and their superscripts $u$ and $d$ indicate the $+1/2$ and $-1/2$ hypercharges, respectively. The tree-level (GUT-scale) quark and lepton mass matrices then take the following form:
    \begin{align}
    \label{eq:MU}
    M_U &= Y_{10}\; v^u_{10} + \tilde{Y}_{10}\; v^{d~\ast}_{10} + Y_{126}\; v^{u}_{126},\\
    \label{eq:MD}
    M_D &= Y_{10}\; v^d_{10} + \tilde{Y}_{10}\; v^{u~\ast}_{10} + Y_{126}\; v^d_{126},\\
    \label{eq:ME}
    M_E &= Y_{10}\; v^d_{10} + \tilde{Y}_{10}\; v^{u~\ast}_{10} - 3Y_{126}\;v^d_{126},\\
    \label{eq:MnuD}
    M^D_\nu &= Y_{10}\; v^u_{10} + \tilde{Y}_{10}\; v^{d~\ast}_{10} - 3Y_{126}\; v^u_{126},
    \end{align}
where $U$, $D$ and $E$ refer, respectively, to the up, down and charged lepton sectors, while $M^D_\nu$ is the Dirac type neutrino mass matrix. The EW VEVs are normalized such that
\begin{align}
    v_{SM}^{2}&= |v_{10}^{u}|^{2}+|v_{10}^{d}|^{2}+|v_{126}^{u}|^{2}+|v_{126}^{d}|^{2},
\end{align}
where $v_{SM}\doteq 174\,\mathrm{GeV}$. Furthermore, the seesaw type I and II Majorana mass matrices read
    \begin{align}
    \label{eq:MnuM}
    M^{M,\text{type I}}_\nu & \propto Y_{126} \;\sigma,
    & M^{M,\text{type II}}_\nu& \propto Y_{126}\; w,
    \end{align}
where
    \begin{align}
    w &= \left \langle (1,3,+1)_{\overline{126}} \right \rangle
    \end{align}
is the induced EW VEV of the scalar $SU(2)_L$ triplet relevant for the type II seesaw.

Renormalizable Yukawa sectors of $\mathrm{SO}(10)$ GUTs have been studied in great detail in the literature, see e.g.~\cite{Babu:1992ia,Bajc:2005zf,Joshipura:2011nn,Altarelli:2013aqa,Dueck:2013gca,Babu:2015bna,Boucenna:2018wjc,Ohlsson:2019sja,Mummidi:2021anm,Meloni:2014rga,Babu:2016bmy,Meloni:2016rnt,Ohlsson:2018qpt,Fukuyama:2019zun}. Let us just summarize the salient points below:
\begin{itemize}
    \item Due to the $\mathrm{SO}(10)$ decomposition 
        $
        \mathbf{16} \otimes \mathbf{16} = \mathbf{10} \oplus \mathbf{120} \oplus \mathbf{126}
        $ \cite{Slansky:1981yr},
        the most general renormalizable Yukawa sector includes scalar  representations $\mathbf{126}$, $\mathbf{10}$ and/or $\mathbf{120}$, with the latter requiring an anti-symmetric Yukawa matrix. Since at least 2 Yukawa matrices are required for a successful fit of fermion masses and mixing, the Higgs model $\mathbf{45}\oplus\mathbf{126}$ requires an upgrade to become potentially realistic; the addition of an extra $\bf 10$ represents the obvious option here\footnote{The extension with an additional $\mathbf{10}$ is clearly minimal in terms of extra degrees of freedom, but it also has the advantage of the symmetry of the corresponding Yukawa matrices. The anti-symmetric Yukawa of the $\mathbf{120}$ with only one more symmetric Yukawa matrix of the $\mathbf{126}$ does not lead to realistic fermion mass matrices~\cite{Joshipura:2011nn}.}.
    \item The added $\mathbf{10}$ should better be complex (or equivalently, two copies of a real $\mathbf{10}$ should be considered), otherwise the relation $v^u_{10} = v^d_{10}{}^{\ast}$ among the SM VEV projections emerges, and the fit of SM fermion masses and mixing is not possible, since it clashes with the phenomenological requirement~\cite{Bajc:2005zf} 
        \begin{align}
	       \left \vert \frac{v^u_{10}}{v^d_{10}} \right \vert \approx \frac{m_t}{m_b} \gg 1.
	    \end{align}
    \item Two different Yukawa terms can be constructed with the complex $\mathbf{10}_\mathbb{C}$ (in the non-SUSY case), so the resulting theory has $3$ Yukawa matrices in total and a good fermion-sector fit is  easily attained. In fact, such a fit is already possible with a single Yukawa matrix associated to a real $\mathbf{10}$ along with the one associated to $\mathbf{126}$, which is a scenario arising in the SUSY case~\cite{Bertolini:2006pe,Deppisch:2018flu} or if a Peccei-Quinn-like (PQ) global $\mathrm{U}(1)$ symmetry is introduced~\cite{Babu:1992ia,Bajc:2005zf,Joshipura:2011nn,Altarelli:2013aqa,Dueck:2013gca,Babu:2015bna,Boucenna:2018wjc,Ohlsson:2019sja,Mummidi:2021anm}. In the latter case the typical assignment of the PQ charges reads
        \begin{align}
        [\mathbf{16}_F]_{PQ}&=+1, &
        [\mathbf{45}]_{PQ}&=0, &
        [\mathbf{126}]_{PQ}&=+2, &
        [\mathbf{10}]_{PQ}&= -2. \label{eq:PQ-assignments}
        \end{align}
    These charges forbid the $\tilde{Y}_{10}$ Yukawa coupling in Eq.~\eqref{eq:YukawaLagrangian}, thus increasing the predictivity of the fit. Furthermore, the PQ symmetry forbids a number of couplings in the scalar potential~\eqref{eq:scalar-potential-0}--\eqref{eq:scalar-potential-Vtildemix} which, naively, may look as a good way to greatly simplify the current analysis. Unfortunately,
    any non-trivial assignment of the PQ charge to $\Sigma$ necessarily forbids the same $\gamma_{2}$ term in Eq.~\eqref{eq:scalar-potential-V126} that was found to be necessary for curing the tachyonicity of PGB states in the Higgs model $\mathbf{45}\oplus\mathbf{126}$, cf.~\cite{Jarkovska:2021jvw}, so we shall not entertain this option at all.
    \item A successful fermion mass fit accomplished via the tree-level relations of Eqs.~\eqref{eq:MU}-\eqref{eq:MnuD} requires a significant admixture of the weak doublets from both $\mathbf{10}_\mathbb{C}$ and $\mathbf{126}$ within the SM Higgs doublet, otherwise either $v^{u,d}_{126} \approx 0$  or $v^{u,d}_{10}\approx 0$, for which the Yukawa fit does not work\footnote{Note that the case $v^{u,d}_{126}\approx 0$ implies a (GUT-scale) $\SU(5)$-like relation
	 $M_D \approx M_E$, while the case $v^{u,d}_{10}\approx 0$ yields $M_D \approx -3 M_E$, none of which is realistic.}. 
\end{itemize}

\subsubsection{The structure of the doublet mass matrix \label{sec:doublet-requirements}}
The last point above has very important ramifications for the structure of the doublet mass matrix of Eq.~(\ref{eq:doubletmassmatrix}). 
In the basis of Eq.~\eqref{eq:basis-doublets} it can be schematically written as
\begin{align}
\label{eq:doubletmassstructure}
M^2(1,2,+\tfrac{1}{2}) = \begin{pmatrix}
M^2_{126} &M^2_{mix}\\
M^{2 \dagger}_{mix} & M^2_{10}
\end{pmatrix},
\end{align}
where the $2\times 2$ block structures $M^2_{126}$, $M^2_{10}$ and $M^2_{mix}$ encompass contributions of different $\mathrm{SO}(10)$ origin. 
To this end, the former two mix doublets solely within the $\mathbf{126}$ and $\mathbf{10}_\mathbb{C}$, respectively, while the last one controls their cross-talk. Since no mixing among the  $\mathbf{126}$ and $\mathbf{10}_\mathbb{C}$ is possible until the intermediate $G_{421}$ symmetry gets broken, the $M^2_{mix}$ block must vanish for $\sigma\to 0$. Furthermore, the expected magnitudes of entries in the $M^2_{126}$ and $M^2_{10}$ blocks are ${\cal O}(\omega_{R }^2)$ while  $M^2_{mix}$ is expected at the $|\sigma|^2$ level\footnote{
    There can be no $\sigma \omega_{R}$ contribution in $M^2_{mix}$ due to gauge symmetry. This can be seen by considering the $\mathrm{SU}(4)_C$ factor, under which the doublet/anti-doublet product transforms as a $15$-plet, while $\sigma\sim \overline{10}$ and $\omega_R\sim 1$.}, 
with $|\sigma|^2 \ll \omega_{R }^2$ assumed.  This should be the case at all perturbative orders, and can be explicitly verified for the tree-level expression in Eq.~\eqref{eq:doubletmassmatrix}.

The requirement of accommodating the SM Higgs doublet implies that one eigenvalue of the matrix in Eq.~\eqref{eq:doubletmassstructure} should be
fine-tuned to about $(125\,\mathrm{GeV})^2$. On top of that, the doublet-admixture condition of Sect.~\ref{sec:Yukawa-sector} requires the SM Higgs doublet to have a sufficiently large projection to both the $\mathbf{126}$ and $\mathbf{10}_\mathbb{C}$ sectors. Since, however, $M^2_{mix}$ is of the order of $|\sigma|^{2}$, this boils down to a need for an initial {\em pre-tuning} of at least one eigenvalue in both $M^2_{126}$ and $M^2_{10}$ from their natural scale of $\omega_R^2$ down to roughly $|\sigma|^{2}$. 

The {\em pre-tuning} requirement can be brought into sharper quantitative form by estimating the required weight of the $\mathbf{126}$-components within the SM Higgs. The corresponding VEV ratio obeys
\begin{align}
\label{eq:126-component-size}
\sqrt{\frac{|v^u_{126}|^2  + |v^d_{126}|^2}{v_{SM}^2}} = \frac{1}{4} \frac{|\mathrm{Tr}\,M_D - \mathrm{Tr}\,M_E| }{ v_{SM}} \frac{\sqrt{1 + |r|^2 |s|^2}}{|\mathrm{Tr}\,Y_{126}|}\gtrsim \frac{\left|\mathrm{Tr}\,M_D - \mathrm{Tr}\,M_E\right| }{ v_{SM}}\frac{\sqrt{1 + |r|^2 |s|^2}}{4},  
\end{align}
where $M_D$ and $M_E$ are defined in Eqs.~\eqref{eq:MD} and \eqref{eq:ME}, $r$ and $s$ stand for the usual ratios~\cite{Ohlsson:2019sja}
\begin{align}
r&:= \frac{v^u_{10}}{v^d_{10}}, &
s&:= \frac{v^u_{126}}{v^d_{126}} \frac{v^d_{10}}{v^u_{10}},
\end{align}
and perturbativity has been assumed for $Y_{126}$ in the form of $\mathrm{Tr}\left[Y_{126}\right] \lesssim 1$.
Inspecting the existing Yukawa sector fits like~\cite{Ohlsson:2019sja}, one obtains a conservative estimate\footnote{
    Taken strictly, the fermion fit in~\cite{Ohlsson:2019sja} assumes a PQ-like setting with only $2$ Yukawa matrices,
    but it is still to a large degree applicable to Eq.~\eqref{eq:126-component-size} in the more general $3$ Yukawa case. Namely, the size of $M_{D}-M_{E}$ is enforced by the difference $y_b-y_\tau $ at the GUT scale (unrelated to $\tilde{Y}_{10}$), while only a large $|rs|$ can 
    simultaneously achieve a suitably large $M_{U}$ (large top Yukawa) and a small $M_{\nu}^{D}$. Note that the see-saw scale $|\sigma|\sim 10^{11}\,\mathrm{GeV}$ is rather small in the current case, therefore within type I see-saw a small $M_{\nu}^{D}$ is indeed needed to suppress the enhanced inverse of $M_{\nu}^{M,\text{type I}}$ for suitable neutrino masses.
    } 
\begin{align}
\label{eq:126-comp-bound}
\sqrt{\frac{|v^u_{126}|^2  + |v^d_{126}|^2}{v_{SM}^2}} \gtrsim 3\times 10^{-2} \equiv \epsilon.
\end{align}
  
In terms of the eigenvalues of the two diagonal blocks (denoted by $m_{126\pm}^{2}$ for $M^{2}_{126}$ and $m_{10\pm}^{2}$ for $M^{2}_{10}$, with $+$ referring to the larger and $-$ to the smaller ones), the {\em pre-tuning} conditions can be recast as  
\begin{align}
    m_{10-}^{2}&\sim |\sigma|^{2}, &
    m^{2}_{126-} & \sim |\sigma|^{2}/\epsilon,
    \label{eq:doublet-fine-tuning-condition}
\end{align}
with $\epsilon=3\times 10^{-2}$. For better numerical intuition, the $\mathbf{45}\oplus\mathbf{126}$ Higgs model analysis~\cite{Jarkovska:2021jvw} suggests \hbox{$\omega_{R}\gtrsim 10^{14.8}\,\mathrm{GeV}$} and $|\sigma|< 10^{11.8}\,\mathrm{GeV}$, and hence
$
    |\sigma|^{2} \lesssim 10^{-6} \omega_{R}^{2}
$,
which translates the conditions in Eq.~\eqref{eq:doublet-fine-tuning-condition} into\footnote{Note that in~\cite{Jarkovska:2021jvw} the estimates for $\omega_R$ and $|\sigma|$ have been obtained in a simplified setting without $\mathbf{10}_{\mathbb{C}}$; however, since these follow from the unification analysis where the presence/absence of the extra $\mathbf{10}_{\mathbb{C}}$ plays a marginal role, one can expect very similar constraints to be relevant also for the full-model case, see Appendix~\ref{app:full-vs-Higgs}.
}
\begin{align}
    m_{10-}^{2}&\sim 10^{-6}\,\omega_{R}^{2}, &
    m^{2}_{126-} & \sim 3.3\cdot 10^{-5}\,\omega_{R}^{2}.
    \label{eq:doublet-fine-tuning-condition-omegaR}
\end{align}

The $M^2_{10}$ block is the only part of the tree-level scalar spectrum that depends on the new couplings $\kappa_0',\kappa_2'$ and $\tau'$. The {\em pre-tuning} of $m_{10-}^{2}$ is thus easily achieved via these parameters.

On the other hand, it is namely the condition on $m^{2}_{126-}$ in Eq.~\eqref{eq:doublet-fine-tuning-condition} that turns out to be difficult to conform to in the potentially realistic and perturbative parts of the model parameter space, thus heavily disfavoring the $\mathbf{45}\oplus\mathbf{126}\oplus\mathbf{10}_\mathbb{C}$ setting as a basis for a robust and calculable GUT scenario. This {\em pre-tuning} issue is demonstrated in Section~\ref{sec:tree-level-fine-tuning} at tree-level analytically, and then in Section~\ref{sec:one-loop-fine-tuning} at one-loop level numerically.

To summarize, the arrangement of a SM-compatible Higgs doublet mass in the  $M^2(1,2,+\tfrac{1}{2})$ mass matrix first requires a {\em pre-tuning} to the intermediate scale in each of the $M^{2}_{126}$ and $M^{2}_{10}$ blocks, before the full fine-tuning down to the EW scale should be performed. This procedure brings in an additional set of constraints on the scalar potential parameters. The obstructions to {\em pre-tuning} in the $\mathbf{126}$-sector and the associated implications are the main focus of the remainder of this study.

%%%%%%%%%%%%%%%%%%%%%%%%%%
\section{Doublet fine-tuning at tree level --- analytically}\label{sec:tree-level-fine-tuning}
%%%%%%%%%%%%%%%%%%%%%%%%%%

It is instructive to first consider the {\em pre-tuning} obstruction for doublets in $M^{2}_{126}$ of Eq.~\eqref{eq:doubletmassstructure} at tree-level, since one has analytic grip on the tree-level mass expressions, see Eq.~\eqref{eq:doubletmassmatrix}. A possible reservation of the reader may be that the model's first consistent perturbative order is only at one loop, but note that these intricacies are intimately connected only to the PGB sector of scalars. Quantitatively, the quantum corrections thus play no bigger role in the non-PGB sectors, e.g.~in the doublet sector, than in the usual perturbative situation with a consistent tree order. A tree-level consideration of doublets will thus offer valuable insight, while the quantum nature of the model will percolate into the analysis only through limitations on the parameter space, as will become apparent in due course.

The tree-level eigenvalues $m^2_{126\pm,(0)}$ of the $M_{126}^{2}$ block are 
    \begin{align}
    m^2_{126\pm,(0)} &= \left(\frac{\beta_4}{2} - 5\beta_4' - 2a_2 \chi \pm \sqrt{\left(4\beta_4' + a_2\chi\right)^2 + 4|\gamma_2|^2}\right)\omega_{R }^2,\label{eq:tree-level-doublet-eingevalues}
    \end{align}
where only the dominant contributions are considered given the VEV hierarchy $|\sigma| \ll |\omega_{R}|$. Doublet {\em pre-tuning} demands that $m^{2}_{126-,(0)}$ is brought down roughly to $|\sigma|^{2}/\epsilon$, cf.~Eq.~\eqref{eq:doublet-fine-tuning-condition}.

The {\em pre-tuning} of $m^2_{126-,(0)}$ is most conveniently viewed as a constraint on $|\gamma_{2}|$, since this parameter is present at tree-level only in the mass expressions for multiplets $(15,2,+\tfrac{1}{2})$ and $(6,1,0)$ of the $G_{421}$ intermediate symmetry, to which the SM doublets $(1,2,+\tfrac{1}{2})$ and triplets $(\overline{3},1,+\tfrac{1}{3})$ respectively belong, cf.~Tables VII and IX in~\cite{Jarkovska:2021jvw}. Consequently, the constraint on $|\gamma_{2}|$ is directly entangled only with the non-tachyonicity of triplets, which thus warrants special attention. 

Non-tachyonicity of the SM triplets requires the mass-square matrix in Eq.~\eqref{eq:tripletmassmatrix} to be positive (semi-) definite. Applying Sylvester's criterion for positive-definiteness requires the leading principal minors (the determinants of the $1\times 1$, $2\times 2$, etc.~upper-left submatrices) of $M^2(\overline{3},1,+\tfrac{1}{3})_{\text{tree}}$ to be positive. The relevant constraints are already given by the first three minors, which involve the triplet states from the representation $\mathbf{126}$ only: 
    \begin{align}
	0&\quad<\quad \mathcal{M}^2_{1\times 1}= \left(\beta_4 - 4\beta_4' - 2a_2 \chi\right) \omega_{R }^2, \label{eq:minor-constraints-begin}\\
	%%%%%%%%%%%%%%%
	0&\quad<\quad \mathcal{M}^2_{2 \times 2}= \theta ~\omega_R^4, \\
	%%%%%%%%%%%%%%
	0&\quad<\quad \mathcal{M}^2_{3 \times 3}= \theta \left(\beta_4 - 4\beta_4' - 2a_2 \chi\right) \omega_R^6 + \mathcal{O}\left(|\sigma|^2 \omega_R^2\right),\label{eq:minor-constraints-end}
	\end{align}
where the entire $\gamma_2$-dependence is conveniently packaged into a single parameter $\theta$ defined by
	\begin{align}
	\theta  &\equiv \left(\beta_4 - 4\beta_4' - 2a_2 \chi\right)^2 - 16 |\gamma_2|^2.
	\end{align}
The triplet non-tachyonicity conditions in Eqs.~\eqref{eq:minor-constraints-begin}-\eqref{eq:minor-constraints-end} (allowing now for the semi-definite case) simplify to
	\begin{align}
	\theta & \geq 0, &
	%%%%%%
	\beta_4 - 4\beta_4' - 2a_2 \chi & \geq 0,\label{eq:nontachconstraint}
	\end{align}
and these need to be amended by the non-tachyonicity requirements from all other non-PGB scalar fields at tree-level (cf.~Table~IX in~\cite{Jarkovska:2021jvw}): 
    \begin{align}
		4\beta_4' + a_2 \chi &\leq 0, &
		2\beta_4' + a_2 \chi &\leq 0, &
		\beta_4 - 2\beta_4' - a_2 \chi &\geq 0, &
		\beta_4 - 10\beta_4' - 4a_2 \chi &\geq 0. \label{eq:Higgs-model-parameter-space}
	\end{align}
Furthermore, the comprehensive analysis of the $\mathbf{45} \oplus \mathbf{126}$ $\mathrm{SO}(10)$ Higgs model in \cite{Jarkovska:2021jvw} gives the viable ranges for these parameters:
	\begin{align}
		\beta_4 &\in \left[0.2,1\right],  & \beta_4' &\in \left[-0.2,-0.01\right], &
		a_2 &\in \left[-0.05,0.05\right], & \chi &\in \left[-1,1\right]. \label{eq:viable-box}
	\end{align}
The viable region must therefore lie in the multi-dimensional box defined by Eq.~\eqref{eq:viable-box}. Note that restricting to this box
is precisely where the necessary requirements for PGB non-tachyonicity (at one-loop level) have slipped in. Interestingly, the ranges of Eq.~\eqref{eq:viable-box}
make all inequalities in Eq.~\eqref{eq:Higgs-model-parameter-space} except for the 2nd one redundant.

We now return to the quantity $m^2_{126-,(0)}$ that we wish to tune, and rewrite it in terms of $\theta$ instead of $|\gamma_{2}|$:
    \begin{align}
	m^2_{126-,(0)}(\theta) = \left(\frac{\beta_4}{2} - 5\beta_4' - 2a_2 \chi -\frac{1}{2} \sqrt{4\left(4\beta_4' + a_2\chi\right)^2 + \left(\beta_4 - 4\beta_4' - 2a_2 \chi\right)^2 - \theta}\right)\omega_{R }^2.
	\end{align}
Notice that $m^2_{126-,(0)}(\theta)$ is an increasing function of $\theta$, i.e. 
	\begin{align}
	\frac{\partial m^2_{126-,(0)}}{\partial \theta} = \frac{1}{4} \frac{\omega_{R }^2}{\sqrt{4\left(4\beta_4' + a_2\chi\right)^2 + \left(\beta_4 - 4\beta_4' - 2a_2 \chi\right)^2 - \theta}} >0.
	\end{align}
Triplet non-tachyonicity in Eq.~\eqref{eq:nontachconstraint} requires a non-negative $\theta$, and thus
	\begin{align}
	m^2_{126-,(0)}(\theta) & \quad \geq \quad  m^2_{126-,(0)}(0) = \left(\frac{\beta_4}{2} - 5\beta_4' - 2a_2 \chi -\frac{1}{2} \sqrt{4\left(4\beta_4' + a_2\chi\right)^2 + \left(\beta_4 - 4\beta_4' - 2a_2 \chi\right)^2 }\right)\omega_{R }^2
	\end{align}
for any fixed choice of $\beta_4, \beta_4'$ and $a_2 \chi$. As a numerical exercise, we can then find the lowest numerical value for $m^{2}_{126-,(0)}$ at $\theta=0$ given the parameter restrictions of Eq.~\eqref{eq:nontachconstraint}--\eqref{eq:viable-box}:
	\begin{align}
	\min_{\beta_4, \beta_4', a_2 \chi } \left[ m^2_{126-,(0)}(0)\right] &\gtrsim 0.008 \,\omega_{R }^2, \label{eq:doublet-fine-tuning-tree}
	\end{align}
which is obtained for $\beta_4 = 0.2$, $\beta_4' = -0.01$, $a_2 \chi = 0.02$. Comparing with the requirement of Eq.~\eqref{eq:doublet-fine-tuning-condition-omegaR}, we see that doublet {\em pre-tuning} cannot be directly achieved at tree-level. 

Incidentally, solving $\theta=0$ for the minimal point gives $|\gamma_{2}|=0.05$, which is smaller than the rough lower bound $|\gamma_{2}|\gtrsim 0.1$ from \cite{Jarkovska:2021jvw}, indicating the estimate in Eq.~\eqref{eq:doublet-fine-tuning-tree} to actually be conservative and making the fine-tuning situation worse. An attainable value is e.g.
    \begin{align}
        \text{for}\ (\beta_{4},\beta_{4}',a_{2}\chi,|\gamma_{2}|)=(0.4,-0.01,0.02,0.1): \qquad m^{2}_{126-}&\approx 0.009 \,\omega_{R}^{2}. \label{eq:Higgs-model-tuned-point}
    \end{align}

To summarize, we demonstrated at tree-level that doublet {\em pre-tuning} in the $\mathbf{126}$-block is obstructed by the non-tachyonicity constraints of other scalars. This, however, does not conclude the analysis; the tree-level value in Eq.~\eqref{eq:Higgs-model-tuned-point} is sufficiently suppressed that quantum corrections (shown to be easily at $10\,\%$ of $\omega_{R}^{2}$ or larger in Section~\ref{sec:one-loop-fine-tuning}) may bridge the {\em pre-tuning} gap, provided they properly align for the doublets and the obstruction states. 

Ironically, the burden of making the model realistic is thus again placed on quantum corrections, as it was for attaining a suitable vacuum. We show in the next section that the obstruction to doublet {\em pre-tuning} is instead significantly reinforced at one-loop level, leading to bleak prospects for the model.

%%%%%%%%%%%%%%%%%%%%%%%%%%
\section{Doublet fine-tuning at one loop --- numerically}\label{sec:one-loop-fine-tuning}
%%%%%%%%%%%%%%%%%%%%%%%%%%

In the tree-level analysis of Section~\ref{sec:tree-level-fine-tuning} it was made clear that a successful {\em pre-tuning} in the $\mathbf{126}$-sector may happen only if
loop-corrections to various fields properly align. To asses the feasibility of such a situation, a complete one-loop analysis along the lines of \cite{Jarkovska:2021jvw} is required for the full model. The usual perturbativity and non-tachyonicity constraints are, however, to be amended by the {\em pre-tuning} requirements. 

Such an analysis involves numeric scans of the parameter space, for which at every considered point the tree-level and one-loop scalar spectrum is computed. The one-loop calculation is performed by numerically evaluating the one-loop regularized effective mass-squares.\footnote{
    The regularized effective scalar mass is the scalar mass computed from the one-loop effective potential in which IR diverging logarithmic contributions are tamed by mimicking the shift from the effective to the physical mass. The associated techniques are given in~\cite{Jarkovska:2021jvw}.
    }

The one-loop analysis in this section is organized as follows: the technical requirements of {\em pre-tuning} in the context of perturbation theory are discussed and formulated in Section~\ref{sec:fine-tuning}, the necessary criteria for a parameter point to be considered viable are summarized in Section~\ref{sec:viability-constraints}, the parameter inputs of the analysis are listed in Section~\ref{sec:inputs}, and the results of the parameter space scans are presented in Section~\ref{sec:results}.

%%%%%%%%%%%%%%%%%%%%%%%%%%
\subsection{Fine-tuning of doublets in perturbation theory \label{sec:fine-tuning}}
%%%%%%%%%%%%%%%%%%%%%%%%%%

The block structure of the doublet mass matrix in Eq.~\eqref{eq:doubletmassstructure} and the required level of {\em pre-tuning} in Eq.~\eqref{eq:doublet-fine-tuning-condition} as described in Section~\ref{sec:doublet-requirements}  refer to quantities that would ideally be computed to all orders in perturbation theory (with an optimal truncation due to the asymptotic nature of this series). In practice, however, one only has access to the few lowest orders, while the effect of higher-order corrections, despite not being readily accessible, should nonetheless be anticipated. The goal of the discussion below is to formulate a technical criterion for the desired {\em pre-tuning} in the $\mathbf{126}$ sector in the context of perturbation theory.

In Section~\ref{sec:doublet-requirements} the eigenstates in the $\mathbf{126}$- and $\mathbf{10}$-blocks of the scalar doublet mass matrix (\ref{eq:doubletmassstructure}) were labeled by $m^{2}_{126\pm}$ and $m^{2}_{10\pm}$, respectively. In full analogy, let us label the $n$-loop estimates for these quantities respectively by $m^{2}_{126\pm,(n)}$ and $m^{2}_{10\pm,(n)}$, where the tree-level case corresponds to $n=0$. Furthermore, the pure $n$-loop corrections contributing to these quantities will be denoted by $\delta m^{2}_{126\pm,(n)}$ and $\delta m^{2}_{10\pm,(n)}$ for $n\geq 1$, so that
\begin{align}
    m^{2}_{126\pm,(n+1)}&= m^{2}_{126\pm,(n)} + \delta m^{2}_{126\pm,(n+1)},\qquad m^{2}_{10\pm,(n+1)}= m^{2}_{10\pm,(n)} + \delta m^{2}_{10\pm,(n+1)}.
\end{align}
As long as loop corrections to the smaller eigenvalues are larger than the required threshold of {\em pre-tuning} in Eq.~\eqref{eq:doublet-fine-tuning-condition}, a milder {\em pre-tuning} condition may be formulated. At any given loop order $n$, {\em pre-tuning} needs to be performed only down to the (estimated) next-order corrections: 
\begin{align}
    \left|m^{2}_{126-,(n)}\right| & \lesssim \left|\delta m^{2}_{126-,(n+1)}\right|, & 
    \left|m^{2}_{10-,(n)}\right| & \lesssim \left|\delta m^{2}_{10-,(n+1)}\right|.  \label{eq:perturbative-tuning-definition-1}
\end{align}
This is only a necessary and not a sufficient condition at any given perturbative order, since one relies on the correct alignment of the $(n+1)$-th-order (and higher) corrections. We refer to Eq.~\eqref{eq:perturbative-tuning-definition-1} as $n$-PL {\em pre-tuning}, where ``$n$-PL'' stands for ``$n$-th perturbative level''. To this end, We relied on $0$-PL {\em pre-tuning} in Section~\ref{sec:tree-level-fine-tuning} to keep the hope alive beyond tree-level. If for any given $n$ (assuming $n$ is low enough that $n$-loop corrections are larger than the tuning threshold) $n$-PL {\em pre-tuning} fails, then all-order {\em pre-tuning} is not possible.

A practical limitation of implementing the $n$-PL {\em pre-tuning} consists in the fact that the left-hand side of Eq.~\eqref{eq:perturbative-tuning-definition-1} typically corresponds to the highest really computed order, so the corrections of  even higher orders on the right-hand side are not readily accessible. The $(n+1)$-th order corrections must thus be estimated in some way.

To keep track of perturbative orders in a mass-square $m^{2}_{x}$, we introduce a \textit{loop expansion parameter} $\eta_x$ which should encode the characteristic size of the loop corrections to $m^{2}_{x}$. Needless to say, thus defined $\eta_x$ depends on the specific parameter-space point as well as on the field $x$ under consideration. The $n$-loop correction to $m^{2}_{x}$ should then obey
\begin{align}
    \left| \delta m^{2}_{x,(n)}/m^{2}_{x,(0)} \right| &\sim \eta^{n}_{x}. \label{eq:definition-eta}
\end{align}
Note, however, that this estimate makes sense only when $m^{2}_{x,(0)}$ is not spuriously small or tuned, i.e.~a generic-sized tree-level mass $m^{2}_{x,(0)}$ is required. In the case of the lighter doublet in each block, which are to be {\em pre-tuned}, the heavier eigenstate tree-level mass may be (and will be) used as a replacement.

Focusing from now on the shape of the one-loop spectrum (i.e. setting $n=1$) the 1-PL {\em pre-tuning} must happen down to the level of two-loop corrections whose sizes can be estimated by
\begin{align}
    \left| \delta m^2_{126-,(2)} \right| &\approx \eta_{126}^{2} \;\left|m^{2}_{126+,(0)}\right|, &
    \left| \delta m^2_{10-,(2)} \right| &\approx \eta_{10}^{2} \;\left|m^{2}_{126+,(0)}\right|, \label{eq:two-loop-estimate}
\end{align}
where the loop-suppression factors $\eta_{126}$ and $\eta_{10}$ in the spirit of Eq.~\eqref{eq:definition-eta} are taken as separate quantities for the blocks $M^2_{126}$ and $M^2_{10}$ (due to different scalar parameters involved in those blocks). 

Notice that $m^{2}_{126+,(0)}$ was used as the tree-level quantity of comparison in both cases, since this is the only eigenvalue that is necessarily of a generic size and, thus, suitable for the task. Namely, the smaller eigenvalues $m^{2}_{126-,(0)}$ and $m^{2}_{10-,(0)}$ are {\em pre-tuned} and hence inappropriate, while {\em pre-tuning} in the $M^{2}_{10}$ block might give a smaller value\footnote{Unlike $m^{2}_{126+,(0)}$, a suppressed $m^{2}_{10+,(0)}$ is possible due to the large number of only loosely restricted new parameters that were introduced through the addition of $\mathbf{10}_\mathbb{C}$.} also for $m^{2}_{10+,(0)}$.

The loop suppression factors $\eta_{126}$ and $\eta_{10}$ relevant for the $M^2_{126}$ and $M^2_{10}$ blocks, respectively, can be estimated by comparing the one-loop corrections to the generic tree-level contribution identified above: 
\begin{align}
    \eta_{126} &\eqsim \left| \frac{\delta m^{2}_{126+,(1)}}{m^{2}_{126+,(0)}}\right| =\left|\frac{m^2_{126+,(1)} - m^2_{126+,(0)}}{m^2_{126+,(0)}}\right|, \label{eq:eta-126}\\
    \eta_{10} &\eqsim \left| \frac{\delta m^{2}_{10+,(1)}}{m^{2}_{126+,(0)}}\right| =\left|\frac{m^2_{10+,(1)} - m^2_{10+,(0)}}{m^2_{126+,(0)}}\right|.\label{eq:eta-10}
\end{align}
It is also convenient to define suppression ratios $R_{126}$ and $R_{10}$ of the smaller eigenvalues $m^2_{126-}$ and $m^2_{10-}$ at one-loop with respect to the generic tree-level contribution by
\begin{align}
    R_{126} &:= \left|\frac{m^2_{126-,(1)}}{m^2_{126+,(0)}}\right|, & R_{10} &:= \left|\frac{m^2_{10-,(1)}}{m^2_{126+,(0)}}\right|, \label{eq:R-126-10}
\end{align} 
where the mass $m^2_{126+,(0)}$ was again used as a proxy for the generic (non-tuned) tree-level mass-square. 

Finally, the above insights are combined in the following measures of the quality/strictness of the $1$-PL {\em pre-tuning} used in the parameter space scans (smaller $S$'s indicating better {\em pre-tuning} levels achieved):
\begin{align}
    S_{126} &:= \frac{R_{126}}{\eta_{126}^2}, &
    S_{10} &:= \frac{R_{10}}{\eta_{10}^2 \cdot 10^{-2}}. \label{eq:S-126-10}
\end{align}
Note the safety factor $10^{-2}$  introduced into the denominator of $S_{10}$; the artificially increased strictness here is meant to compensate for the use of $m^{2}_{126+,(0)}$ from the $\mathbf{126}$-sector as the generic tree-level quantity. Since there is no obstruction to attaining doublet {\em pre-tuning} in the $M^{2}_{10}$ block (as it is achieved with parameters not directly involved anywhere else in tree-level mass formulae) there is no danger from overly strict tuning demands potentially imposed on $S_{10}$ on other parts of the analysis. 

The access to the tree-level and one-loop spectrum allows for explicit numeric evaluation of the $\eta$- and $R$-quantities, and thus of the measures $S_{126}$ and $S_{10}$. A successful $1$-PL {\em pre-tuning}, i.e. a situation compatible with the basic requirements in Eq.~\eqref{eq:perturbative-tuning-definition-1}, is then characterised by $S_{126},S_{10}\lesssim 1$, which is a direct consequence of Eqs.~\eqref{eq:eta-126}--\eqref{eq:R-126-10}. This is also the technical form in which the doublet 1-PL {\em pre-tuning} is imposed as a scan constraint, as reiterated in Section~\ref{sec:viability-constraints}.

%%%%%%%%%%%%%%%%%%%%%%%%%%
\subsection{Viability and fine-tuning constraints at one-loop \label{sec:viability-constraints}}
%%%%%%%%%%%%%%%%%%%%%%%%%%

Before any doublet fine-tuning considerations, we list the general \textit{viability constraints} that valid parameter points must satisfy. The conditions are analogous to the analysis in \cite{Jarkovska:2021jvw}, which also provides further technical details, while we only give a summary and describe any updates to the conditions below:

\begin{enumerate}
	\item \underline{Non-tachyonicity of the scalar spectrum} \label{item:nontach-criterion}\\[6pt]
	The broken phase of the model has to be developed around a vacuum in which all physical scalar masses are non-tachyonic. This criterion is checked on the
  numerically evaluated one-loop regularized effective masses-squared. The computation is updated from the previous study~\cite{Jarkovska:2021jvw} by including the contributions from the $\mathbf{10}_\mathbb{C}$, both inside the loops and in the outer legs of diagrams.
	\item \underline{Gauge coupling unification} \label{item:gauge-unification-criterion} \\[6pt]
	In GUTs, the SM gauge couplings are required to unify at some high energy scale --- the unification scale. To check this, a top-down analysis is performed, where the $\mathrm{SO}(10)$ gauge coupling $g$ at the GUT scale is run down to the $Z$-boson mass scale through a sequence of effective QFTs  via renormalization group equations (RGEs). There, we demand the SM gauge couplings to have computed values within $\chi^{2}<9$ of their experimental ones.  
	In contrast to the previous study~\cite{Jarkovska:2021jvw}, the unification analysis is upgraded to two-loop RGEs with one-loop threshold corrections, see Appendix~\ref{app:two-loop-gauge-running} for further technical details.
	\item \underline{Perturbativity} \label{item:perturb-criterion}\\[6pt]
	All calculations we perform rely on perturbative methods, so perturbativity criteria every viable point must pass have to be devised. This, in turn, ensures the self-consistency of the computation as a whole. %in that specific parameter point.
	\begin{itemize}
		\item The \textit{global mass-perturbativity test} (GMP test) restricts the relative size of one-loop corrections with respect to the average tree-level mass of heavy scalars.\footnote{Heavy scalar fields are those whose masses are not $|\sigma|-$proportional and they do not belong among the would-be Goldstone bosons or pseudo-Goldstone bosons.} As in~\cite{Jarkovska:2021jvw}, it is imposed by requiring $\overline{\Delta}<1$, where the GMP measure $\overline{\Delta}$ is defined by
		\begin{align}
		\label{eq:Delta}
		\overline{\Delta} := \frac{\max_{i,j \in \text{heavy fields}}\left[|M^2_{ij,\text{one-loop}} - M^2_{ij,\text{tree}}|\right]}{\overline{M^2_{\text{heavy}}}}.
		\end{align}
		%%%%%%%%%%%%%%%%%%%%%%
		\item The next perturbativity requirement demands \textit{stability under RG running}, since the one-loop effective scalar masses, and consequently the GMP test, possess residual renormalization-scale dependence. Stability of the scalar spectrum is studied via the complete set of one-loop RGEs for dimensionless couplings, see Appendix~\ref{app:scalar-beta-functions}. The associated perturbativity measure $\bar{t}$ is defined by 
        \begin{align}
            \overline{t} &:= \sqrt{t_+ t_-}, &
		      t_\pm &:= \log_{10} \frac{\mu_{R\pm}}{\mu_R},
            \label{eq:t}
		\end{align}
		where $\mu_R$ is the initial renormalization scale and $\mu_{R\pm}$ are the upper/lower renormalization scales where a dimensionless scalar coupling hits a Landau-pole instability. We impose $t_{+}>0.5$ on all points,  i.e.~one can run the points at least half an order of magnitude up in energy without any parameter blowing up. Note that some datasets considered in the following sections have stricter additional requirements implemented on $\bar{t}$. 
		%%%%%%%%%%%%%%%%%%%%%%%%%%
		\item We compute the one-loop $(\mu^{2},\nu^{2},\tau)$-position of the vacuum and determine the shift from its initial tree-level position. The test is repeated after running the scalar parameters up half an order of magnitude in scale with the RGEs in Appendix~\ref{app:scalar-beta-functions}. As in \cite{Jarkovska:2021jvw}, we require sufficient \textit{stability of VEVs}, because their shift translates to shifts in one-loop effective scalar masses. Since such a perturbativity test is fast to evaluate, it is used as an initial perturbativity check before the full one-loop spectrum is computed.
		%%%%%%%%%%%%%%%%%
		\item The calculation of regularized effective masses (see Appendix~A in~\cite{Jarkovska:2021jvw} for details) is not reliable if the PGB masses are overly-sensitive to the mass used for these same PGBs inside the loops. We test for this instability by performing an iterative computation of PGB masses, and then by comparing the converged (fixed-point) PGB mass to the initial value. The criterion for \textit{iterative convergence of PGBs} is technically the same as in~\cite{Jarkovska:2021jvw}.
	\end{itemize}
\end{enumerate}
If a point in the parameter space satisfies all the aforementioned \textit{viability criteria}, it is referred to as \textit{viable}.
While the constraints~\ref{item:nontach-criterion} and \ref{item:perturb-criterion} are purely theoretical, i.e.~required for mathematical consistency or computational stability, constraint~\ref{item:gauge-unification-criterion} is phenomenological in nature. Note also that since there is no unambiguous measure of ``perturbativity'', the constraints under point~\ref{item:perturb-criterion} are to some degree arbitrary. 

In addition to viability constraints, we also impose a limit on the quality of the \textit{pre-tuning} in the doublet sector, cf. Sect.~\ref{sec:doublet-requirements}. Specifically, the lightest one-loop eigenvalue in each of the blocks $M^2_{126,\text{one-loop}}$ and $M^2_{10,\text{one-loop}}$ of matrix~\eqref{eq:doubletmassstructure} is required to be at most of the size of two-loop corrections, as was discussed in Section~\ref{sec:fine-tuning}. Technically, the relevant $1$-PL {\em pre-tuning} conditions are imposed by demanding 
\begin{align}
    S_{126}  &\leq 1, & S_{10} &\leq 1, \label{eq:doublet-inequality}
\end{align}
where the $S$ tuning measures were defined in Eq.~\eqref{eq:S-126-10}.

Taken all together, a candidate parameter point is considered \textit{SM-compatible} if it is viable (satisfies the general viability criteria of cases \ref{item:nontach-criterion}--\ref{item:perturb-criterion}) and its doublets are $1$-PL {\em pre-tuned} (satisfying Eq.~\eqref{eq:doublet-inequality}). We emphasize that all the aforementioned criteria actually represent necessary and not sufficient conditions: the perturbativity requirements ($\overline{\Delta}<1$ and $t_{+}>0.5$) are rather mild in the sense that any parameter point under ``reasonable'' perturbative control should easily fulfil them, and the discarded points are only those that are badly non-perturbative, while the {\em pre-tuning} condition tacitly assumes that two-loop corrections properly align. As we shall see from the results of numeric scans, these necessary conditions already limit the parameter space to very specific regions, which however are in strong tension with stricter perturbativity requirements one might find reasonable to impose (especially in the $\bar{t}$ measure).

%%%%%%%%%%%%%%%%%%%%%%%%%%
\subsection{Input parameters and their ranges \label{sec:inputs}}
%%%%%%%%%%%%%%%%%%%%%%%%%%
A point in the parameter space of the model under consideration is fully specified by providing the values of the following set of quantities: 
\begin{align}
\{ g,\; \omega_{R},\; |\sigma|,\; \chi,\; \lambda_{\mathbb{R}},\;
|\lambda_{\mathbb{C}}|,\;
\arg\lambda_{\mathbb{C}},\; \tau',\; \xi^{2},\; |\xi'^{2}|,\; \arg\xi'^{2} \},\label{eq:input-parameters}
\end{align}
where $g$ is the unified gauge coupling, $\omega_R$, $|\sigma|$ and $\chi$ encode the three relevant symmetry-breaking VEVs via Eqs.~(\ref{eq:VEV-definitions}) and (\ref{eq:chi-definition}), 
$\lambda_{\mathbb{R}}$ and $\lambda_{\mathbb{C}}$ are generic symbols for the 15 real and 14 complex dimensionless couplings in the scalar potential of Eq.~(\ref{eq:scalar-potential-0}) and $\tau'$, $\xi$ and $\xi'$ are their dimensionful counterparts related to the extra $\mathbf{10}_\mathbb{C}$, cf. Table~\ref{tab:potential-parameters}. The other potential parameters $\mu^{2}$, $\nu^{2}$ and $\tau$ are related to those in the list of Eq.~(\ref{eq:input-parameters}) via the stationarity conditions in Eqs.~(\ref{eq:mu})-(\ref{eq:tau}). 
In total, the list represents a real vector in a 51-dimensional space to be explored. 

Besides trivial limits on the values assumed by the periodic parameters 
$
    0 \leq \arg(\lambda_{\mathbb{C}}),\arg(\xi'^{2}) < 2\pi
$,
the following perturbativity conditions have been imposed:
\begin{align}
|\chi|,|\lambda_{\mathbb{R}}|,|\lambda_{\mathbb{C}}| &< 1, 
&
1\,\mathrm{GeV} < |\tau'|,|\xi|,|\xi'| &< 10\,\omega_{R},  \label{eq:input-parameter-ranges}
\end{align}
along with the correct hierarchy of the physical scales $M_Z<|\sigma|<\omega_{R}<M_{\text{Pl}}$ (with $M_{\text{Pl}}$ and $M_{Z}$ denoting the Planck and the Z-boson mass, respectively).

%%%%%%%%%%%%%%%%%%%%%%%%%%
\subsection{Results of the numerical analysis \label{sec:results}}
%%%%%%%%%%%%%%%%%%%%%%%%%%
The searches for   \textit{viable} parameter points, as defined in Section~\ref{sec:viability-constraints}, with various degrees of \textit{pre-tuning} (as measured by $S_{126}$ and $S_{10}$) 
were performed using the same methods as in~\cite{Jarkovska:2021jvw}, namely the stochastic version of the differential evolution algorithm~\cite{diffevolalg} (version ``DE/rand/1''~with randomly selected $F \in (0.5,2)$ for every point). Every undesired feature or restriction on a parameter point was implemented as a non-negative penalization. The algorithm then iteratively improves upon generations of points by constructing new candidates from the preceding generation. One can use the algorithm in two modes: (i) the minimization mode, in which one tries to find the best possible point with certain features, and (ii) the search mode, where one accepts all newly found points with zero penalization and thus explores the parameter space of points with desired features. 

%%%%%%%%%%%%%%%%%%%%%%%%
\subsubsection{A SM-compatible benchmark point \label{sec:results-best}}
%%%%%%%%%%%%%%%%%%%%%%%%%%

We first present in Table~\ref{tab:benchmark-point} a \textit{SM-compatible} benchmark point (BP), i.e.~a point that is both \textit{viable} and satisfies the 1-PL \textit{pre-tuning} requirements $S_{10}\leq 1$ and $S_{126}\leq 1$ of Eq.~\eqref{eq:doublet-inequality}. It was obtained as the best point (with the lowest $S_{126}$) after a considerable computational effort of about $10^5$ generations using differential evolution. While the basic {\em viability} (as defined in Sect.~\ref{sec:viability-constraints}) and the $S_{10}\leq 1$ criteria were quickly satisfied in the course of minimization, it turned out to be very difficult to satisfy $S_{126}\leq 1$. Ultimately, the minimal attained value corresponds to the BP at  
\begin{align}
    S_{126}^{\text{BP}} &= 0.915 . \label{eq:s126-best}
\end{align} 
We remind the reader that $S_{126}$ defines the level of $1$-PL \textit{pre-tuning} in the $M^{2}_{126}$ block of the doublet mass matrix in Eq.~\eqref{eq:doubletmassstructure}. The $1$-PL \textit{pre-tuning} difficulty in the $\mathbf{126}$-sector, however, was not entirely unexpected given the obstructions arising already for $0$-PL \textit{pre-tuning} in Section~\ref{sec:tree-level-fine-tuning}. 
Note that the value in Eq.~(\ref{eq:s126-best}) is not claimed to be the absolute minimum, since the minimization was stopped manually
for practical (time) reasons. Nevertheless, given the clear convergence trends in the spread of parameter values in successive generations and the exponentially increasing difficulty in finding better points as the search progressed, any further improvement of the value quoted above -- if it exists -- should be very small. 
\begin{table}[htb]     \caption{Parameter input values for the \textit{SM-compatible} benchmark point. For the values of dimensionful parameters, the ``sign-log'' function $\slog(x):=\sign(x) \cdot \log_{10}(|x|/\mathrm{GeV})$ has been employed. 
}	\label{tab:benchmark-point}
	\vspace{0.1cm}
    \centering
	\begin{tabular}{rr}
		\toprule
		parameter&value\\
		\midrule
		$ a_0$ & $\phantom{+}0.096$ \\
		$ a_2$ & $\phantom{+}0.023$ \\
		$\lambda_0$ & $\phantom{+}0.139$ \\
		$\lambda_2$ & $\phantom{+}0.613$ \\
		$\lambda_4$ & $-0.735$ \\
		$\lambda_4^\prime$ & $\phantom{+}0.040$ \\
		$\alpha$ & $-0.050$ \\
		$\beta_4$ & $\phantom{+}0.739$ \\
		$\beta_4^\prime$ & $-0.153$ \\
		$|\gamma_2|$ & $\phantom{+}0.286$ \\
		$\arg \gamma_2$ & $\phantom{+}1.978$ \\
		$|\eta_2|$ & $\phantom{+}0.355$ \\
		$\arg \eta_2$ & $\phantom{+}0.709$ \\
		\bottomrule	
	\end{tabular}
 \hspace{0.01\linewidth}
	\begin{tabular}{rr}
		\toprule
		parameter&value\\
		\midrule
  		$h_2$ & $-0.379$ \\
		$h_2^\prime$ & $\phantom{+}0.301$ \\
		$\kappa_0$ & $\phantom{+}0.212$ \\
		$\kappa_2$ & $-0.457$ \\
		$\rho_0$ & $-0.564$ \\
		$\rho_2$ & $-0.707$ \\
		$|h_3|$ & $\phantom{+}0.481$ \\
		$\arg h_3$ & $\phantom{+}3.472$ \\
		$|h_4|$ & $\phantom{+}0.496$ \\
		$\arg h_4$ & $\phantom{+}0.104$ \\
		$|\kappa_0^\prime|$ & $\phantom{+}0.005$ \\
		$\arg \kappa_0^\prime$ & $\phantom{+}1.369$ \\
		$|\kappa_2^\prime|$ & $\phantom{+}0.008$ \\
		\bottomrule	
	\end{tabular}
 \hspace{0.01\linewidth}
	\begin{tabular}{rr}
		\toprule
		parameter&value\\
		\midrule
        $\arg \kappa_2^\prime$ & $\phantom{+}1.917$ \\
		$|\zeta|$ & $\phantom{+}0.483$ \\
		$\arg \zeta$ & $\phantom{+}0.440$ \\
		$|\zeta^\prime|$ & $\phantom{+}0.432$ \\
		$\arg \zeta^\prime$ & $\phantom{+}1.502$ \\
		$|\rho_0^\prime|$ & $\phantom{+}0.076$ \\
		$\arg \rho_0^\prime$ & $\phantom{+}1.841$ \\
		$|\psi_0|$ & $\phantom{+}0.238$ \\
		$\arg \psi_0$ & $\phantom{+}0.003$ \\
		$|\psi_1|$ & $\phantom{+}0.177$ \\
		$\arg\psi_1$ & $\phantom{+}0.741$ \\
		$|\psi_2|$ & $\phantom{+}0.658$ \\
		$\arg \psi_2$ & $\phantom{+}1.502$ \\
		\bottomrule	
	\end{tabular}
\hspace{0.01\linewidth}
	\begin{tabular}{rr}
		\toprule
		parameter&value\\
		\midrule
        $|\varphi|$ & $\phantom{+}0.016$ \\
		$\arg \varphi$ & $\phantom{+}0.756$ \\
		$|\varphi^\prime|$ & $\phantom{+}0.160$ \\
		$\arg \varphi^\prime$ & $\phantom{+}1.647$ \\
        $g$& $\phantom{+}0.550$\\
        $\chi$ & $-0.873$ \\
        $\slog \omega_{R }$ & $\phantom{+}14.752$ \\
		$\slog \sigma$ & $\phantom{+}11.082$ \\
		$\slog \tau^\prime$ & $-15.706$ \\
		$\slog |\xi|$ & $\phantom{+}0.189$ \\
        $\sign \xi^2$ & $+1$ \\
		$\slog |\xi^{\prime }|$ & $\phantom{+}10.853$ \\
		$\arg \xi^{\prime}$ & $\phantom{+}2.409$ \\
		\bottomrule	
	\end{tabular}
\end{table}

In any case, the benchmark point of Table~\ref{tab:benchmark-point} represents an existence proof that \textit{SM-compatible} points do in principle exist. However, it is important to recall that the imposed viability criteria of Section~\ref{sec:viability-constraints} are not very strict regarding perturbativity; indeed, the corresponding $\bar{t}$ perturbativity measure~(\ref{eq:t}) attains a rather small  value at the BP, namely
\begin{align}
    \bar{t}\,{}^{\text{BP}}&= 0.38.
\end{align}
The RG flow of the scalar parameters thus hits a pole within $0.38$ orders of magnitude around the GUT scale (up-down average), which hardly provides confidence in the perturbative stability of such a point. 
As it turns out, this is actually a rather general pattern: \textit{attainability of the desired level of fine-tuning in the doublet sector strongly correlates with the regions of parameter space which are perturbatively disfavored}.

%%%%%%%%%%%%%%%%%%%%%%%%
\subsubsection{Tension between doublet fine-tuning and perturbativity \label{sec:results-shape}}
%%%%%%%%%%%%%%%%%%%%%%%%%%
As the claim above is the central point of the current study, it deserves a closer look and a more detailed analysis.
For that sake, we have used the DE algorithm in search mode, i.e.~as a tool for exploring the parameter space with different penalization functions corresponding to various thresholds of strictness in the imposed constraints, cf. Table~\ref{tab:datasets}.

All points accepted into the datasets are \textit{viable} as defined in Section~\ref{sec:viability-constraints}, but differ in the acceptance threshold either for the level of doublet fine-tuning or for the level of their perturbative stability. To this end, the dataset $\mathcal{D}$ in Table~\ref{tab:datasets} is a mere reference sample with no further imposed constraints on $S_{10}$, $S_{126}$ and $\bar{t}$, and as such represents a general map of the \textit{viable} part of parameter space in the $\mathbf{45}\oplus \mathbf{126}\oplus \mathbf{10}_\mathbb{C}$  model.\footnote{The $\mathcal{D}$ dataset makes it also possible to compare the viable parameter space regions between the current $\mathbf{45}\oplus \mathbf{126}\oplus \mathbf{10}_\mathbb{C}$ model and the simpler $\mathbf{45}\oplus \mathbf{126}$ setting studied previously in ref.~\cite{Jarkovska:2021jvw}, see Appendix~\ref{app:full-vs-Higgs}.} The $\mathcal{S}^{(x)}$ datasets for $x=2.0$, $1.5$ and $1.0$ conform to the $1$-PL \textit{pre-tuning} requirement in the $M^{2}_{10}$ block of Eq.~\eqref{eq:doubletmassstructure} imposed by $S_{10}\leq 1$, along with an increasingly demanding constraint on the level of $1$-PL \textit{pre-tuning} in the $M^{2}_{126}$ block implemented via the corresponding $S_{126}\leq x$ condition. In essence, such $\mathcal{S}^{(x)}$ datasets explore smaller and smaller neighborhoods around the benchmark point of Table~\ref{tab:benchmark-point}. Finally, the  dataset $\mathcal{T}$ defined by $\bar{t} \geq 1$ explores the parameter-space domains with extended RG-flow perturbativity/stability.

\begin{table}[htb]
	\centering
	\caption{The computed datasets of parameter points along with the constraints imposed on top of \textit{viability} in each case. The last column provides the resulting ranges of $\bar{t}$ in the datasets.  \label{tab:datasets}}
	\vskip 0.1cm
	\begin{tabular}{l@{$\quad$}r@{$\qquad$}l@{$\qquad$}l}
		\toprule
		Dataset&$\#$ of points&constraints&$\overline{t}$ range\\
		\midrule
		$\mathcal{D}$ & 20000 & unconstrained & $0.36-1.06$\\
		$\mathcal{S}^{(2.0)}$ & 2000 & $S_{10}\leq 1$, $S_{126} \leq 2$ & $0.34 - 0.49$\\
		$\mathcal{S}^{(1.5)}$ & 2000& $S_{10}\leq 1$,  $S_{126} \leq 1.5$ & $0.34 - 0.52$\\
        $\mathcal{S}^{(1.0)}$ & 2000& $S_{10}\leq 1$,  $S_{126} \leq 1$ & $0.34 - 0.41$\\
        $\mathcal{T}$ & 20000 & $\overline{t} \geq 1$ & $1.00 -1.20$\\
		\bottomrule
	\end{tabular}
\end{table}

Even a quick glance at Table~\ref{tab:datasets} already reveals a tension between the $\mathcal{T}$ dataset and the $\mathcal{S}^{(x)}$ datasets. 
Indeed, by increasing the demands on the level of \textit{pre-tuning} in $M^2_{126}$ (corresponding to the progression from $x=2.0$ to $x=1.0$ in $\mathcal{S}^{(x)}$), the maximum attainable $\bar{t}$ (reflecting the level of perturbativity) is diminishing and never reaches above
$\bar{t}\sim 0.52$. 

\begin{figure}[htb]
	\centering
    \mbox{
    \includegraphics[width=12cm]{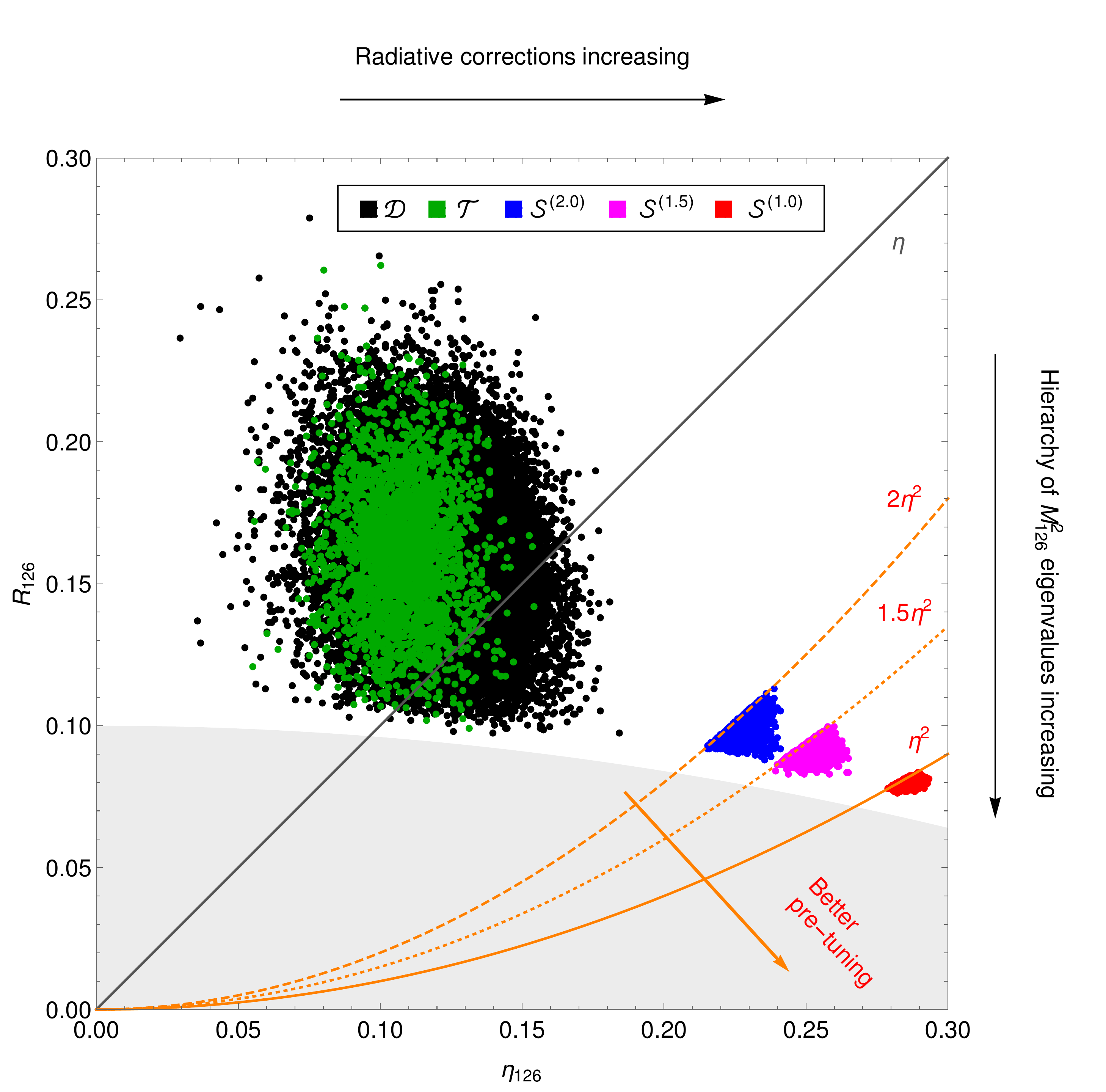}
    }
    \vspace{-0.2cm}
	\caption{
        The plot compares the one-loop tuning ability $R_{126}$ to the loop expansion parameter $\eta_{126}$ for points in the datasets of Table~\ref{tab:datasets}.
        On the vertical axis the ratio $R_{126}$ determines the relative size of the smaller eigenvalue (i.e. the one to be suppressed as much as possible upon {\em pre-tuning}) of the one-loop $M_{126}^2$ block of the mass matrix in Eq.~\eqref{eq:doubletmassstructure} as compared to the  larger one. Physically, the lower limit on $R_{126}$ at around 0.08 indicates that the two eigenvalues of $M_{126}^2$ cannot be separated arbitrarily, i.e. points cannot enter the grey region, and hence the only way to attain the desired level of \textit{pre-tuning} is to resort to large loop corrections.  
        The estimate of the relative size of such loop corrections (encoded in the $\eta_{126}$ parameter) for any point in the plot is quantified by its horizontal coordinate.
        The typical non-tuned points are around the grey line $R_{126}=\eta_{126}$, while the properly tuned points satisfying $S_{126}< 1$ are below the orange-colored $\eta^{2}$ curve.
        The plot makes it clear that the more pre-tuning in $M_{126}^2$ is imposed, the more to the right one has to push, and hence larger loop corrections have to be invoked.   
	\label{fig:R126eta126}}
\end{figure}

The same effect is easily seen in Figure~\ref{fig:R126eta126}, where each point from the datasets is positioned according to its one-loop ``tuning capacity'' $R_{126}$ of Eq.~\eqref{eq:R-126-10} and the loop expansion parameter $\eta_{126}$ of Eq.~\eqref{eq:eta-126}. 
The color coding for points from different datasets of Table~\ref{tab:datasets} in Figure~\ref{fig:R126eta126} and later plots is as follows: the points of the unconstrained dataset $\mathcal{D}$ are colored \textbf{black}, the dataset $\mathcal{T}$ with enhanced perturbativity is colored \textbf{\textcolor{plotGreen}{green}}, while the datasets $\mathcal{S}^{(2.0)}$, $\mathcal{S}^{(1.5)}$ and $\mathcal{S}^{(1.0)}$ with increased tuning attained in the doublet sector are colored \textbf{\textcolor{plotBlue}{blue}}, \textbf{\textcolor{plotMagenta}{magenta}} and \textbf{\textcolor{plotRed}{red}}, respectively.

Intuitively, the horizontal axis in Figure~\ref{fig:R126eta126} indicates how large in relative terms the loop corrections are, while the vertical axis quantifies how much the one-loop mass of the light doublet in the $M^{2}_{126}$ block is tuned compared to the corresponding heavy eigenstate. The straight grey line in the figure (corresponding to $R_{126}=\eta_{126}$) represents the points where the one-loop mass of the light doublet is of the size of a typical one-loop contribution, while the orange curve labeled by $\eta^{2}$ defines the upper boundary of the region
where the light doublet in $M_{126}^2$ has been successfully $1$-PL \textit{pre-tuned} down to the typical scale of two-loop effects. One can see that with decreasing $x$ the points of the $\mathcal{S}^{(x)}$ datasets are receding further and further away from the black bulk of the unconstrained dataset $\mathcal{D}$, in accordance with the $S_{126}\leq x$ limits depicted by the relevant $x\cdot\eta^2$ curves,  cf.~definition in Eq.~\eqref{eq:S-126-10}.  Notice also that for smaller $x$ the corresponding $\mathcal{S}^{(x)}$ conditions are fulfilled in gradually shrinking domains moving to the right rather than down in the plot, which means that the tuning of the light doublet mass in the $M_{126}^2$ down to the desired level of typical two-loop effects is achieved by enlarging the expected size of the loop corrections $\eta_{126}$ rather than suppressing the light eigenstate mass. As a consequence, the desired level of fine-tuning in $M_{126}^2$ is attained only for relatively large $\eta_{126}\gtrsim 0.28$, i.e., at the expense of significantly reduced perturbativity (i.e., small $\bar t$) of the corresponding points. On the contrary, the points with improved perturbative behaviour (corresponding to the $\mathcal{T}$ dataset with $\bar t>1$) generally cluster in the $\eta_{126}<0.15$ domain, hence indicating a fundamental clash among the two criteria.

\begin{figure}[h!]
	\centering
\includegraphics[width=16cm]{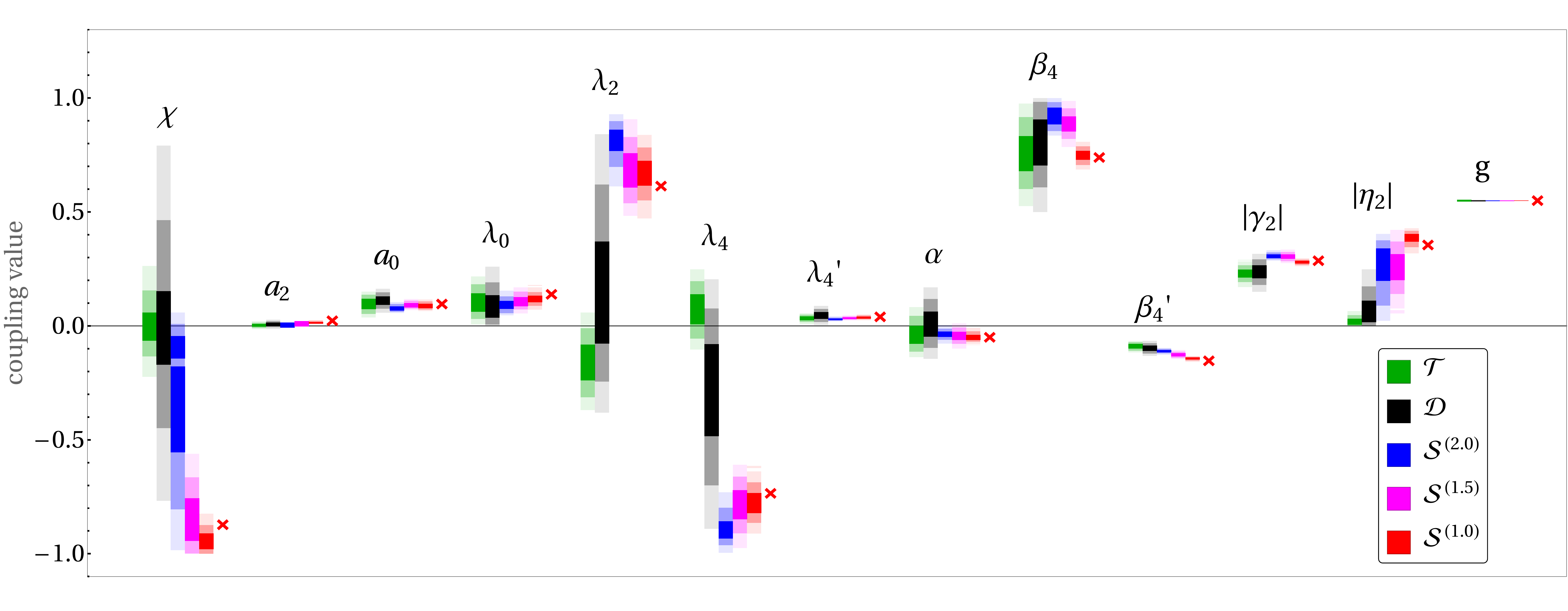}
    \vspace{-0.4cm}
	\caption{
		The ranges of the ``old'' dimensionless scalar parameters from Table~\ref{tab:potential-parameters} (i.e.~those appearing already in the $\mathbf{45}\oplus \mathbf{126}$ Higgs model of Ref.~\cite{Jarkovska:2021jvw}) along with the universal VEV ratio $\chi$ of Eq.~\eqref{eq:chi-definition} for different datasets defined in Table~\ref{tab:datasets}. 
		The same general color scheme as in Figure~\ref{fig:R126eta126} has been adopted, with decreasing color opacities representing the $1$-, $2$- and $3$-$\sigma$ \textit{highest density intervals} (HDIs) for each quantity. In the Bayesian interpretation these would correspond to HDIs of marginal probability distributions in a single parameter when the sampling of acceptable regions is performed by the differential evolution algorithm.
        The values corresponding to the benchmark point in Table~\ref{tab:benchmark-point} are marked by a red cross.
    \label{fig:old-dimless-scalar-parameters-finetuned}
  }
\end{figure}

\begin{figure}[h!]
	\centering
    \includegraphics[width=16cm]{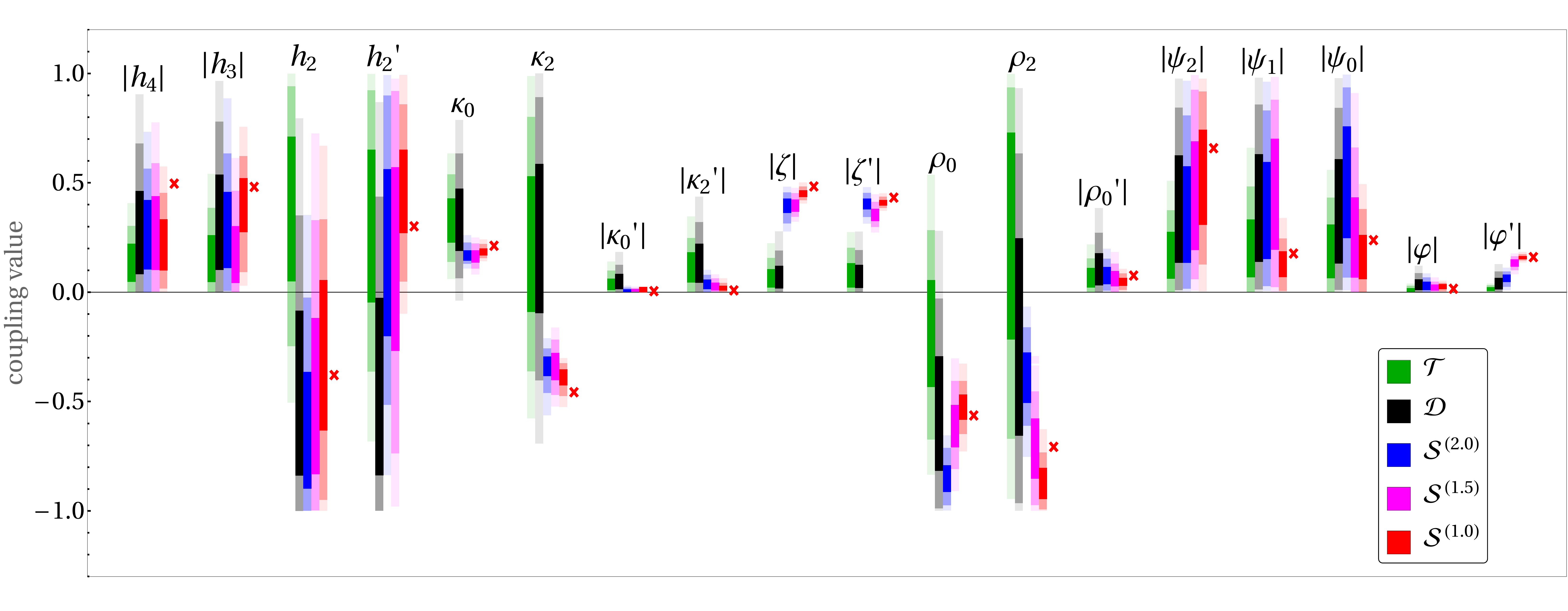}
    \vspace{-0.4cm}
	\caption{ 
    The HDIs (same conventions as in Figure~\ref{fig:old-dimless-scalar-parameters-finetuned})
  for the new dimensionless scalar parameters specific to the model with the added $\mathbf{10}_\mathbb{C}$, cf. Table~\ref{tab:potential-parameters}.
    \label{fig:new-dimless-parameters-finetuned}
	}
\end{figure}

\begin{figure}[h!]
    \includegraphics[width=8cm]{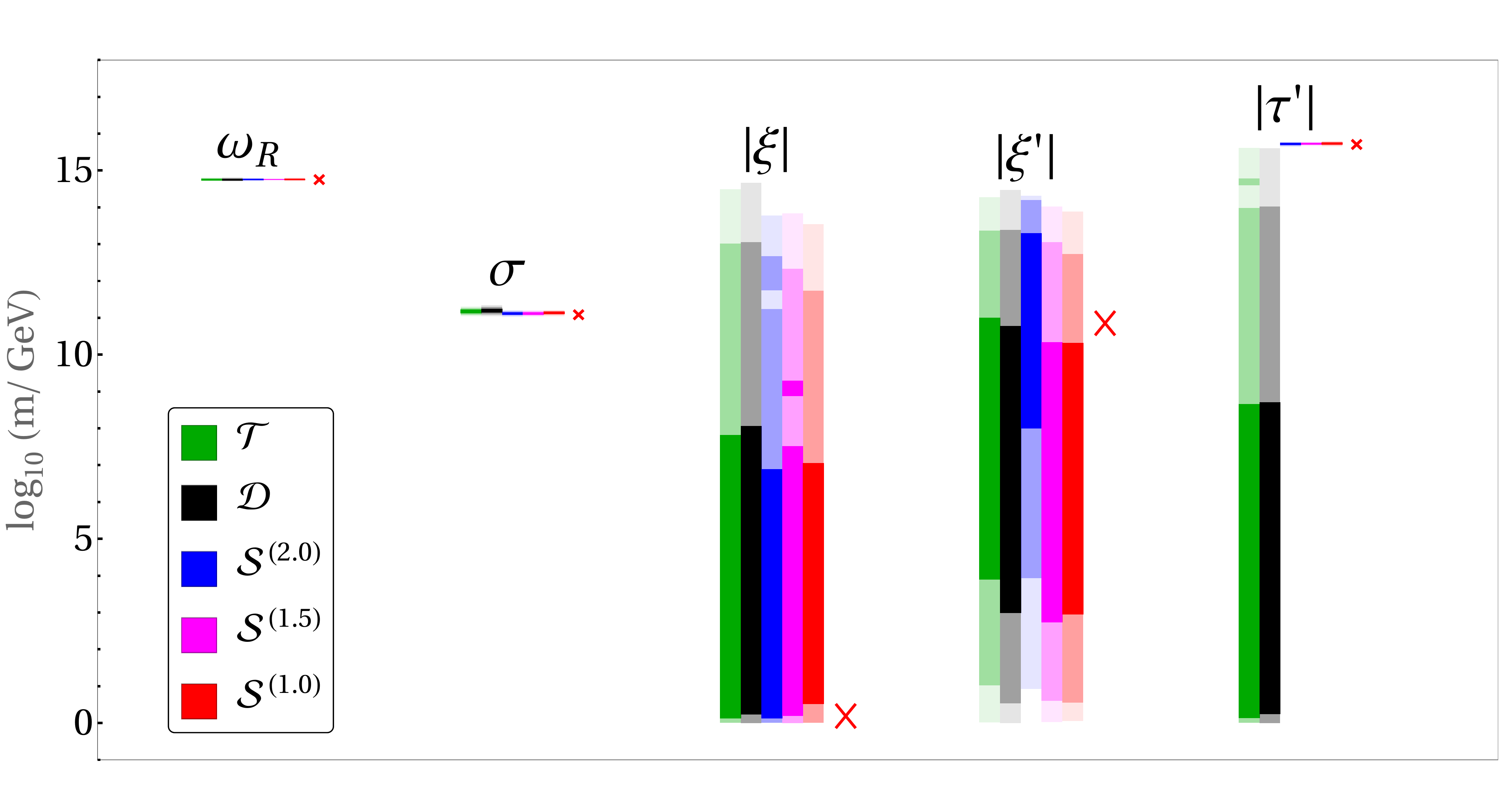}
    \vspace{-0.4cm}
	\caption{ 
    The HDIs (same conventions as in Figure~\ref{fig:old-dimless-scalar-parameters-finetuned})
  for the dimensionful input parameters. 
    \label{fig:dimful-parameters-finetuned}
	}
\end{figure}

The non-compatibility of  $\mathcal{S}^{(x)}$ with $\mathcal{T}$ is also apparent from the shape of the parameter space regions that support each of these datasets, see Figures~\ref{fig:old-dimless-scalar-parameters-finetuned}--\ref{fig:dimful-parameters-finetuned} .  
In particular, there is a clear tension between the values of $\lambda_2$, $\lambda_4$ and $|\eta_2|$, which for the $\mathcal{S}$-type datasets are generally driven towards relatively large (absolute) values, while the same couplings tend to be suppressed across the $\mathcal{T}$ dataset, cf. Figure~\ref{fig:old-dimless-scalar-parameters-finetuned}. This pattern can be understood qualitatively as a consequence of the complexity of the doublet-sector fine-tuning, which generally requires a non-trivial conspiracy among certain  parameters; these, in turn, tend to assume values away from their ``natural'' perturbativity domains in the vicinity of $0$.

%%%%%%%
%%%%%%%

%%%%%%%
%%%%%%%%

%%%%%%%
%%%%%%%%

\begin{figure}[htb]
	\centering
    \includegraphics[width=8cm]{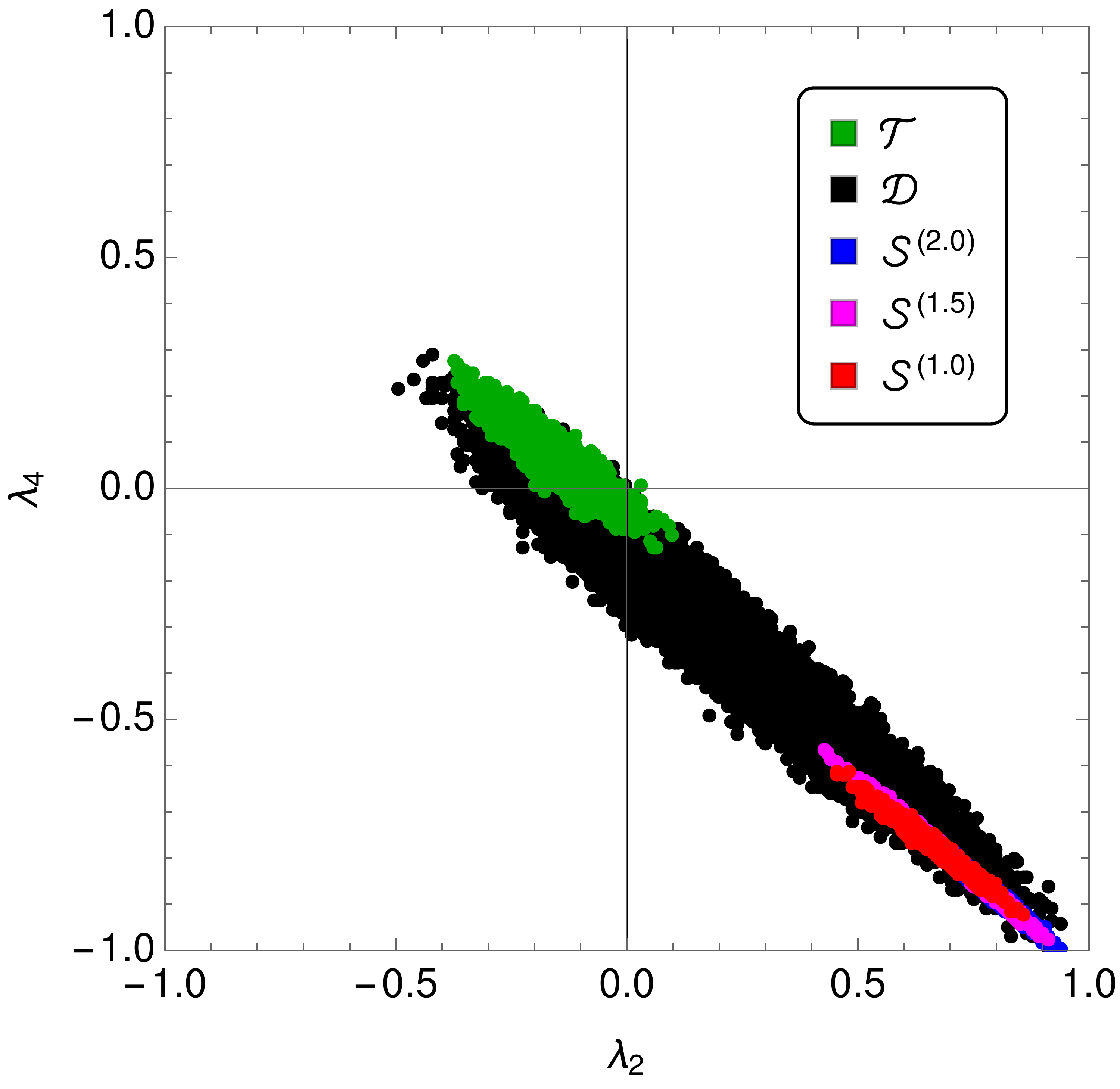}
    \vspace{-0.4cm}
	\caption{
		The correlation plot in the inputs $\lambda_2$ and $\lambda_4$ for the datasets of Table~\ref{tab:datasets}. Enhanced perturbativity (the $\mathcal{T}$ dataset) and $1$-PL {\em pre-tuning} (the $\mathcal{S}^{(1.0)}$ dataset) are incompatible.
	\label{fig:lambda2-lambda4}
 }
\end{figure}

The tension is further illustrated in the $\lambda_{2}$-$\lambda_{4}$ plane in Figure~\ref{fig:lambda2-lambda4}, which not only depicts the strong anti-correlation of the two parameters across all datasets (a feature that has been observed already in the simplified ${\bf 45}\oplus {\bf 126}$ case of Ref.~\cite{Jarkovska:2021jvw}), but makes it very clear that the $\mathcal{T}$ and $\mathcal{S}$ regions are fundamentally incompatible.

%%%%%%%%
%%%%%%%%

Let us conclude this section with a few more comments concerning the results presented above, which may elucidate some further details of the underlying numerical analysis:  
\begin{itemize}
    %%%%%%%%%%%%%%%%%%%%%%%%%%%%%%%%%%%
    \item Gauge unification constraints along with perturbativity push  
    $g$, $\omega_{R}$ and $\sigma$ into very narrow ranges, cf.
    Figures~\ref{fig:old-dimless-scalar-parameters-finetuned} and \ref{fig:dimful-parameters-finetuned}, thus fixing the unification scale and to a large degree also the spectrum of the GUT-scale gauge leptoquarks. A similar effect is observed for the majority of the ``old'' dimensionless parameters of the ${\bf 45}\oplus {\bf 126}$ Higgs model of Ref.~\cite{Jarkovska:2021jvw}, cf. Figure~\ref{fig:old-dimless-scalar-parameters-finetuned}; indeed, these parameters dominate the leading contributions to the masses of most scalar fields in the model, and are hence restricted by viability constraints even in the generic dataset $\mathcal{D}$. 
    \item
    On the other hand, most of the parameters specific to the $\mathbf{10}_\mathbb{C}$ extension in Figure~\ref{fig:new-dimless-parameters-finetuned} stretch over relatively large domains\footnote{This behaviour, however, is not entirely universal, as some of the new parameters such as $\varphi$ and $\varphi'$ still turn out to be quite constrained.} ---  this should not be surprising, since these parameters
    influence only a small part of the tree-level spectrum (only the doublet and triplet mass matrices of~Eqs.~\eqref{eq:doubletmassmatrix} and \eqref{eq:tripletmassmatrix} are affected). Similarly, the limited number of additional degrees of freedom the extra $\mathbf{10}_\mathbb{C}$ brings in should not have a big impact on the shape of the RGEs for the ``old'' scalar parameters, which play a major role in the determination of the perturbative stability domains. 
    %%%%%%%%%%%%%%%%%%%%%%%%%%%%%%%%%%%
	\item 
    The tension between the \textit{viable} parameter space domains favoured by the dataset $\mathcal{T}$ with enhanced perturbative stability and the fine-tuned datasets $\mathcal{S}^{(x)}$, cf.~Figs.~\ref{fig:old-dimless-scalar-parameters-finetuned}--\ref{fig:dimful-parameters-finetuned}, is not limited to the aforementioned $\lambda_2$--$\lambda_4$--$|\eta_2|$ subset. Indeed, a similar (though perhaps less pronounced) behaviour is observed also in $\chi$,  $\beta'_{4}$, $\kappa_{2}$, $|\zeta|$, $|\zeta'|$ and $\varphi'$.  
	%%%%%%%%%%%%%%%%%%%%%%%%%%%%%%%%%%%
    \item There is a relatively large qualitative difference between the behaviour of the $\tau'$ parameter in the $\mathcal{T}$ and $\mathcal{D}$ datasets as compared to the $\mathcal{S}^{(x)}$ ones, in which the doublet {\em pre-tuning} has been implemented, cf. Fig.~\ref{fig:dimful-parameters-finetuned}. This is a mere consequence of the additional algebraic constraint implemented in the $M^2_{10}$ block of the doublet mass matrix of Eq.~\eqref{eq:doubletmassstructure} in the latter case.
    %
    %%%%%%%%%%%%%%%%%%%%%%%%%%%%%%%%%%%%%%%
    \item
    The phases of the complex parameters (such as $\eta_2$, $\gamma_2$, etc.) are not displayed in Figures~\ref{fig:old-dimless-scalar-parameters-finetuned} and \ref{fig:new-dimless-parameters-finetuned}, because the values they assume are not critical to the key observations made above. Nevertheless, as shown in Figure~\ref{fig:phase-scatter-plots}, they are not  entirely insensitive to the stringent {\em pre-tuning} constraints imposed in the doublet sector and interesting correlations may emerge, especially for the $\mathcal{S}^{(x)}$ datasets with small $x$.     
    \begin{figure}[h!]
        \centering
        \includegraphics[width=6cm]{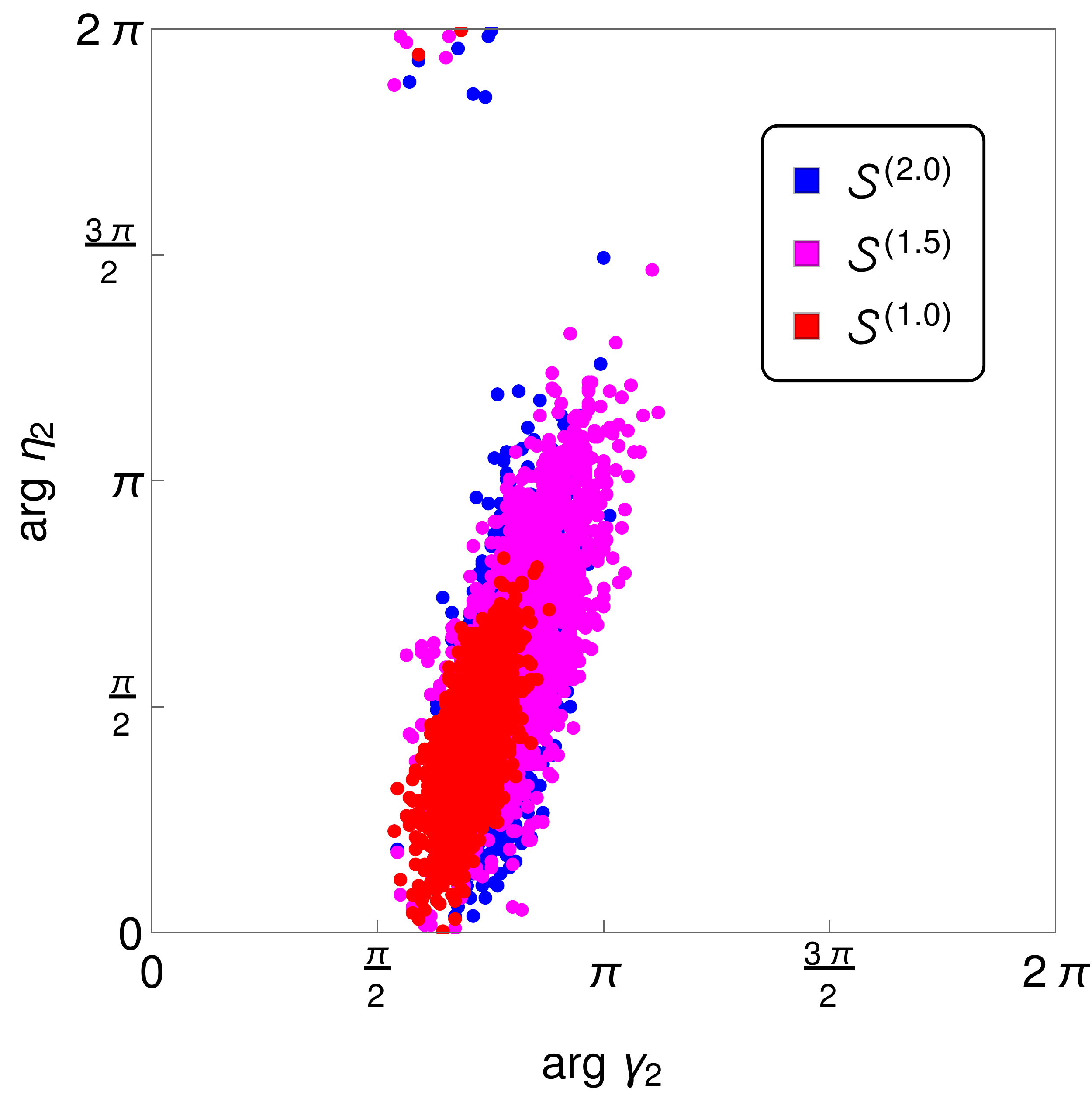}
        \hspace{0.5cm}
        \includegraphics[width=6cm]{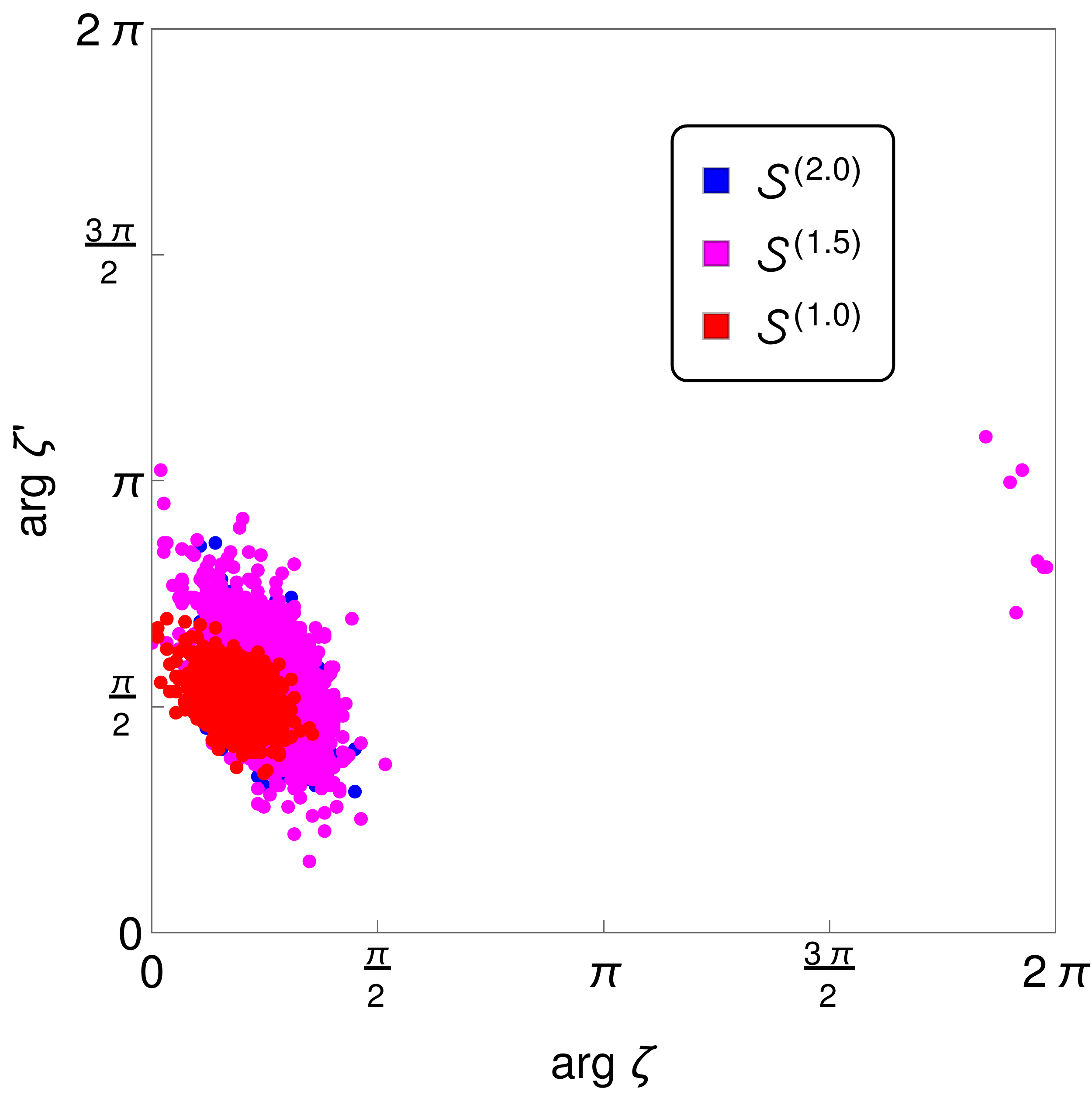}
        \vspace{-0.2cm}
        \caption{
        Correlations of the phases of two selected pairs of complex parameters ($\gamma_{2}$-$\eta_{2}$ on the left and $\zeta$-$\zeta'$ on the right) in the $\mathcal{S}^{(x)}$ datasets of Table~\ref{tab:datasets}. Note that points from the datasets $\mathcal{D}$ and $\mathcal{T}$ are not shown, since the phases there are uniformly distributed over the entire $[0,2\pi)$ interval and no interesting correlations emerge.  
        \label{fig:phase-scatter-plots}
        }
    \end{figure}
    %%%%%%%%%%%%%%%%%%%%%%%%%%%%%%%%%%%
\end{itemize}

%%%%%%%%%%%%%%%%%%%%%%%%%%
\section{Conclusions \label{sec:conclusions-and-outlook}}
%%%%%%%%%%%%%%%%%%%%%%%%%%

The renormalizable $\mathrm{SO}(10)$ model with the $\mathbf{45}\oplus\mathbf{126}\oplus\mathbf{10}_\mathbb{C}$ Higgs sector (with $\mathbb{C}$ denoting complexification of the relevant representation) has all the features to be considered minimal of its kind, and at the same time potentially realistic. The $\mathbf{45}\oplus\mathbf{126}$ part of its scalar sector is sufficient to implement the desired GUT symmetry breaking down to the SM level, whilst realistic Yukawa sector fits are known to be possible with $\mathbf{126}\oplus\mathbf{10}_{\mathbb{C}}$ at play.

In practice, however, the model is rather complicated to work with because of the general tachyonic instability of its tree-level scalar spectrum encountered along all physically viable symmetry breaking chains; this in turn makes
it purely quantum in nature and calls for a detailed approach with loop corrections taken into account. Moreover, the recent thorough one-loop analysis~\cite{Jarkovska:2021jvw} (focusing on the $\mathbf{45}\oplus\mathbf{126}$ part of the scalar sector) revealed that in major parts of its parameter space the underlying Higgs model suffers from perturbativity issues, which can be alleviated only in a very specific regime featuring an intermediate  $\mathrm{SU}(4)_C\times\mathrm{SU}(2)_L\times\mathrm{U}(1)_R$ symmetry stage with the GUT scale localized below $10^{15}$~GeV.

Remarkably enough, even with such a low scale governing the $d=6$ baryon number violating processes, one should be very careful with discarding the $\mathbf{45}\oplus\mathbf{126}$-type of models on the basis of their alleged incompatibility with existing proton lifetime limits for at least three reasons: in~\cite{Jarkovska:2021jvw}
(i) the position of the GUT scale quoted above has been estimated only at the lowest non-trivial order of perturbative expansion; (ii) the possible effects of the additional $\mathbf{10}_\mathbb{C}$ present in the minimal potentially realistic setup have not been taken into account; (iii) little attention has been paid to the special regions in the parameter space which may support accidentally light thresholds in the GUT desert with a potentially large impact on $M_{GUT}$, cf.~\cite{Bertolini:2013vta,Kolesova:2014mfa}.   

In the current analysis, we carefully addressed points (i) and (ii) above and revealed a remarkable feature of the full $\mathbf{45}\oplus\mathbf{126}\oplus\mathbf{10}_\mathbb{C}$ setting that makes the concern (iii) entirely obsolete. In particular, we argued that in the perturbative regime the model cannot accommodate  a sufficiently large component of the ${\bf 126}$ scalar admixed within the SM Higgs doublet; hence, {\em realistic fits of the quark and lepton masses and mixings are inconceivable at the perturbative level in the $\mathbf{45}\oplus\mathbf{126}\oplus\mathbf{10}_\mathbb{C}$  scenario}. This important observation, hinted at semi-analytically already by the doublet mass matrix at tree-level (once the one-loop parameter ranges from~\cite{Jarkovska:2021jvw} are taken into account), is fully supported by a detailed numerical analysis of its quantum-level structure.

Beyond the specific GUT model under consideration, this study further illustrates the general importance 
of quantum corrections in models with a large number of degrees of freedom. Perturbativity considerations of the type implemented in this paper greatly restrict the domain where the model is calculable, and may in fact be incompatible with the set of other demands or limitations imposed on the model (such a fine-tuning).

%%%%%%%%%%%%%%%%%%%%%%%%%%%%%%%%%%%%%%%%%%%%%%%%%%%%%%%
\section*{Acknowledgments}
%%%%%%%%%%%%%%%%%%%%%%%%%%%%%%%%%%%%%%%%%%%%%%%%%%%%%%%
The authors acknowledge financial support from the Grant Agency of the Czech Republic (GA\v{C}R) via Contract No.~20-17490S, as well as from the Charles University Research Center Grant No.~UNCE/SCI/013. K.J.~was also supported by GAUK, Project No.~1558119. Special thanks goes to Timon Mede for discussions in the early stages of this project.

\appendix

%%%%%%%%%%%%%%%%%%%%%%%%%%
\section{Comparing the full model with the Higgs model\label{app:full-vs-Higgs}}
%%%%%%%%%%%%%%%%%%%%%%%%%%

In this Appendix, we compare the shape of the viable parameter space of the $\mathbf{45} \oplus \mathbf{126}$ Higgs model from~\cite{Jarkovska:2021jvw}  with that of the 
full $\mathrm{SO}(10)$ model $\mathbf{45} \oplus \mathbf{126} \oplus \mathbf{10}_{\mathbb{C}}$ studied in this paper. 

The full model includes an additional complexified representation $\mathbf{10}_{\mathbb{C}}$ in the scalar sector, and hence new couplings in the scalar potential; for a full list of parameters see Table~\ref{tab:potential-parameters}. Moreover, unlike in the previous study~\cite{Jarkovska:2021jvw}, where the gauge coupling RG analysis has been done at the one-loop level, the unification constraints in the current study have been implemented at two loops\footnote{Note that in this two-loop unification analysis, the $SU(2)_L$ doublets are assumed to have been successfully fine-tuned, and thus their masses are set to correspond to the physically desired setup: one light eigenstate is identified with the SM Higgs, another lies at around the $\sigma$-scale, while the remaining two have masses near the GUT scale. }. Nevertheless, since $\mathbf{10}_\mathbb{C}$ is a small representation that is not involved in the symmetry breaking, and since the contributions driven by the parameters which are shared by both variants (namely, those of the  $\mathbf{45}\oplus \mathbf{126}$ sector, cf.~Table~\ref{tab:potential-parameters}) dominate most of the tree-level scalar masses, it has always been implicitly assumed in previous analyses that the corresponding regions of the viable parameter space of the two model variants is very similar, with perhaps only minor quantitative differences. With the results of the current study at hand, this proposition can now be tested.

In practice we compare the dataset $\mathcal{B}$ composed of the \textit{viable} parameter-space points of the $\mathbf{45} \oplus \mathbf{126}$ Higgs model~\cite{Jarkovska:2021jvw} (labeled $B_+$ therein) with the dataset $\mathcal{D}$ of the \textit{viable} points of the full $\mathbf{45} \oplus \mathbf{126} \oplus \mathbf{10}_{\mathbb{C}}$ model discussed in Section~\ref{sec:results-shape}. At the technical level, these two datasets are compared in Table~\ref{tab:dataset-difference}.
\begin{table*}[htb]
	\centering
	\caption{Comparison of the $\mathcal{B}$ and $\mathcal{D}$ datasets containing viable points within the two variants of the model as discussed in~\cite{Jarkovska:2021jvw} and in the current study.  \label{tab:dataset-difference}}
	\vskip 0.1cm
	\begin{tabular}{l@{$\quad$}r@{$\qquad$}l@{$\qquad$}l}
		\hline
		Dataset&$\#$ of points&Scalar content&Gauge RGEs\\
		\hline
		$\mathcal{B}$ of~\cite{Jarkovska:2021jvw}& 30000& $\mathbf{45}\oplus \mathbf{126}$ & One-loop \\
		$\mathcal{D}$& 20000& $\mathbf{45}\oplus \mathbf{126} \oplus \mathbf{10}_{\mathbb{C}}$ & Two-loop \\
		\hline
	\end{tabular}
\end{table*}

\begin{figure*}[htb]
	\centering
	\mbox{
    \includegraphics[width=11cm]{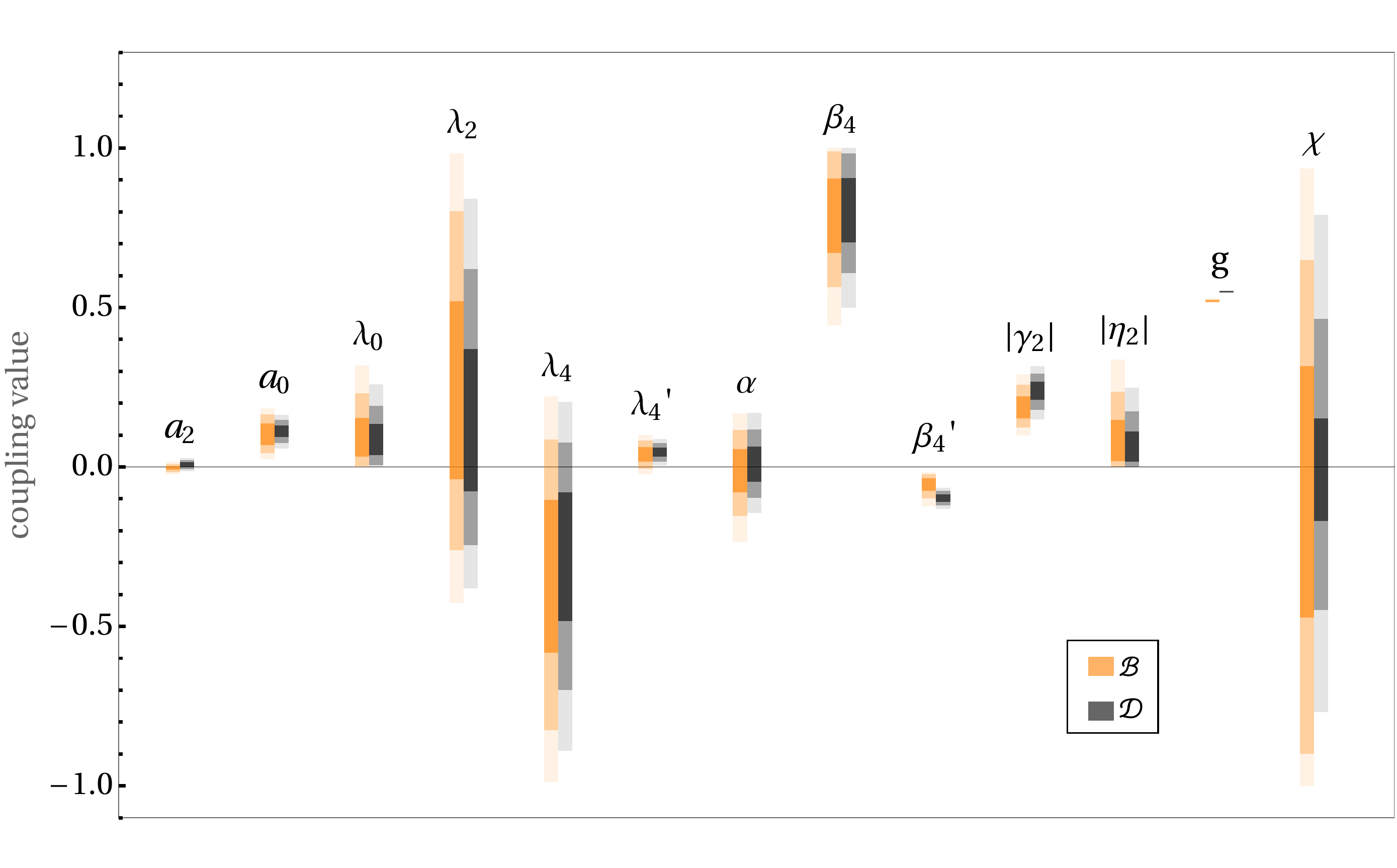}
    \includegraphics[width=6.7cm]{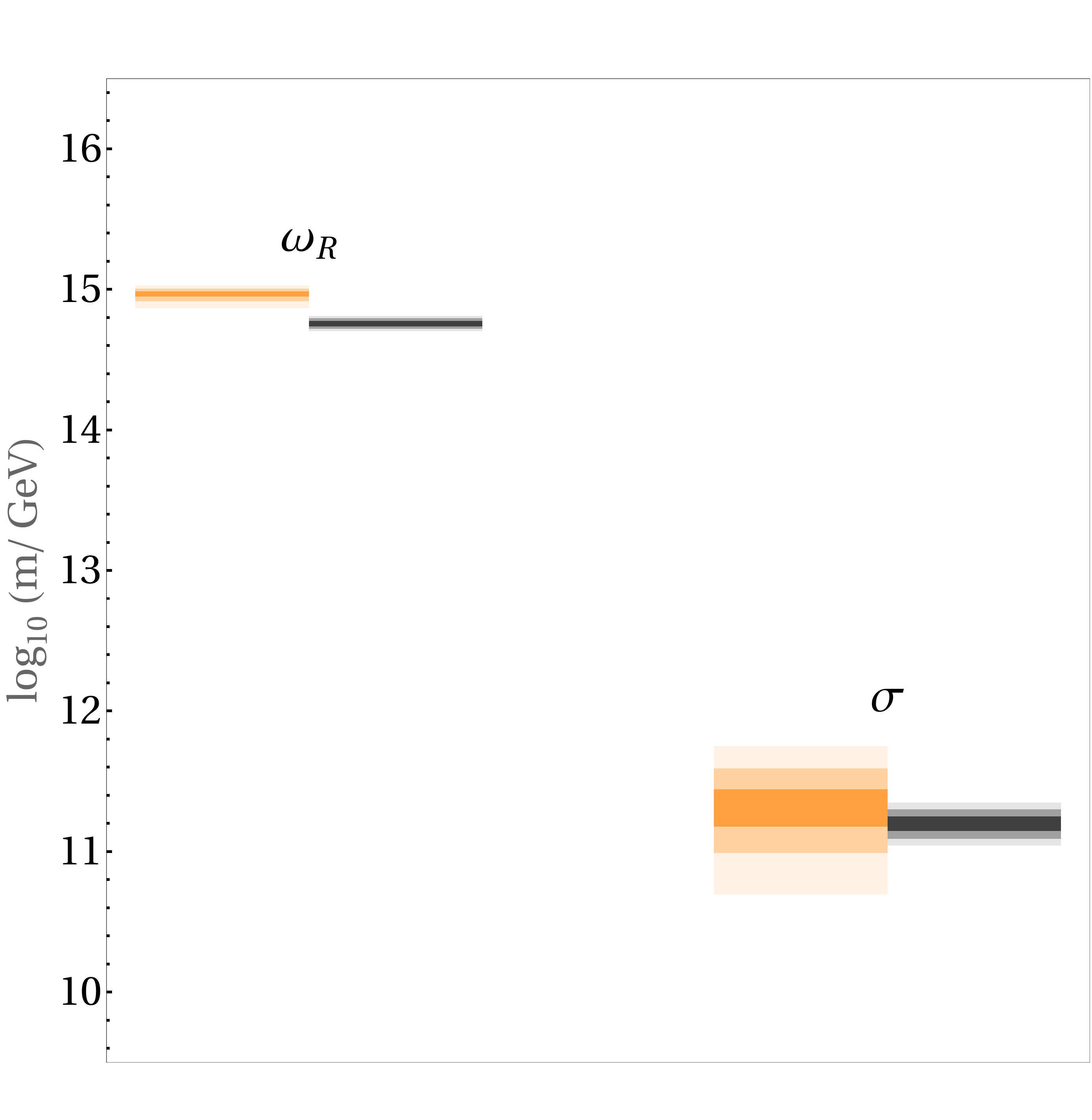}
    }
    \vspace{-1.0cm}
	\caption{
		The $1$-, $2$- and $3$-$\sigma$ HDI ranges (distinguished by decreasing opacity) of the old dimensionless scalar parameters with ratio $\chi$ (left panel) and VEVs $\omega_R$, $\sigma$ (right panel) for datasets $\mathcal{B}$ and $\mathcal{D}$.
	}
	\label{fig:old-dimless-scalar-parameters}
\end{figure*}

In analogy to Section~\ref{sec:results-shape}, we display the HDI ranges for the common parameters of the two scenarios of interest (i.e., those parameters denoted as ``old'' in Table~\ref{tab:potential-parameters} along with the ratio $\chi$, and the VEVs $|\sigma|$ and $\omega_R$) in Figure~\ref{fig:old-dimless-scalar-parameters}. Note that one can view the $\mathbf{45} \oplus \mathbf{126}$ Higgs model as a special case of the full $\mathbf{45} \oplus \mathbf{126} \oplus \mathbf{10}_{\mathbb{C}}$ setting with all the ``new'' parameters set to zero. 

One can see that the salient features of the new dataset $\mathcal{D}$ are practically identical to those of the old dataset $\mathcal{B}$; hence, their understanding developed in~\cite{Jarkovska:2021jvw} is readily applicable also to the model with $\mathbf{10}_{\mathbb{C}}$ of this study. 

There are, however, some minor differences that are worth commenting on:
%%%
\begin{itemize}
	\item \underline{The VEVs $\omega_R$ and $\sigma$ and coupling $g$:}
    The quantities $\omega_R$, $|\sigma|$ and $g$ are essentially determined by the gauge coupling unification pattern, so small shifts in these parameters between the $\mathcal{B}$ and $\mathcal{D}$ dataset are expected due to the upgrade from one- to two-loop RGEs. According to~\cite{Bertolini:2009qj}, the GUT scale (and thus $\omega_{R}$) is expected to decrease for the two-loop case, which is indeed seen in Figure~\ref{fig:old-dimless-scalar-parameters}.
    Interestingly, the slight increase in $|\sigma|$ asserted in~\cite{Bertolini:2009qj} is not reproduced here; rather a slight tendency for its decrease is observed. This can be tracked down to the threshold effects associated to the {\em pre-tuned} doublet at the intermediate scale of the $\mathcal{D}$ dataset, which counteract the generic behaviour predicted by \cite{Bertolini:2009qj}.
	%%%%%%%%%%%%%%%%%%%%%%
	\item \underline{The ``old'' parameters:}
	Interestingly, the spread in all the ``old'' dimensionless parameters of Table~\ref{tab:potential-parameters} as well as in the ratio $\chi$ and the VEV $|\sigma|$ tend to shrink in the $\mathcal{D}$ dataset (as compared to $\mathcal{B}$ of~\cite{Jarkovska:2021jvw}), cf.~Figure~\ref{fig:old-dimless-scalar-parameters}.
 This may sound a bit counterintuitive, since the new couplings, due to the extra $\mathbf{10}_\mathbb{C}$ in the full model, might be expected to help with countering large perturbative corrections, thus presumably extending these ranges. It turns out though that two-loop gauge unification in combination with tachyonicity and perturbativity constraints actually leads to more stringent restrictions. The central values of the ``old'' parameters remain largely unchanged, small shifts are seen only in  $a_2$, $\beta_4'$, and $\gamma_2$.
   
	%%%%%%%%%%%% 
	\item \underline{Phases of complex parameters:}
	The distributions of the phases $\arg\eta_{2}$ and $\arg\gamma_{2}$ across both the $\mathcal{B}$ and $\mathcal{D}$ datasets, cf.  Tables~\ref{tab:datasets} and~\ref{tab:dataset-difference}, are almost uniform over the entire $[0,2\pi)$ domain and, hence, they have been omitted from the plots. 
	%%%%%%%%%%
\end{itemize}

%%%%%%%%%%%%%%%%%%%%%%%%%%
\section{Two-loop gauge unification constraints} \label{app:two-loop-gauge-running}
%%%%%%%%%%%%%%%%%%%%%%%%%%

The requirement of unification of gauge couplings at $M_{GUT}$ is implemented for any given point in the parameter space via the standard goodness-of-fit $\chi^{2}$ measure applied to the values of the three SM running gauge couplings evolved from the common GUT-scale value $g$ down to the electroweak scale. 
This evolution is governed by the relevant two-loop RG flow passing through several thresholds corresponding to a sequence of effective theories, which in the $\omega_{BL}\to 0$ limit (and  for $|\sigma| \ll |\omega_R|$) can be written as
\begin{align}
\mathcal{T}_{10} \;\xrightarrow{M_{GUT}}\;
\mathcal{T}_{421+} \;\xrightarrow{M_{PGB}}\;
\mathcal{T}_{421} \;\xrightarrow{M_{\sigma\phantom{P}}}\;
\mathcal{T}_{321},
\label{eq:eff-field-theory-scheme}
\end{align}
where $M_{GUT}> M_{PGB} > M_{\sigma}$ denote the three crucial matching scales around which various heavy fields are integrated out, cf. Table~\ref{tab:theories}. Note that the gauge structure changes only at $M_{GUT}$ and 
$M_{\sigma}$, where the heavy gauge fields effectively disappear, but there is no such change at $M_{PGB}$, where only the scalar PGBs drop from the spectrum. The details of the entire procedure are summarized below in three sub-sections:  the structure of the four effective theories of Eq.~\eqref{eq:eff-field-theory-scheme} and the RGEs for the corresponding gauge couplings are given in Section~\ref{sec:subsection-RGE}, the matching conditions in Section~\ref{sec:subsection-matching}, and the details of their practical implementation in Section~\ref{sec:subsection-stepbystep}.

\subsection{Two-loop RGEs for gauge couplings \label{sec:subsection-RGE}}

In an effective field theory with the gauge symmetry composed of $N$ factors $G_1 \times G_2 \times \ldots \times G_N$ (with at most one $U(1)$ involved), while the fermions and scalars are in irreducible representations $F_{f}$ and $S_{s}$, the two-loop evolution equations for the gauge couplings $g_i$ ($i=1\ldots N$) can be written as \cite{Ellis:2015jwa,Bertolini:2009qj}
%%%%%%%%%%
\begin{align}
\label{eq:alpha-inv-RGE}
\frac{d}{dt}\alpha^{-1}_{i}&= -\frac{1}{2\pi} \Big(a_{i}+\frac{1}{4\pi} \sum_{j=1}^N b_{ij} \alpha_{j}-\frac{1}{16 \pi^2}\sum_{f}\eta_{F_{f}}Y_{4}^{(i)}(F_{f})\Big)\,,
\end{align}
where $\alpha_{i}=g_{i}^{2}/4\pi$ and  
\begin{align}
t = \log \left(\frac{\mu_R}{M_Z}\right), \label{eq:definition-t}
\end{align}
with $\mu_R$ denoting the renormalization scale. The one- and two-loop coefficients $a_{i}$ and $b_{ij}$ are defined as 
\begin{align}
\label{eq:ai}
a_{i}&=-\tfrac{11}{3} C_{i}(G_{i}) + \tfrac{1}{3} \sum_{s}\eta_{S_{s}}\, D_{i}(S_{s}) + \tfrac{2}{3} \sum_{f}\eta_{F_{b}}\, D_{i}(F_{f}),\\
\begin{split}
\label{eq:bij}
b_{ij}&=4\sum_{s} \eta_{S_{s}}\, D_{i}(S_{s})\,C_{j}(S_{s})+2\sum_{f} \eta_{F_{f}}\,D_{i}(F_{f})\,C_{j}(F_{f})+\\
&\quad + \delta_{ij}\left[-\tfrac{34}{3} C_{i}(G_{i})^2 + \tfrac{2}{3}\sum_{s} \eta_{S_{s}}\,D_{i}(S_{s})\;C_{i}(G_{i})+\tfrac{10}{3}\sum_{f} \eta_{F_{f}}\,D_{i}(F_{f})\;C_{i}(G_{i})\right].
\end{split}
\end{align}
In the sums above, the indices $f$ and $s$ run over all fermionic and scalar representations involved, $D_i$ and $C_i$ label the Dynkin index and quadratic Casimir with respect to the gauge factor $G_i$ of the representation in their arguments, $\eta_S = \tfrac{1}{2}$ or $1$ for real or complex scalars and $\eta_F = \tfrac{1}{2}$ or $1$ for Weyl or Dirac fermions, respectively. The specific values of $a_{i}$ and $b_{ij}$ for different effective theories in the chain of Eq.~\eqref{eq:eff-field-theory-scheme} are given in Table~\ref{tab:theories}, with a few accompanying definitions~\eqref{eq:matter-content-definitions-begin}--\eqref{eq:matter-content-definitions-end} provided below:
\def\ADDSPACE{\addlinespace[6pt]}
\begin{table}[htb]
    \centering
    \caption{The sequence of different effective-theory stages defining the two-loop RG evolution of gauge couplings from the GUT scale down to the EW scale. For each stage the labels of the corresponding gauge couplings are listed along with the relevant fermionic and scalar degrees of freedom (cf. Eqs.~\eqref{eq:matter-content-definitions-begin}--\eqref{eq:matter-content-definitions-end}), as well as the one- and two-loop beta coefficients $a_{i}$ and $b_{ij}$. \label{tab:theories}}
    \vskip 0.2cm
    \begin{tabular}{l@{\quad}l@{\quad}l@{\quad}l@{\quad}l@{\quad}ll}
         \toprule
         theory & symmetry & couplings & fermions & scalars & $a_i$ & $b_{ij}$\\
         \midrule
         $\mathcal{T}_{10}$&
            $\mathrm{SO}(10)$&
            $(\alpha_{10})$&
            $3\times \mathbf{16}_F$&
            $\mathbf{45}_{\mathbb{R}}\oplus \mathbf{126}_{\mathbb{C}}\oplus \mathbf{10}_{\mathbb{C}}$&
            $\left(-12\right)$&
            $\left(\tfrac{3223}{2}\right)$\\
        \ADDSPACE
         $\mathcal{T}_{421+}$&
            $\SU(4)_{C}\!\times\!\SU(2)_L \!\times\!\mathrm{U}(1)_R $&
             $(\alpha_{C},\alpha_{L},\alpha_{R})$&
            $3\times\mathbf{F}_{421}$&
            $\mathbf{S}_{421}\oplus \mathbf{S}_{421+}$&
            $\left(-\tfrac{19}{3},-\tfrac{1}{3},10\right)$&
            $\begin{pmatrix}
	           \tfrac{1019}{6} & \tfrac{57}{2} & \tfrac{43}{2} \\
	           \tfrac{285}{2} & \tfrac{143}{3} & 8 \\
	           \tfrac{645}{2} & 24 & 51 \\
	           \end{pmatrix}$\\
         \ADDSPACE   
         $\mathcal{T}_{421}$&
            $\SU(4)_{C} \!\times\! \SU(2)_L \!\times\! \mathrm{U}(1)_R $&
            $(\alpha_{C},\alpha_{L},\alpha_{R})$&
            $3\times\mathbf{F}_{421}$&
            $\mathbf{S}_{421}$&
            $\left(-7,-\tfrac{2}{3},10\right)$&
            $\begin{pmatrix}
	           \tfrac{265}{2} & \tfrac{57}{2} & \tfrac{43}{2} \\
	           \tfrac{285}{2} & \tfrac{115}{3} & 8 \\
	            \tfrac{645}{2} & 24 & 51 \\
	           \end{pmatrix}$\\
        \ADDSPACE
         $\mathcal{T}_{321}$&
            $\SU(3)_c \!\times\! \SU(2)_L \!\times\! \mathrm{U}(1)_{Y}$&
            $(\alpha_{c},\alpha_{L},\alpha_{1})$&
            $3\times\mathbf{F}_{321}$&
            $\mathbf{S}_{321}$&
            $\left(-7,-\tfrac{19}{6},\tfrac{41}{10}\right)$&
            $\begin{pmatrix}
	           -26 & \tfrac{9}{2} & \tfrac{11}{10} \\
	           12 & \tfrac{35}{6} & \tfrac{9}{10} \\
	           \tfrac{44}{5} & \tfrac{27}{10} & \tfrac{199}{50} \\
	           \end{pmatrix}$\\
         \bottomrule
    \end{tabular}
\end{table}

\begin{align}
    \mathbf{F}_{421}&\equiv (4,2,0) \oplus (\overline{4},1,+\tfrac{1}{2}) \oplus (\overline{4},1,-\tfrac{1}{2}), \label{eq:matter-content-definitions-begin} \\
    \mathbf{F}_{321}&\equiv (3,2,\tfrac{1}{6}) \oplus (\overline{3},1,-\tfrac{2}{3}) \oplus (\overline{3},1,+\tfrac{1}{3}) \oplus (1,2,-\tfrac{1}{2}) \oplus (1,1,+1), \label{eq:matter-content-definitions-2}\\
    \mathbf{S}_{421}&\equiv (\overline{10},1,-1)_{\mathbb{C}}\oplus (15,2,+\tfrac{1}{2})_{\mathbb{C}} \oplus (1,2,+\tfrac{1}{2})_{\mathbb{C}} \label{eq:matter-content-definitions-3},\\
    \mathbf{S}_{421+}&\equiv (1,3,0)_{\mathbb{R}} \oplus (15,1,0)_{\mathbb{R}},\\
    \mathbf{S}_{321}&\equiv (1,2,+\tfrac{1}{2})_{\mathbb{C}}. \label{eq:matter-content-definitions-end}
\end{align}

The Yukawa contribution $Y_{4}^{(i)}(F_{f})$ in Eq.~\eqref{eq:alpha-inv-RGE} is computed via \cite{Bertolini:2009qj}
\begin{align}
Y_{4}^{(i)}(F_{f})	&=\sum_{y} \mathrm{Tr}\left[C_{i}(F_{f})\;\mathbf{Y}_y\mathbf{Y}_y^\dagger\right]/\mathrm{dim}(G_{i}),
\end{align}
where the sum runs over all Yukawa matrices in the theory (indexed by $y$), and the trace has to be taken over all gauge and family indices. To avoid unnecessary complication with Yukawa matching, and assuming the full dominance of the top Yukawa coupling $y_t$, the last term in Eq.~\eqref{eq:alpha-inv-RGE} is approximated by 
\begin{align}
\sum_{f} \eta_{F_f }Y^{(i)}_4 (F_f) &= \tilde{b}_i y_t^2, \label{eq:rge-Yukawa-approximation}
\end{align}
with $\tilde{b}_i = \lbrace 2, \tfrac{3}{2},\tfrac{17}{10}\rbrace$
 for the ${\cal T}_{321}$ stage and $\tilde{b}_i = \lbrace 2, 2,2 \rbrace$
 for the ${\cal T}_{421}$ and ${\cal T}_{421+}$ stages. For the same reason the running of the $y_t$ parameter is approximated by a purely SM-like evolution 
\begin{align}
\label{eq:yt-rge}
\frac{d}{dt} y_t &= \frac{1}{16 \pi^2} \frac{9}{2} y_t^3 - \frac{1}{4\pi} y_t \sum_{i\in \{c,L,1\}} \tilde{a}_i \alpha_i,
\end{align}
where $\{\alpha_c,\alpha_L,\alpha_1\}$ are the three SM gauge coupling factors (calculated in the ${\cal T}_{421}$ and ${\cal T}_{421+}$ stages from the values of $\{\alpha_C,\alpha_L,\alpha_R\}$ as $\alpha_1^{-1}=\tfrac{3}{5}\alpha_R^{-1}+\tfrac{2}{5}\alpha_C^{-1}$ and $\alpha_c^{-1}=\alpha_C^{-1}$) 
and
$
%\label{eq:aitilde-bitilde}
\{\tilde{a}_c,\tilde{a}_L,\tilde{a}_1\}= \lbrace 8,\tfrac{9}{4},\tfrac{17}{20}\rbrace
$.
Several additional remarks are noteworthy here: 
\begin{itemize}
    \item Since the procedure of top-down unification starts at $M_{GUT}$, the $a$- and $b$-factors for the unified theory $\mathcal{T}_{10}$ are not needed. We provide them regardless for completeness.
    %%%%%%%%%%%%%%%%%%%%%%%%%%%%%%%%%%%%%%%%
    \item The representations listed in Eqs.~\eqref{eq:matter-content-definitions-begin}--\eqref{eq:matter-content-definitions-end} are written in notation appropriate for the symmetry at that stage. The index $\mathbb{R}$ or $\mathbb{C}$ for a scalar representation denotes whether it carries real or complex degrees of freedom, respectively.
    %%%%%%%%%%%%%%%%%%%%%%%%%%%%%%%%%%%%%%%%
    \item The presence of the scalar set $\mathbf{S}_{421}$ (cf.~Eq.~\eqref{eq:matter-content-definitions-3}) in the $\mathcal{T}_{421}$ and $\mathcal{T}_{421+}$ stages is consistent with the \textit{extended survival hypothesis}. Indeed, the breakdown of $\SU(4)_C$ proceeds through the first ($\SU(2)_L$-singlet) representation therein, whilst the latter two multiplets ($\SU(2)_L$-doublets) accommodate the SM Higgs doublet, cf.~Section~\ref{sec:Yukawa-sector}. The extra $\mathbf{S}_{421+}$ scalars present at the $\mathcal{T}_{421+}$ stage are the PGBs.
    %%%%%%%%%%%%%%%%%%%%%%%%%%%%%%%%%%%%%%%%
    \item $\mathcal{T}_{321}$ is the usual Standard Model that includes one Higgs doublet. Note that the hypercharges in Eqs.~\eqref{eq:matter-content-definitions-2} and \eqref{eq:matter-content-definitions-end} refer to the values of $Y$ in the usual SM normalization (with $Q=T_L^3+Y$), while the traditional GUT normalization with $\alpha_{1}=\tfrac{5}{3}\alpha_{Y}$ is used for the $a$- and $b$-coefficients in Table~\ref{tab:theories}.
    %%%%%%%%%%%%%%%%%%%%%%%%%%%%%%%%%%%%%%%%
    \item The approximate treatment of Yukawa contributions via Eqs.~\eqref{eq:rge-Yukawa-approximation} and \eqref{eq:yt-rge} is justified by noting their minuscule effect on unification. In particular, in a top-down fit of the low energy gauge coupling data, switching on separately the gauge and Yukawa part of the $2$-loop RGE induces a relative change of $1.5\,\%$ and $10^{-4}$, respectively, in the log of the GUT scale (in the quantity $\log_{10}(M_{GUT}/\mathrm{GeV})$) compared to the fit with $1$-loop RGE. The Yukawa contribution in Eq.~\eqref{eq:alpha-inv-RGE} is thus a negligible part of the $2$-loop RGE upgrade, and hence the details of its treatment have little impact on the current analysis.
\end{itemize}

\subsection{Matching conditions for gauge couplings\label{sec:subsection-matching}}

At each of the symmetry-breaking steps defined in Table~\ref{tab:theories}, the gauge couplings associated to the higher-scale gauge symmetry $G=G_1 \times G_2 \times \ldots \times G_N$ (encoded in $\alpha_j\equiv g_j^2/4\pi$ with $j=1\ldots N$) must be matched to those of the lower-scale effective theory with the gauge structure $\tilde{G}=\tilde{G}_1 \times \tilde{G}_2 \times \ldots \times \tilde{G}_M$ (and couplings $\tilde \alpha_i$, $i=1\ldots M$).  
The corresponding matching conditions then read\footnote{Here it is implicitly assumed that there is at most one Abelian factor among the $G_j$'s, and at most one among $\tilde{G}_i$'s. This ensures that there is no kinetic mixing between different $\mathrm{U}(1)$, which simplifies the situation significantly.} \cite{Weinberg:1980wa,Hall:1980kf,Ellis:2015jwa} 
\begin{align}
\label{eq:matching-cond}
\tilde{\alpha}^{-1}_i (\mu_R) = \sum_{j=1}^N c_{ij}\,\alpha^{-1}_j(\mu_R) - 4\pi \lambda_i(\mu_R),
\end{align}
where $\mu_R$ is the matching scale, the $c_{ij} \in \mathbb{R}$ coefficients are determined by the embedding of $\tilde{G}$ into $G$ (see Table~\ref{tab:matching}), and the one-loop threshold factors $\lambda_{i}$ are defined as
\begin{align}
\label{eq:thr-corr}
\lambda_i (\mu_R) &= \frac{1}{8 \pi^2} \left[\sum_{g}\tilde{D}_i(V_g) \left(\frac{1}{6} - \frac{11}{3} \log \frac{M_{V_g}}{\mu_R}\right) + \frac{4}{3}\sum_{f} \eta_{F_f}\tilde{D}_i(F_f)\log \frac{M_{F_f}}{\mu_R} + \frac{1}{3} \sum_{s} \eta_{S_s}\tilde{D}_i(S_s) \log \frac{M_{S_s}}{\mu_R}\right].
\end{align}
Here the sums over $g$, $f$ and $s$ respectively run over all irreducible representations (of the $\tilde{G}$ symmetry) containing massive gauge bosons $V_g$, fermions $F_f$ and scalars $S_s$ that are integrated out at $\mu_R$. The $\eta_{F_f}$- and $\eta_{S_s}$-coefficients are defined as in Eqs.~\eqref{eq:ai}--\eqref{eq:bij} and $\tilde{D}_{i}$ are Dynkin indices relative to the $\tilde{G}_{i}$ factor in $\tilde{G}$. The $M_{V_g}$, $M_{F_f}$ and $M_{S_s}$ quantities in arguments of logs denote the (tree-level) masses of the corresponding fields; the would-be Goldstone bosons are formally included with a mass equal to the mass of the associated gauge fields. 
More specific information on the elements entering the formulae above  at different matching scales is given in Table~\ref{tab:matching}. 
\begin{table}[htb]
    \centering
    \caption{The inputs of the matching conditions in Eqs.~\eqref{eq:matching-cond} and~\eqref{eq:thr-corr} when transitioning from the theory $\mathcal{T}$ to $\mathcal{\tilde{T}}$ at the matching scale $\mu_R$. The $c_{ij}$ coefficients entering Eq.~\eqref{eq:matching-cond} and the fields to be integrated out at $\mu_R$ (i.e.~those that contribute to the threshold corrections in Eq.~\eqref{eq:thr-corr}) are given for each transition in the effective-theory chain of Eq.~\eqref{eq:eff-field-theory-scheme}. The index $\mathbb{C}$/$\mathbb{R}$ reflects the complex/real nature of the scalars in the relevant representation, while  $G_{\mathbb{C}}$/$G_{\mathbb{R}}$ and $WGB_\mathbb{C}$/$WGB_\mathbb{R}$ indicate complexity/reality of gauge bosons and the corresponding scalar WGBs, respectively. \label{tab:matching}}
    \vspace{0.1cm}
    \begin{tabular}{l@{\quad}l@{$\;\to\quad$}l@{\quad}l@{\quad}p{10cm}}
        \toprule
         $\mu_R$ & $\mathcal{T}$&$\tilde{\mathcal{T}}$ & $c_{ij}$ & fields to integrate out\\
         \midrule
         $M_{GUT}$ & $\mathcal{T}_{10}$&$\mathcal{T}_{421+}$ & \phantom{xx}\par
            $\begin{pmatrix}
                1\\ 1\\ 1\\
                \end{pmatrix}$&  
            \vspace{-0.8cm}\par
            $(6,2,+\tfrac{1}{2})_{G_{\mathbb{C}}} \oplus (1,1,+1)_{G_{\mathbb{C}}}\oplus%$,\par$
            (6,2,+\tfrac{1}{2})_{WGB_{\mathbb{C}}} \oplus (1,1,+1)_{WGB_{\mathbb{C}}}$,\newline
            $(\overline{10},1,+1)_{\mathbb{C}} \oplus (\overline{10},1,0)_{\mathbb{C}} \oplus (10,3,0)_{\mathbb{C}} \oplus (15,2,-\tfrac{1}{2})_{\mathbb{C}} \oplus (6,1,0)_{\mathbb{C}} \oplus (6,1,0)_{\mathbb{C}} \oplus (1,2,+\tfrac{1}{2})_{\mathbb{C}}$
            \\
        \ADDSPACE
         $M_{PGB}$ & $\mathcal{T}_{421+}$&$ \mathcal{T}_{421}$ & 
            $\begin{pmatrix}
                1 &0&0\\ 0&1&0\\ 0&0&1\\
                \end{pmatrix}$& 
            $(1,3,0)_{\mathbb{R}} \oplus (15,1,0)_{\mathbb{R}}$\\
        \ADDSPACE
         $M_{\sigma}$ & $\mathcal{T}_{421}$&$ \mathcal{T}_{321}$ & 
            $\begin{pmatrix}
                1 &0&0\\ 0&1&0\\ \tfrac{2}{5}&0&\tfrac{3}{5}\\
                \end{pmatrix}$&    
            \vspace{-0.8cm}\par
            $(3,1,+\tfrac{2}{3})_{G_{\mathbb{C}}}\oplus (1,1,0)_{G_{\mathbb{R}}} \oplus%$,\par$
            (3,1,+\tfrac{2}{3})_{WGB_{\mathbb{C}}}\oplus (1,1,0)_{WGB_{\mathbb{R}}}$,\par
            $(\overline{6},1,-\tfrac{1}{3})_{\mathbb{C}} \oplus (1,1,0)_{\mathbb{R}} \oplus (8,2,+\tfrac{1}{2})_{\mathbb{C}} \oplus (\overline{3},2,-\tfrac{1}{6})_{\mathbb{C}}\oplus (3,2,+\tfrac{7}{6})_{\mathbb{C}} \oplus (1,2,\tfrac{1}{2})_{\mathbb{C}}$
            \\
         \bottomrule
    \end{tabular}
\end{table}

Note that the threshold corrections $\lambda_{i}$ effectively take into account the details of the spectrum of the fields integrated out around the matching scale $\mu_R$, which is in principle arbitrary (as long as it is not far from the typical mass scale of the fields being integrated out), since its specific position influences the result only at next order in the perturbative expansion. The details of  determination of the matching scales $M_{GUT}$, $M_{PGB}$ and $M_{\sigma}$ are given in Section~\ref{sec:subsection-stepbystep}.

\subsection{Implementation of the two-loop gauge unification constraints \label{sec:subsection-stepbystep}}
For the sake of explicitness, a short note on the practical implementation of two-loop gauge unification constraints is provided in the form of a step-by-step procedure, which is applied to every parameter-space point from the numerical scans described in Section~\ref{sec:one-loop-fine-tuning}:
\begin{enumerate}
	\item
	The unification-scale value of the unified gauge coupling is called $g$ and it is one of the inputs. Then $M_{GUT}$ is calculated as the barycenter of the $\omega_R$-proportional dominant terms of the tree-level gauge boson masses:
	\begin{align}
	M_{GUT} = \sqrt{\frac{8}{13}}\, g\, \omega_{R }.
	\end{align}
	  The GUT-scale initial condition for the top quark Yukawa coupling is always taken to be
	\hbox{$
	y_t(M_{GUT}) = 0.494,
	$}
    which is the value obtained from the bottom-up $y_t$ running in the SM (at one loop) to $10^{15}\,\mathrm{GeV}$.
	%%%%%%%%%%%%%%%%%%%%%%
	\item Match $\mathcal{T}_{10}\to\mathcal{T}_{421+}$ at $M_{GUT}$, including threshold corrections.
	%%%%%%%%%%%%%%%%%%%%
	\item Run $(\alpha_{C},\alpha_L,\alpha_R)$ and $y_t$ in the  $\mathcal{T}_{421+}$ framework from $M_{GUT}$ to the pseudo-Goldstone-boson scale $M_{PGB}$, which is defined as the weighted average of the PGB masses:
	\begin{align}
	M_{PGB} = \frac{\sum_{X} d_X M_X}{\sum_{X} d_X},
	\end{align} 
    where the sum goes over all PGB multiplets $X$ (cf.~row 2 of Table~\ref{tab:matching}), while $d_X$ and $M_X$ denote the size (in terms of real degrees of freedom) and mass of $X$, respectively.
	%%%%%%%%%%%%%%%%%%
	\item Match $\mathcal{T}_{421+}\to\mathcal{T}_{421}$ at $M_{PGB}$, including threshold corrections.
	%%%%%%%%%%%%%%%%%%%%%%%%%%%%%
	\item Run $(\alpha_{C},\alpha_L,\alpha_R)$ and $y_t$ in the theory $\mathcal{T}_{421}$ from $M_{PGB}$ to the seesaw scale $M_{\sigma}$. The scale $M_{\sigma}$ is determined as the barycenter of $\sigma$-proportional tree-level gauge boson masses, i.e. 
	\begin{align}
	M_{\sigma} = \sqrt{\frac{22}{7}}\, g\, |\sigma|.
	\end{align} 
	%%%%%%%%%%%%%%%%%%
	\item Match $\mathcal{T}_{421}\to\mathcal{T}_{321}$ at $M_{\sigma}$, including threshold corrections.
	%%%%%%%%%%%%%%%%%%%%%%%%%%%%%
	\item Run $(\alpha_{c},\alpha_L,\alpha_{1})$ and $y_t$ in the theory $\mathcal{T}_{321}$ from $M_{\sigma}$ to the $Z$-boson mass $M_{Z} = 91.19\,\mathrm{GeV}$ \cite{ParticleDataGroup:2018ovx}. 
    % 91.1876 \pm 0.0021
	%%%%%%%%%%%%%%%
	\item Calculate $\chi^2$ based on the computed gauge couplings $\alpha_{c}(M_Z),\alpha_L(M_Z),\alpha_{1}(M_Z)$ compared to their experimental values \cite{Marciano:1991ix,Kuhn:1998ze,Martens:2010nm,Sturm:2013uka,ParticleDataGroup:2020ssz}
	\begin{align}
    \alpha^{-1}_{c}(M_Z) &= 8.550 \pm 0.065,\\
	\alpha^{-1}_L (M_Z)&=29.6261 \pm 0.0051,\\
    \alpha^{-1}_{1}(M_Z) &= 59.1054 \pm 0.0031.
	\end{align}
\end{enumerate}

A value of $\chi^2 \leq 9$ is required for the parameter space point to pass the unification test without penalization. 

Note that the procedure above assumes a spectrum where \textit{doublet fine-tuning} has been hypothetically achieved along the lines of Section~\ref{sec:fine-tuning}. In that case, out of the four SM-doublet scalars $(1,2,+\tfrac{1}{2})$ in the model, two 
have masses around the unification scale $M_{GUT}$, one has a mass around the seesaw scale $M_{\sigma}$, and the remaining mass is at the level of the SM Higgs. For that sake, we fix the corresponding masses by hand to 
\begin{align}
M_S(1,2,+ \tfrac{1}{2})_4 &= M_S(1,2,+ \tfrac{1}{2})_3 = M_{GUT},\\
M_S(1,2,+ \tfrac{1}{2})_2 &= M_\sigma\;10^{z/2},\\
M_S(1,2,+ \tfrac{1}{2})_1 &= 125\,\mathrm{GeV},
\end{align}
where the newly introduced parameter $z$ (assuming random values in the (-1,1) interval) models the possible offsets of the mass of the $(1,2,+ \tfrac{1}{2})_2$ scalar around the $M_\sigma$ scale. The specific choice of $z$ does not influence the results of the current study in any appreciable way.

%%%%%%%%%%%%%%%%%%%%%%%%%%
\section{$\beta$-functions of scalar couplings of the $\mathbf{45} \oplus \mathbf{126} \oplus \mathbf{10}_\mathbb{C}$ model}\label{app:scalar-beta-functions}
%%%%%%%%%%%%%%%%%%%%%%%%%%

The list of one-loop $\beta$-functions of all dimensionless couplings governing the scalar potential of the $\mathbf{45} \oplus \mathbf{126} \oplus \mathbf{10}_\mathbb{C}$ model (i.e.~the $D=0$ couplings of Table~\ref{tab:potential-parameters}) is given below. The computation was performed using the effective potential method described in  Appendix~C of Ref.~\cite{Jarkovska:2021jvw}.

The $\beta$-functions for the real couplings of the ``old'' setting of Table~\ref{tab:potential-parameters} (i.e. those entering the $V_{45}$, $V_{126}$ and $V_{\rm mix}$ terms of Eq.~\eqref{eq:scalar-potential-0}), extended for the presence of $\mathbf{10}_\mathbb{C}$, read
\def\MINISKIP{\\[8pt]}
\begingroup
\allowdisplaybreaks 
\begin{align}
    \begin{split}
    %%%%%%%%%%%%%%%%% a0
    16 \pi ^2\;\beta_{a_{0}} &= \phantom{+} 126 \alpha ^2+56 \alpha  \beta_{4}+112 \alpha  \beta'_{4}+424 a_{0}^2+152 a_{0} a_{2}+12 a_{2}^2+\tfrac{33}{2} \beta_{4}^2+26 \beta_{4} \beta'_{4}+106 \beta'^{2}_{4} - \\
    &\quad -56 |\gamma_{2}|^{2} +12 |\zeta|^{2} +12 |\zeta'|^{2}+10 \kappa_{0}^2+4 \kappa_{0} \kappa_{2}+40 |\kappa'_{0}|^{2} +8 \kappa'_{0} \kappa'^{\ast}_{2}+8 \kappa'^{\ast}_{0} \kappa'_{2} + \frac{9}{2} g^4 -96 a_0 g^2,\\
    \end{split} \label{eq:couplings-RGE-begin}\MINISKIP
    %%%%%%%%%%%%%%%%% a2
    \begin{split}
        16 \pi ^2 \beta_{a_{2}} &= \phantom{+} 96 a_{0} a_{2}+76 a_{2}^2-5 \beta_{4}^2+60 \beta_{4} \beta'_{4}-100 \beta'^{2}_{4}+560 |\gamma_{2}|^{2} -24 |\zeta|^{2} -24 |\zeta'_{}|^{2} +4 \kappa_{2}^2 +16 |\kappa'_{2}|^{2}+\\
         &\quad+3 g^4-96 a_2 g^2,\\
    \end{split}\MINISKIP
    %%%%%%%%%%%%%%%%% lambda0
    \begin{split}
        16 \pi ^2 \beta_{\lambda_{0}} &= \phantom{+} 90 \alpha ^2+40 \alpha  \beta_{4}+80 \alpha  \beta'_{4}+10 \beta_{4}^2+80 \beta'^{2}_{4}+520 \lambda_{0}^2+2440 \lambda_{0} \lambda_{2}+2680 \lambda_{0} \lambda_{4}+4960 \lambda_{0} \lambda'_{4}+ \\
        &\quad +3460 \lambda_{2}^2+7880 \lambda_{2} \lambda_{4}+12320 \lambda_{2} \lambda'_{4}+4660 \lambda_{4}^2+13280 \lambda_{4} \lambda'_{4}+16960 \lambda'^{2}_{4}+10 \rho_{0}^2+10 \rho_{0} \rho_{2} +\\
        &\quad +40 |\rho'_{0}|^{2} +5 \rho_{2}^2+3840 |\varphi|^2 +3840 |\varphi'_{}|^{2}+\frac{135 }{2} g^4 -150  \lambda_0 g^2,\\
    \end{split}\MINISKIP
    %%%%%%%%%%%%%%%%%lambda2
    \begin{split}
        16 \pi ^2 \beta_{\lambda_{2}} &= -4 \beta_{4}^2-32 \beta'^{2}_{4}-32 |\gamma_{2}|^{2}-1264 |\eta_{2}|^{2}+24 \lambda_{0} \lambda_{2}-180 \lambda_{2}^2-584 \lambda_{2} \lambda_{4}-160 \lambda_{2} \lambda'_{4}-656 \lambda_{4}^2-\\
        &\quad -800 \lambda_{4} \lambda'_{4}-2560 \lambda'^{2}_{4}-\rho_{2}^2-4 |\psi_{0}|^{2}-2 |\psi_{1}|^{2}-4 |\psi_{2}|^{2}-1408 |\varphi|^{2}-1408 |\varphi'|^{2} -24 g^4-\\
        &\quad-150 \lambda_2 g^2,\\
    \end{split}\MINISKIP
    %%%%%%%%%%%%%%%%% lambda4
    \begin{split}
        16 \pi ^2 \beta_{\lambda_{4}} &= \phantom{+} 2 \beta_{4}^2+16 \beta'^{2}_{4}+32 |\gamma_{2}|^{2}+1328 |\eta_{2}|^{2}+24 \lambda_{0} \lambda_{4}+16 \lambda_{2}^2+112 \lambda_{2} \lambda_{4}+128 \lambda_{2} \lambda'_{4}+268 \lambda_{4}^2+\\
        &\quad + 640 \lambda_{4} \lambda'_{4}+1408 \lambda'^{2}_{4}+4 |\psi_{0}|^{2}+2 |\psi_{1}|^{2}+4 |\psi_{2}|^{2}+768 |\varphi|^{2}+768 |\varphi'|^{2} + 12 g^4-150 \lambda_4 g^2 ,\\
    \end{split}\MINISKIP
    %%%%%%%%%%%%%%%%% lambda4p
    \begin{split}
        16 \pi ^2 \beta_{\lambda'_{4}} &= \phantom{+} 4 \beta_{4} \beta'_{4}-4 \beta'^{2}_{4}+32 |\eta_{2}|^{2}+24 \lambda_{0} \lambda'_{4}-4 \lambda_{2}^2-8 \lambda_{2} \lambda_{4}-16 \lambda_{2} \lambda'_{4}+4 \lambda_{4}^2+112 \lambda_{4} \lambda'_{4}-240 \lambda'^{2}_{4}-\\
        &\quad -416 |\varphi|^{2}-416 |\varphi'|^{2} -3 g^4-150  \lambda_4' g^2,\\
    \end{split}\MINISKIP
    %%%%%%%%%%%%%%%%% alpha
    \begin{split}
        16 \pi ^2 \beta_{\alpha} &= \phantom{+} 8 \alpha ^2+508 \alpha  \lambda_{0}+1220 \alpha  \lambda_{2}+1340 \alpha  \lambda_{4}+2480 \alpha  \lambda'_{4}+376 \alpha  a_{0}+80 a_{0} \beta_{4}+160 a_{0} \beta'_{4}+76 \alpha  a_{2}+\\
        &\quad +16 a_{2} \beta_{4}+32 a_{2} \beta'_{4}+4 \beta_{4}^2+16 \beta_{4} \beta'_{4}+112 \beta_{4} \lambda_{0}+272 \beta_{4} \lambda_{2}+288 \beta_{4} \lambda_{4}+512 \beta_{4} \lambda'_{4}+144 \beta'^{2}_{4}+\\
        &\quad +224 \beta'_{4} \lambda_{0}+544 \beta'_{4} \lambda_{2}+576 \beta'_{4} \lambda_{4}+1024 \beta'_{4} \lambda'_{4}+64 |\gamma_{2}|^{2}+20 \kappa_{0} \rho_{0}+10 \kappa_{0} \rho_{2}+\\
        &\quad +40 \kappa'_{0} \rho'^{\ast}_{0}+40 \kappa'^{\ast}_{0} \rho'_{0}+4 \kappa_{2} \rho_{0}+2 \kappa_{2} \rho_{2}+8 \kappa'_{2} \rho'^{\ast}_{0}+8 \kappa'^{\ast}_{2} \rho'_{0} +12 g^4-123 \alpha  g^2,\\
    \end{split}\MINISKIP
    %%%%%%%%%%%%%%%%% beta4
    \begin{split}
        16 \pi ^2 \beta_{\beta_{4}} &= \phantom{+} 16 \alpha  \beta_{4}+16 a_{0} \beta_{4}+16 a_{2} \beta'_{4}+48 \beta_{4}^2+80 \beta_{4} \beta'_{4}+4 \beta_{4} \lambda_{0}-8 \beta_{4} \lambda_{2}+32 \beta_{4} \lambda_{4}+16 \beta_{4} \lambda'_{4}+\\
        &\quad +16 \beta'^{2}_{4}+ 16 \beta'_{4} \lambda_{2}+48 \beta'_{4} \lambda_{4}+640 \beta'_{4} \lambda'_{4}+64 |\gamma_{2}|^{2}+24 |\zeta|^{2}+96 \zeta  \varphi^{\ast}_{}+96 \zeta^{\ast}_{} \varphi +24 |\zeta'|^{2} +\\
        &\quad +96 \zeta'_{} \varphi'^{\ast}_{}+96 \zeta'^{\ast}_{} \varphi'_{} + 12 g^4-123 \beta_4 g^2,\\
    \end{split}\MINISKIP
    %%%%%%%%%%%%%%%%% beta4p
    \begin{split}
        16 \pi ^2 \beta_{\beta'_{4}} &= \phantom{+} 16 \alpha  \beta'_{4}+16 a_{0} \beta'_{4}+2 a_{2} \beta_{4}-4 a_{2} \beta'_{4}-\beta_{4}^2-28 \beta_{4} \beta'_{4}+2 \beta_{4} \lambda_{2}+6 \beta_{4} \lambda_{4}+80 \beta_{4} \lambda'_{4}-\\
        &\quad -124 \beta'^{2}_{4}+4 \beta'_{4} \lambda_{0}-12 \beta'_{4} \lambda_{2}+20 \beta'_{4} \lambda_{4}-144 \beta'_{4} \lambda'_{4}+16 |\gamma_{2}|^{2}-48 \zeta  \varphi^{\ast}_{}-48 \zeta^{\ast}_{} \varphi -\\
        &\quad -48 \zeta'_{} \varphi'^{\ast}_{}-48 \zeta'^{\ast}_{} \varphi'_{}-3 g^4-123 \beta_4' g^2,\\
    \end{split}
    %%%%%%%%%%%%%%%%%
\end{align}
\endgroup

%%%%%%%%%%%%%%%%%%%%%%%%%%%%%%%%%%%%%%%%%%%%%%%%%%%%%%%%%%%%%%%%%%%%
\noindent
while those for the ``old'' complex couplings get extended to
\begin{align}
    %%%%%%%%%%%%%%%%% gamma2
    \begin{split}
        16 \pi ^2 \beta_{\gamma_{2}} &= \phantom{+} 16 \alpha  \gamma_{2}+16 a_{0} \gamma_{2}+36 a_{2} \gamma_{2}+28 \beta_{4} \gamma_{2}+56 \beta'_{4} \gamma_{2}+4 \gamma_{2} \lambda_{0}+40 \gamma_{2} \lambda_{2}+180 \gamma_{2} \lambda_{4}+\\
        &\quad +160 \gamma_{2} \lambda'_{4}+440 \gamma^{\ast}_{2} \eta_{2}+12 \zeta  \zeta'_{}+2 \kappa_{2} \psi_{1}+4 \kappa'_{2} \psi_{0}+4 \kappa'^{\ast}_{2} \psi_{2} -123 \gamma_2 g^2,\\
    \end{split}\MINISKIP
    %%%%%%%%%%%%%%%%% eta2
    \begin{split}
        16 \pi ^2 \beta_{\eta_{2}} &= \phantom{+} 16 \gamma_{2}^2+24 \eta_{2} \lambda_{0}+160 \eta_{2} \lambda_{2}+600 \eta_{2} \lambda_{4}+640 \eta_{2} \lambda'_{4}+4 \psi_{0} \psi_{2}+\psi_{1}^2 -150 \eta_2 g^2 .\\
    \end{split}
    %%%%%%%%%%%%%%%%%
\end{align}
%%%%%%%%%%%%%%%%%%%%%%%%%%%%%%%%%%%%%%%%%%%%%%%%%%%%%%%%%%%%%%%%%%%%
\noindent
For the ``new'' real couplings of Table~\ref{tab:potential-parameters} one has
\begingroup
\allowdisplaybreaks 
\begin{align}
    %%%%%%%%%%%%%%%%% h2
    \begin{split}
        16 \pi ^2 \beta_{h_{2}} &= \phantom{+} 56 h_{2}^2+16 h_{2} h'_{2}+16 h'^{2}_{2}+72 |h_{3}|^2+64 |h_{4}|^2+90 \kappa_{0}^2+36 \kappa_{0} \kappa_{2}+10 \kappa_{2}^2+32 |\kappa'_{2}|^{2}+\\
        &\quad +126 \rho_{0}^2+126 \rho_{0} \rho_{2}+49 \rho_{2}^2+140 |\psi_{0}|^{2}+56 |\psi_{1}|^{2}+140 |\psi_{2}|^{2} + \frac{15 }{8} g^4-54 h_2 g^2 ,\\
    \end{split}\MINISKIP
    %%%%%%%%%%%%%%%%% h2p
    \begin{split}
        16 \pi ^2 \beta_{h'_{2}} &= \phantom{+} 24 h_{2} h'_{2}+40 h'^{2}_{2}+36 |h_{3}|^2+224 |h_{4}|^2+180 |\kappa'_{0}|^{2}+36 \kappa'_{0} \kappa'^{\ast}_{2}+36 \kappa'^{\ast}_{0} \kappa'_{2}+8 \kappa_{2}^2+\\
        &\quad +4 |\kappa'_{2}|^{2}+252 |\rho'_{0}|^{2}-\tfrac{35}{2} \rho_{2}^2-14 |\psi_{0}|^{2}+70 |\psi_{1}|^{2}-14 |\psi_{2}|^{2} +\frac{3 }{2}g^4-54 h_2' g^2 ,\\
    \end{split}\MINISKIP
    %%%%%%%%%%%%%%%%% kappa0
    \begin{split}
        16 \pi ^2 \beta_{\kappa_{0}} &= \phantom{+} 252 \alpha  \rho_{0}+126 \alpha  \rho_{2}+376 a_{0} \kappa_{0}+72 a_{0} \kappa_{2}+76 a_{2} \kappa_{0}+8 a_{2} \kappa_{2}+56 \beta_{4} \rho_{0}+28 \beta_{4} \rho_{2}+\\
        &\quad +112 \beta'_{4} \rho_{0}+56 \beta'_{4} \rho_{2}-28 \gamma_{2} \psi^{\ast}_{1}-28 \gamma^{\ast}_{2} \psi_{1}+84 |\zeta|^{2}+84 |\zeta'|^{2}+44 h_{2} \kappa_{0}+8 h_{2} \kappa_{2}+\\
        &\quad +8 h'_{2} \kappa_{0}+48 h_{3} \kappa'^{\ast}_{0}+8 h_{3} \kappa'^{\ast}_{2}+48 h^{\ast}_{3} \kappa'_{0}+8 h^{\ast}_{3} \kappa'_{2}+8 \kappa_{0}^2+32 |\kappa'_{0}|^{2}+4 \kappa_{2}^2+16 |\kappa'_{2}|^{2} +\\
        &\quad +\frac{3 }{2}g^4-75 g^2 \kappa_0,\\
    \end{split}\MINISKIP
    %%%%%%%%%%%%%%%%% kappa2
    \begin{split}
        16 \pi ^2 \beta_{\kappa_{2}} &= \phantom{+} 16 a_{0} \kappa_{2}+36 a_{2} \kappa_{2}+140 \gamma_{2} \psi^{\ast}_{1}+140 \gamma^{\ast}_{2} \psi_{1}-84 |\zeta|^{2}-84 |\zeta'|^{2}+4 h_{2} \kappa_{2}+8 h'_{2} \kappa_{2}+\\
        &\quad +8 h_{3} \kappa'^{\ast}_{2}+8 h^{\ast}_{3} \kappa'_{2}+16 \kappa_{0} \kappa_{2}+32 \kappa'_{0} \kappa'^{\ast}_{2}+32 \kappa'^{\ast}_{0} \kappa'_{2}+20 \kappa_{2}^2+80 |\kappa'_{2}|^{2} +\frac{9 }{2}g^4-75 \kappa_2 g^2 ,\\
    \end{split}\MINISKIP
    %%%%%%%%%%%%%%%%% rho0
    \begin{split}
        16 \pi ^2 \beta_{\rho_{0}} &= \phantom{+} 180 \alpha  \kappa_{0}+36 \alpha  \kappa_{2}+40 \beta_{4} \kappa_{0}+8 \beta_{4} \kappa_{2}+80 \beta'_{4} \kappa_{0}+16 \beta'_{4} \kappa_{2}+24 |\zeta|^{2}+44 h_{2} \rho_{0}+\\
        &\quad +20 h_{2} \rho_{2}+8 h'_{2} \rho_{0}+8 h'_{2} \rho_{2}+48 h_{3} \rho'^{\ast}_{0}+48 h^{\ast}_{3} \rho'_{0}+508 \lambda_{0} \rho_{0}+252 \lambda_{0} \rho_{2}+1220 \lambda_{2} \rho_{0}+\\
        &\quad +692 \lambda_{2} \rho_{2}+1340 \lambda_{4} \rho_{0}+788 \lambda_{4} \rho_{2}+2480 \lambda'_{4} \rho_{0}+1232 \lambda'_{4} \rho_{2}+4 \rho_{0}^2+16 |\rho'_{0}|^{2}+4 \rho_{2}^2+\\
        &\quad +24 |\psi_{1}|^{2}+96 |\psi_{2}|^{2}+9984 |\varphi|^{2}+1536 |\varphi'|^{2} + \frac{15 }{2} g^4-102 \rho_0 g^2 ,\\
    \end{split}\MINISKIP
    %%%%%%%%%%%%%%%%% rho2
    \begin{split}
        16 \pi ^2 \beta_{\rho_{2}} &= -24 |\zeta|^{2}+24 |\zeta'|^{2}+4 h_{2} \rho_{2}-8 h'_{2} \rho_{2}+4 \lambda_{0} \rho_{2}-164 \lambda_{2} \rho_{2}-236 \lambda_{4} \rho_{2}+16 \lambda'_{4} \rho_{2}+\\
        &\quad +8 \rho_{0} \rho_{2}+4 \rho_{2}^2+96 |\psi_{0}|^{2}-96 |\psi_{2}|^{2}-8448 |\varphi|^{2}+8448 |\varphi'|^{2}-102 \rho_2 g^2 ,\\
    \end{split}
    %%%%%%%%%%%%%%%%%
\end{align}
\endgroup
%%%%%%%%%%%%%%%%%%%%%%%%%%%%%%%%%%%%%%%%%%%%%%%%%%%%%%%%%%%%%%%%%%%%
\noindent
and, similarly, for the ``new'' complex ones
\begingroup
\allowdisplaybreaks 
\begin{align}
    %%%%%%%%%%%%%%%%% h4
    \begin{split}
        16 \pi ^2 \beta_{h_{4}} &= \phantom{+} 24 h_{2} h_{4}+96 h'_{2} h_{4}+18 h_{3}^2+90 \kappa'^{2}_{0}+36 \kappa'_{0} \kappa'_{2}+18 \kappa'^{2}_{2}+126 \rho'^{2}_{0}+126 \psi^{\ast}_{0} \psi_{2}-54 h_4 g^2 ,\\
    \end{split}\MINISKIP
    %%%%%%%%%%%%%%%%% h3
    \begin{split}
        16 \pi ^2 \beta_{h_{3}} &= \phantom{+} 72 h_{2} h_{3}+72 h'_{2} h_{3}+144 h^{\ast}_{3} h_{4}+180 \kappa_{0} \kappa'_{0}+36 \kappa_{0} \kappa'_{2}+36 \kappa'_{0} \kappa_{2}+36 \kappa_{2} \kappa'_{2}+252 \rho_{0} \rho'_{0}+\\
        &\quad +126 \rho'_{0} \rho_{2}+126 \psi^{\ast}_{0} \psi_{1}+126 \psi^{\ast}_{1} \psi_{2}-54 h_3 g^2 ,\\
    \end{split}\MINISKIP
    %%%%%%%%%%%%%%%%% kappa0p
    \begin{split}
        16 \pi ^2 \beta_{\kappa'_{0}} &= \phantom{+} 252 \alpha  \rho'_{0}+376 a_{0} \kappa'_{0}+72 a_{0} \kappa'_{2}+76 a_{2} \kappa'_{0}+8 a_{2} \kappa'_{2}+56 \beta_{4} \rho'_{0}+112 \beta'_{4} \rho'_{0}-28 \gamma_{2} \psi^{\ast}_{0}-\\
        &\quad -28 \gamma^{\ast}_{2} \psi_{2}+84 \zeta  \zeta'^{\ast}_{}+4 h_{2} \kappa'_{0}+40 h'_{2} \kappa'_{0}+8 h'_{2} \kappa'_{2}+24 h_{3} \kappa_{0}+4 h_{3} \kappa_{2}+96 h_{4} \kappa'^{\ast}_{0}+\\
        &\quad +16 h_{4} \kappa'^{\ast}_{2}+16 \kappa_{0} \kappa'_{0}+8 \kappa_{2} \kappa'_{2}-75  \kappa_0' g^2,\\
    \end{split}\MINISKIP
    %%%%%%%%%%%%%%%%% kappa2p
    \begin{split}
        16 \pi ^2 \beta_{\kappa'_{2}} &= \phantom{+} 16 a_{0} \kappa'_{2}+36 a_{2} \kappa'_{2}+140 \gamma_{2} \psi^{\ast}_{0}+140 \gamma^{\ast}_{2} \psi_{2}-84 \zeta  \zeta'^{\ast}_{}+4 h_{2} \kappa'_{2}+4 h_{3} \kappa_{2}+16 h_{4} \kappa'^{\ast}_{2}+\\
        &\quad +16 \kappa_{0} \kappa'_{2}+16 \kappa'_{0} \kappa_{2}+40 \kappa_{2} \kappa'_{2}-75  \kappa_2' g^2,\\
    \end{split}\MINISKIP
    %%%%%%%%%%%%%%%%% zeta
    \begin{split}
        16 \pi ^2 \beta_{\zeta}&= \phantom{+} 8 \alpha  \zeta +16 a_{0} \zeta -8 a_{2} \zeta +64 \beta_{4} \zeta +160 \beta_{4} \varphi +48 \beta'_{4} \zeta -640 \beta'_{4} \varphi +96 \gamma_{2} \zeta'^{\ast}_{}+8 \zeta  \kappa_{0}-8 \zeta  \kappa_{2}+\\
        &\quad +2 \zeta  \rho_{0}-4 \zeta  \rho_{2}+24 \zeta^{\ast}_{} \psi_{2}+16 \zeta'_{} \kappa'_{0}-16 \zeta'_{} \kappa'_{2}+4 \zeta'_{} \rho'_{0}+12 \zeta'^{\ast}_{} \psi_{1}-99 \zeta  g^2,\\
    \end{split}\MINISKIP
    %%%%%%%%%%%%%%%%% zetap
    \begin{split}
        16 \pi ^2 \beta_{\zeta'_{}} &= \phantom{+} 8 \alpha  \zeta'_{}+16 a_{0} \zeta'_{}-8 a_{2} \zeta'_{}+64 \beta_{4} \zeta'_{}+160 \beta_{4} \varphi'_{}+48 \beta'_{4} \zeta'_{}-640 \beta'_{4} \varphi'_{}+96 \gamma_{2} \zeta^{\ast}_{}+16 \zeta  \kappa'^{\ast}_{0}-\\
        &\quad -16 \zeta  \kappa'^{\ast}_{2}+4 \zeta  \rho'^{\ast}_{0}+12 \zeta^{\ast}_{} \psi_{1}+8 \zeta'_{} \kappa_{0}-8 \zeta'_{} \kappa_{2}+2 \zeta'_{} \rho_{0}+6 \zeta'_{} \rho_{2}+24 \zeta'^{\ast}_{} \psi_{0}-99 \zeta' g^2,\\
    \end{split}\MINISKIP
    %%%%%%%%%%%%%%%%% rho0p
    \begin{split}
        16 \pi ^2 \beta_{\rho'_{0}} &= \phantom{+} 180 \alpha  \kappa'_{0}+36 \alpha  \kappa'_{2}+40 \beta_{4} \kappa'_{0}+8 \beta_{4} \kappa'_{2}+80 \beta'_{4} \kappa'_{0}+16 \beta'_{4} \kappa'_{2}+12 \zeta  \zeta'^{\ast}_{}+4 h_{2} \rho'_{0}+40 h'_{2} \rho'_{0}+\\
        &\quad +24 h_{3} \rho_{0}+12 h_{3} \rho_{2}+96 h_{4} \rho'^{\ast}_{0}+508 \lambda_{0} \rho'_{0}+1220 \lambda_{2} \rho'_{0}+1340 \lambda_{4} \rho'_{0}+2480 \lambda'_{4} \rho'_{0}+\\
        &\quad +8 \rho_{0} \rho'_{0}+4 \rho'_{0} \rho_{2}+24 \psi^{\ast}_{0} \psi_{1}+24 \psi^{\ast}_{1} \psi_{2}+5760 \varphi  \varphi'^{\ast}_{}-102\rho_0' g^2 ,\\
    \end{split}\MINISKIP
    %%%%%%%%%%%%%%%%% psi2
    \begin{split}
        16 \pi ^2 \beta_{\psi_{2}} &= \phantom{+} 32 \gamma_{2} \kappa'_{2}+12 \zeta ^2+440 \eta_{2} \psi^{\ast}_{0}+4 h_{2} \psi_{2}+4 h_{3} \psi_{1}+16 h_{4} \psi_{0}+4 \lambda_{0} \psi_{2}+40 \lambda_{2} \psi_{2}+180 \lambda_{4} \psi_{2}+\\
        &\quad +160 \lambda'_{4} \psi_{2}+8 \rho_{0} \psi_{2}+8 \rho'_{0} \psi_{1}-16 \rho_{2} \psi_{2}+3840 \varphi^2-102 \psi_2 g^2 ,\\
    \end{split}\MINISKIP
    %%%%%%%%%%%%%%%%% psi1
    \begin{split}
        16 \pi ^2 \beta_{\psi_{1}} &= \phantom{+} 32 \gamma_{2} \kappa_{2}+24 \zeta  \zeta'_{}+440 \eta_{2} \psi^{\ast}_{1}+4 h_{2} \psi_{1}+8 h'_{2} \psi_{1}+8 h_{3} \psi_{0}+8 h^{\ast}_{3} \psi_{2}+4 \lambda_{0} \psi_{1}+40 \lambda_{2} \psi_{1}+\\
        &\quad +180 \lambda_{4} \psi_{1}+160 \lambda'_{4} \psi_{1}+8 \rho_{0} \psi_{1}+16 \rho'_{0} \psi_{0}+16 \rho'^{\ast}_{0} \psi_{2}+4 \rho_{2} \psi_{1}+7680 \varphi  \varphi'_{}-102\psi_1 g^2 ,\\
    \end{split}\MINISKIP
    %%%%%%%%%%%%%%%%% psi0
    \begin{split}
        16 \pi ^2 \beta_{\psi_{0}} &= \phantom{+} 32 \gamma_{2} \kappa'^{\ast}_{2}+12 \zeta'^{2}+440 \eta_{2} \psi^{\ast}_{2}+4 h_{2} \psi_{0}+4 h^{\ast}_{3} \psi_{1}+16 h^{\ast}_{4} \psi_{2}+4 \lambda_{0} \psi_{0}+40 \lambda_{2} \psi_{0}+180 \lambda_{4} \psi_{0}+\\
        &\quad +160 \lambda'_{4} \psi_{0}+8 \rho_{0} \psi_{0}+8 \rho'^{\ast}_{0} \psi_{1}+24 \rho_{2} \psi_{0}+3840 \varphi'^{2}-102 \psi_0 g^2,\\
    \end{split}\MINISKIP
    %%%%%%%%%%%%%%%%% varphi
    \begin{split}
        16 \pi ^2 \beta_{\varphi} &= \phantom{+} \beta_{4} \zeta -4 \beta'_{4} \zeta +12 \lambda_{0} \varphi -8 \lambda_{2} \varphi +40 \lambda_{4} \varphi -656 \lambda'_{4} \varphi +6 \rho_{0} \varphi +12 \rho'_{0} \varphi'_{}-8 \rho_{2} \varphi +\\
        &\quad +48 \psi_{2} \varphi^{\ast}_{}+24 \psi_{1} \varphi'^{\ast}_{}-126 \varphi g^2 ,\\
    \end{split}\MINISKIP
    %%%%%%%%%%%%%%%%% varphip
    \begin{split}
        16 \pi ^2 \beta_{\varphi'_{}} &= \phantom{+} \beta_{4} \zeta'_{}-4 \beta'_{4} \zeta'_{}+12 \lambda_{0} \varphi'_{}-8 \lambda_{2} \varphi'_{}+40 \lambda_{4} \varphi'_{}-656 \lambda'_{4} \varphi'_{}+6 \rho_{0} \varphi'_{}+12 \rho'^{\ast}_{0} \varphi +14 \rho_{2} \varphi'_{}+\\
        &\quad +24 \psi_{1} \varphi^{\ast}_{}+48 \psi_{0} \varphi'^{\ast}_{}-126 \varphi' g^2 .\\
    \end{split}\label{eq:couplings-RGE-end}
    %%%%%%%%%%%%%%%%%
\end{align}
\endgroup
Several remarks may again be worth making here:
\begin{itemize}
    \item In the full $\mathbf{45} \oplus \mathbf{126} \oplus \mathbf{10}_\mathbb{C}$ scenario the $\beta$-functions of the ``old'' couplings relevant for the $\mathbf{45} \oplus \mathbf{126}$ Higgs model of Ref.~\cite{Jarkovska:2021jvw} receive contributions from the ``new'' couplings as well. Note that some minor errors present in the expressions for $\beta_{\lambda_2}$, $\beta_{\lambda_4}$, $\beta_{\gamma_2}$ and $\beta_{\eta_2}$ in \cite{Jarkovska:2021jvw} have been corrected. It has also been verified that the corresponding changes have only very minor quantitative effects and the results of~\cite{Jarkovska:2021jvw} remain unchanged.
    \item The beta functions of all real scalar couplings are manifestly real, as they should be.
    \item The beta functions of complex couplings (associated to the presence of $2$ complex scalar representations $\Sigma$ and $H$) exhibit the standard patterns associated to phase redefinitions of complex representations, which greatly restricts the form the individual terms within might take. To illustrate this, one can consider e.g. the phase redefinition of $\Sigma\to e^{i\epsilon} \Sigma$; absorbing the phases that appear in the scalar potential into the (complex) couplings, one can verify readily that the system of the RGE's above will change consistently with this (unphysical) transformation. A systematic way to track this is to define a $\mathrm{U}(1)_\Sigma$ transformation, under which the charge of $\Sigma$ is {\em defined} to be $+1$, with all other scalar representations remaining neutral and the charges of all couplings (playing the role of spurions) set so that the scalar potential of Eqs.~\eqref{eq:scalar-potential-0}--\eqref{eq:scalar-potential-Vtildemix} is $\mathrm{U}(1)_\Sigma$-invariant. Analogously, one can define another phase transformation $\mathrm{U}(1)_H$ acting solely on $H$ (with the associated charge $+1$) and set the $\mathrm{U}(1)_H$ charges of all couplings so that the scalar potential is again $\mathrm{U}(1)_H$-invariant. The corresponding $\mathrm{U}(1)_\Sigma\times\mathrm{U}(1)_H$ charges of all dimensionless scalar couplings are shown in Table~\ref{tab:U(1)SxU(1)H}.
    \begin{table}[htb]
    \caption{Non-trivial charge assignments of the dimensionless scalar couplings under $\mathrm{U}(1)_\Sigma\times\mathrm{U}(1)_H$. \label{tab:U(1)SxU(1)H}}
\centering
    \begin{tabular}{l@{\quad}r@{\quad}r@{\quad}r@{\quad}r@{\quad}r@{\quad}r@{\quad}r@{\quad}r@{\quad}r@{\quad}r@{\quad}r@{\quad}r@{\quad}r@{\quad}r}
        \toprule
        & $\gamma_{2}$ & $\eta_2$ & $h_4$ &  $h_3$ & $\kappa'_0$ & $\kappa'_{2}$ & $\zeta$ & $\zeta'$ & $\rho'_{0}$ & $\psi_2$ & $\psi_1$ & $\psi_0$ & $\varphi$ &$\varphi'$\\
          \midrule
        $\mathrm{U}(1)_\Sigma$
        &$-2$&$-4$&$+0$&$+0$&$+0$&$+0$&$-1$&$-1$&$+0$&$-2$&$-2$&$-2$&$-1$&$-1$\\
        $\mathrm{U}(1)_H$
        &$+0$&$+0$&$-4$&$-2$&$-2$&$-2$&$-1$&$+1$&$-2$&$-2$&$+0$&$+2$&$-1$&$+1$\\
          \bottomrule
    \end{tabular}
    \end{table}
    Since each beta function $\beta_{\lambda}$ in the list above should have the same $\mathrm{U}(1)_\Sigma\times\mathrm{U}(1)_H$ charge as the corresponding coupling $\lambda$, all terms in $\beta_{\lambda}$ must have that same charge; for instance, all terms of $\beta_{\gamma_{2}}$ must have the \hbox{$\mathrm{U}(1)_\Sigma\times\mathrm{U}(1)_H$} charge $(-2,0)$, as they do. One can indeed confirm that all structures in Eqs.~\eqref{eq:couplings-RGE-begin}--\eqref{eq:couplings-RGE-end} do adhere to this rule. This pattern leads to a number of interesting consequences:
    \begin{itemize}
        \item Given the information in Table~\ref{tab:U(1)SxU(1)H}, the scalar potential (with couplings promoted to fields) is invariant under \hbox{$\mathrm{U}(1)_\Sigma\times\mathrm{U}(1)_H$}, and hence also under any linear combination of the two $\mathrm{U}(1)$ factors. In particular, we can define the PQ charge of a field/coupling $x$ as \begin{align}
            [x]_{PQ}&:=2[x]_{\Sigma}-2[x]_H,
        \end{align}
        which is indeed consistent with Eq.~\eqref{eq:PQ-assignments}. This also implies that the $\beta$-functions above must preserve the PQ charge thus defined. 
        \item  Coming back to the plain scalar potential with couplings interpreted as mere parameters rather than spurions, only those  with a zero PQ charge are allowed if $U(1)_{PQ}$ is supposed to be a good symmetry of the model. In such a case, besides the real couplings, the only allowed complex ones are $\zeta$, $\psi_{2}$ and $\varphi$; at the same time, the system of RGE's consistently reduces to a subsystem for these PQ-allowed couplings only, while the beta functions of all the remaining (PQ-charged) couplings vanish identically. Analogous considerations apply for an alternative PQ-like charge $[x]_{PQ'}:=2[x]_\Sigma+2[x]_H$, or more generally for any linear combination of the primary charges $\mathrm{U}(1)_\Sigma$ and $\mathrm{U}(1)_H$. 
        %%%%%%%%%%%%%%%%%%%%%%%%%%
        \item A coupling has a non-trivial $\mathrm{U}(1)_\Sigma\times\mathrm{U}(1)_H$ charge if and only if it is complex. Since the gauge coupling $g$ is real, it must have charge $(0,0)$, and thus the $g^4$ contributions must vanish in the $\beta$-functions of all complex couplings, as confirmed by the results above. Interestingly, the $g^4$ term also vanishes in the $\beta$-function of the real parameter $\rho_{2}$ with no specific insight from this type of arguments.
        %%%%%%%%%%%%%%%%%%%%%%%%%%
        \item Charge conservation provides non-trivial constraints in all beta functions, including those for real couplings. If a coupling is real, its beta function must be both real and $\mathrm{U}(1)_\Sigma\times\mathrm{U}(1)_H$ neutral, e.g.~a contribution like $\kappa'_{0}\kappa'_{2}{}^{\ast}+c.c.$ is allowed but $\kappa'_{0}\kappa'_{2}+c.c.$ is not. 
    \end{itemize}
\end{itemize}

%%%%%%%%%%%%%%%%%%%%%%%%%%%%%%%%%%%%%%%%%%%%%%%%%%%%%%%%%%%%%%%%%%%%%%%%%%%%%%%%
% % %   BIBLIOGRAPHY
%\bibliographystyle{apsrev4-1}
%\bibliography{bib1}
\input{bib1.bbl}

%%%%%%%%%%%%%%%%%%%%%%%%%%%%%%%%%%%%%%%%%%%%%%%%%%%%%%%%%%%%%%%%%%%%%%%%%%%%%%%%%%
%%%%%%%%%%%%%%%%%%%%%%%%%%%%%%%%%%%%%%%%%%%%%%%%%%%%%%%%%%%%%%%%%%%%%%%%%%%%%%%%%%
\end{document}